\documentclass[twocolumn]{aa}

\usepackage[flushleft]{threeparttable}
\usepackage{graphicx}
\usepackage{amsmath}
\usepackage{subfigure}
\usepackage{amssymb}
\usepackage{tabularx}
\usepackage{natbib}
\usepackage{float}
\usepackage{etoolbox}
\usepackage{amsmath}
\usepackage{color}
\usepackage{appendix}
\usepackage{multirow}
\usepackage{enumerate}
\usepackage[normalem]{ulem}
\usepackage{txfonts}

\def\Msun{\mathrm{M}_\odot}

\newcommand{\bea}{\begin{eqnarray}}
\newcommand{\eea}{\end{eqnarray}}

\definecolor{darkgreen}{rgb}{0, .5, 0}

\newcommand{\argdf}{\texttt{ARGdf}~}
\newcommand{\ARGdf}{\texttt{ARGdf}~}
\newcommand{\archain}{\texttt{ARCHAIN}~}
\newcommand{\Ms}{~{\rm M}_\odot}
\newcommand{\gw}{{\rm GW}}

\newcommand{\au}{~{\rm AU}}

\newcommand{\bhb}{{\rm BHB}}
\newcommand{\der}{{\rm d}}

\date{Started July 10, 2018; Revised to \today}

\graphicspath{{./}{./}}
\begin{document}

   \title{Order in the chaos:}

   \subtitle{eccentric black hole binary mergers in triples formed via strong binary-binary scatterings}

\author{Manuel ~Arca Sedda\inst{1} \and Gongjie ~Li\inst{2} \and Bence ~Kocsis\inst{3}}

\institute{
Astronomisches Rechen-Institut\\
Zentrum f\"{u}r Astronomie der Universit\"{a}t Heidelberg\\
M\"onchhofstrasse 12-14,Heidelberg, D-69120, DE
\email{m.arcasedda@gmail.com}
\and
Center for Relativistic Astrophysics, School of Physics\\
Georgia Institute of Technology\\
Atlanta, GA 30332 (USA)
\and
University of Oxford\\
Rudolf Peierls Centre for Theoretical Physics, Clarendon Laboratory, Parks Road, Oxford OX1 3PU, UK
}

\titlerunning{eccentric black hole binary mergers in non-hierarchical triples}
\authorrunning{Arca-Sedda, Li, and Kocsis}

\abstract
{Black hole (BH) triples represent one of the astrophysical pathways for BH mergers in the Universe detectable by LIGO and VIRGO. We study the formation of BH triples via binary-binary encounters in dense clusters, showing that 
up two third of triples formed through this channel are hierarchical whereas the remaining are in a non-hierarchical, unstable configuration. We build a database of $32,000$ $N$-body simulations to investigate the evolution of BH triples focusing on mildly hierarchical and non-hierarchical unstable configurations. Varying the mutual orbital inclination, the three BH masses and the inner and outer eccentricities, we show that retrograde, nearly planar configurations lead to a significant shrinkage of the inner binary. We find an universal trend of triple systems, namely that they tend to evolve toward prograde configurations and that the orbital flip, driven by the torque exerted on the inner BH binary (BHB) by the outer BH, leads in general to tighter inner orbits. In some cases, the resulting BHB undergoes coalescence within a Hubble time, releasing gravitational waves (GWs). A large fraction of merging BHBs with initial separation $1$ AU enter the $10^{-3}-10^{-1}$ Hz frequency band with large eccentricities, thus representing potential eccentric LISA sources. Mergers originating from initially tighter BHB ($a\sim 0.01$ AU), instead, often have eccentricities above 0.7 in the $1$ Hz band. We find that the mergers' mass distribution in this astrophysical channel maps the original BH binary spectrum. This might have interesting consequences in light of the growing population of BH mergers detected by LIGO and VIRGO, namely that eccentric sources detected in high-frequency detectors are most likely connected with a high-velocity dispersion stellar environment, whereas eccentric sources detected in low-frequency detectors are likely to develop in low-density clusters. 
}


\maketitle

\section{Introduction}

The recent detection of gravitational waves (GWs) produced by the coalescence of two stellar mass black holes (BHs) \citep{abbott16a,abbott16b,abbott16c,abbott17a,abbott17b,abbott19} opened a series of questions about the processes driving these cosmic catastrophic phenomena. 
One of the possible formation channels for merging BHBs is through dynamical interactions taking place in the heart of dense stellar systems. Indeed, encounters between a binary and single stars can lead to the shrinking of the binary separation in crowded stellar environments until the binary reaches a separation where it is driven to merge by GW emission. The \textit{dynamical channel}
provides a merger rate compatible with the value inferred from the recent  observations provided by the LIGO-Virgo-Kagra collaboration (LVC) \citep{rodriguez15,rodriguez16,askar17,Fragione_Kocsis18}. This suggests that some of  the observed merging BHBs might have formed in open clusters \citep{rastello18,kumamoto19,dicarlo19,rastello20}, globular clusters \citep{zwart00,wen03,downing10,rodriguez18b}, young massive clusters \citep{mapelli16,banerjee16,banerjee17} or nuclear clusters \citep{oleary09,antonini16b,bartos16,ASCD17b,Hoang18}. Alternatively, isolated binary evolution also provides an explanation for the observed BHB merger rate \citep{belczynski16,belckzynski17}. Furthermore, BH mergers in AGN also possibly explain the observed mergers \citep{bartos16,Tagawa2019}.
These three scenarios possibly leave different signatures in the properties of observed mergers \citep{OLeary2016,gerosa17,ArcaBen19,Yang2019a,Yang2019b,Tagawa2020,ArcaEtAl20}, although assessing a clear criterion to disentangle isolated and dynamically formed mergers is hard due to the small number of detections. Currently, the database of BHB mergers detected by the LVC \citep{GWTC2} is not sufficiently rich to enable us to distinguish between different channels \citep{GWTC2bis,zevin20b}, although there is already clear evidence of a dynamical origin for some of the sources, like GW170729 \citep{LIGO19} or GW190521 \citep{GWTC2}, the first BHB with one component in the so-called upper mass-gap \citep{ArcaEtAl20,gaya20,kimball20,fragione20b,baibhav20}.

The simplest way to efficiently shrink a binary system in a dense stellar environment is through three- and four- body interactions \citep{heggie75,hut92,pooley03}. These interactions have been studied extensively in the hierarchical limit, which consists of a close binary and a distant perturber \citep[e.g.,][]{Antonini17,Hoang18,Rodriguez18,Grishin18}. In this regime, the Kozai-Lidov (KL) mechanism \citep{kozai62,lidov62} (for a review, see \citealt{naoz16}) can significantly enhance the BHB merger rate. In the case of number densities $\sim 10^5-10^6$ pc$^{-3}$, typical of dense clusters, a population of stable hierarchical BH triples is expected to form over a Hubble time through binary-binary interactions. Indeed, this kind of close encounters have a $20-50\%$ chance to leave behind a hierarchical triple \citep{sigurdsson93,kulkarni93,miller02} and contribute to the ovearall BHB merger rate \citep[e.g.]{antonini16,Zevin18,martinez20,fragione20}. 

Furthermore, transient non-hierarchical triples may be common in dense stellar environments, and may play a major role in the formation of eccentric binary mergers \citep{samsing14,samsing17,samsing18}. For instance, binary-single BH encounters commonly lead to closely separated triple systems in low velocity dispersion environments, such as in stellar clusters \citep{Hut83}. 

While the evolution of hierarchical triples has been widely studied in earlier works, the evolution of non-hierarchical triple systems is still poorly investigated, with a few notable exceptions \citep[e.g.]{antonini16,fragione20,Manwadkar20}. Recently, \citet{Li18} have found that non-hierarchical triple interactions in the outer Solar System can lead to eccentricity excitation and orbital flips of trans Neptunian objects (TNOs) when they are perturbed by a near coplanar Planet Nine, analogous to the hierarchical triple interactions \citep{Li14}. It is not clear how the non-hierarchical triple interactions could affect the BHB merger rates. For a first step in exploring this subject, we focus on the dynamics of non-hierarchical triples in this article.

Assessing the properties of BHB mergers developing in different classes of triples can provide valuable information on their evolutionary pathway. The presence of the third object may be discovered directly with the GWs in some cases \citep{Meiron17}. Alternatively, the BHB eccentricity may carry information on the perturber, which can be measured from the observed GW signal \citep{gondan17,gondan18,gondan19}. It was shown that a BHB perturbed by a farther companion in a hierarchical configuration leads to GWs in the 10 Hz frequency band with a moderately high eccentricity, thus being potentially observable by current ground-based detectors \citep{wen03,OLeary06,antonini12,OLeary16,antonini16,samsing18b,gondan17,gondan18,gondan19, liu20}. For a binary similar to GW150914, high accuracy is expected for eccentricity measurement with LIGO/VIRGO/KAGRA at design sensitivity, of the order of $0.001$--$0.01$ at 10 Hz \citep{gondan19}. Hierarchical triple companions may also be detected through eccentricity oscillations using GWs in the mHz regime with LISA \citep{randall19,hoang19,Deme2020}.  In the coming years, this will allow one to obtain a clean view on the eccentricity distribution of merging stellar BHBs, thus providing an unique tool to constrain their formation history. 

In this paper we use a multi-step approach to study the formation and evolution of hierarchical, mildly hierarchical, and short-lived triples forming in dense star clusters. As a first step, we use the MOCCA database of globular cluster Monte Carlo simulations \citep{giersz15,askar17} to reconstruct the history of 
binary-binary scattering in globular clusters. Using these data, we built a sample of strong binary-binary scattering via few-body simulations, showing that a high percentage produce either long-lived or temporary triples that, in some cases, result in the merger of the inner BHB. We explore the parameter space by means of 32,000 simulations, investigating the impact that inner and outer eccentricities, the orbital inclination, and the triple component's mass has on the formation of BHB mergers. We show that when the triple is in a retrograde configuration the tidal perturbation induced by the outer object typically tends to cause the inner orbit to flip as well as to increase its eccentricity, which may ultimately result in the merger of the inner binary due to GW emission.

The paper is organized in three main parts. In the first, namely Section \ref{sec:GC} we present a sample of 2000 simulated BHB-BHB scattering experiments to determine the efficiency of such interactions in producing BH triples together with their expected parameters. We tailor a sample of 1000 triples based on the scattering experiments and discuss the properties of BHB mergers developing through this channel in Section \ref{sec:BBHfull}. In the second part, Section \ref{sec:an}, we focus on the evolution of non-hierarchical triples, presenting $\sim 3\times 10^4$ numerical simulations and showing that this class of triples has well defined evolutionary features. In the last part, Section \ref{sec:gw}, we discuss whether BHB mergers forming through this channel can be seen with ground based and space based GW observatories, and whether their observations can tell us something about their nursing environments. Section \ref{sec:end} is devoted to summarize the conclusions of the work.

\section{Numerical method}

To explore the formation mechanisms and evolution of BH triples we exploit over 32,000 direct 3- and 4-body simulations gathered in 15 sets that can be classified as:
\begin{itemize}
\item {\bf MOCC}: BHB-BHB strong scattering experiments with initial conditions extracted directly from the MOCCA models;
\item {\bf BIN0-1-2}: BHB-BHB strong scattering experiments that we use to assess the formation probability of triple BHs;
\item {\bf SET0}: BH-BHB triples with initial conditions extracted from one of the BHB-BHB scattering simulations set;
\item {\bf SET1-3}: BH-BHB triples with initial conditions tailored to explore the role of the inner BHB eccentricity and longitude of pericentre;
\item {\bf SET5-8}: BH-BHB triples with initial conditions tailored to explore the role of retro-/pro-grade configuration.
\item {\bf SET4-9-10}: BH-BHB triples with simplified initial conditions that serve to understand the role of component masses and orbital configuration in determining the development of BHB mergers;
\end{itemize}

In section \ref{sec:GC} we show that binary scatterings can: a) form triples with a large efficiency, b) around one third of triples formed via this channel are non-hierarchical, c) non-hierarchical triples contribute to around $10\%$ of BHB mergers formed via this channel, d) around $75\%$ of non-hierarchical mergers have high eccentricity in the LISA band. 

In section \ref{sec:an}, instead, we discuss the impact of the orbital properties of non-hierarchical and mildly hierarchical systems in determining the final fate of the most bound BHB.

All the simulations are performed using \ARGdf \citep{ASCD17b}, a modified version of the \archain code \citep{mikkola08}. The code implements general relativity effect via post-Newtonian formalism up to 2.5 order and treats strong gravitational encounters taking advantage of the algorithmic regularization scheme \citep{mikkola99}. Our modified version allows the user to include, in the particle equations of motion, a dynamical friction term, modelled according to a modification of the classical \cite{Cha43I} treatment \citep[see][]{ASCD17b}, and the acceleration due to the external field of the background system hosting the particles. For our purposes, these two terms are neglected, as we focus on the evolution of triple systems on length scales much smaller than the typical size of star clusters. In all the simulations the accrued relative energy error remains below a level of $10^{-9}$. 

\section{Formation of non-hierarchical triples via binary-binary scattering in globular clusters}
\label{sec:GC}

One of the possible pathways to the formation of triple BHs is via close encounters \citep{hut93} like binary-binary interactions \citep{mikkola84,mcmillan91,miller02}. One interesting class of encounters, called {\it resonant} scattering, has proven to form short-lived triple that 
ultimately break up, ejecting one of the components and leaving behind a binary after 10-100 crossing times \citep{hut93}.
Three-body and binary-binary scattering encounters may commonly take place in dense stellar environments, like the inner regions of globular \citep[GCs, see for instance][]{miller02,Zevin18,fragione20,martinez20} or nuclear clusters \citep[NCs, see for instance][]{miller09,arca20}. The study of these interactions in numerical star cluster models has been limited by the computational complexity of modelling million-body systems. Indeed, while a one-to-one modelling of clusters with $\leq 10^4$ members has been possible for decades using $N$-body simulations, million body simulations became possible for GCs only recently \citep{sippel12,sippel13,contenta13,AS16,wang16,banerjee16}, and numerical models of galactic nuclei have also improved significantly \citep{ASK18,Panamarev18}. 
Monte Carlo models provide an alternative to direct $N$-body models, albeit at the cost of important approximations like spherical symmetry and zero rotation. 
The relatively short computational times led to the development of a large database of Monte Carlo models which have been used to predict the evolution of black holes \citep{Morscher15,Arca18,Askar18,Weatherford18} and the possible formation of BHB mergers \citep{rodriguez15,rodriguez16,askar17,kremer20}. These models allowed to demonstrate that dense GCs can retain a sizeable fraction of BHs over a Hubble time, up to $30-50\%$ \citep{Morscher15,Arca18}. The majority of these BHs are expected to sit in the innermost cluster regions and form a BH subsystem \citep[BHS, see e.g.][]{SPITZER,breen13a,AS16,Arca18,Weatherford18,weatherford20}.

\begin{figure}
\centering
\includegraphics[width=8cm]{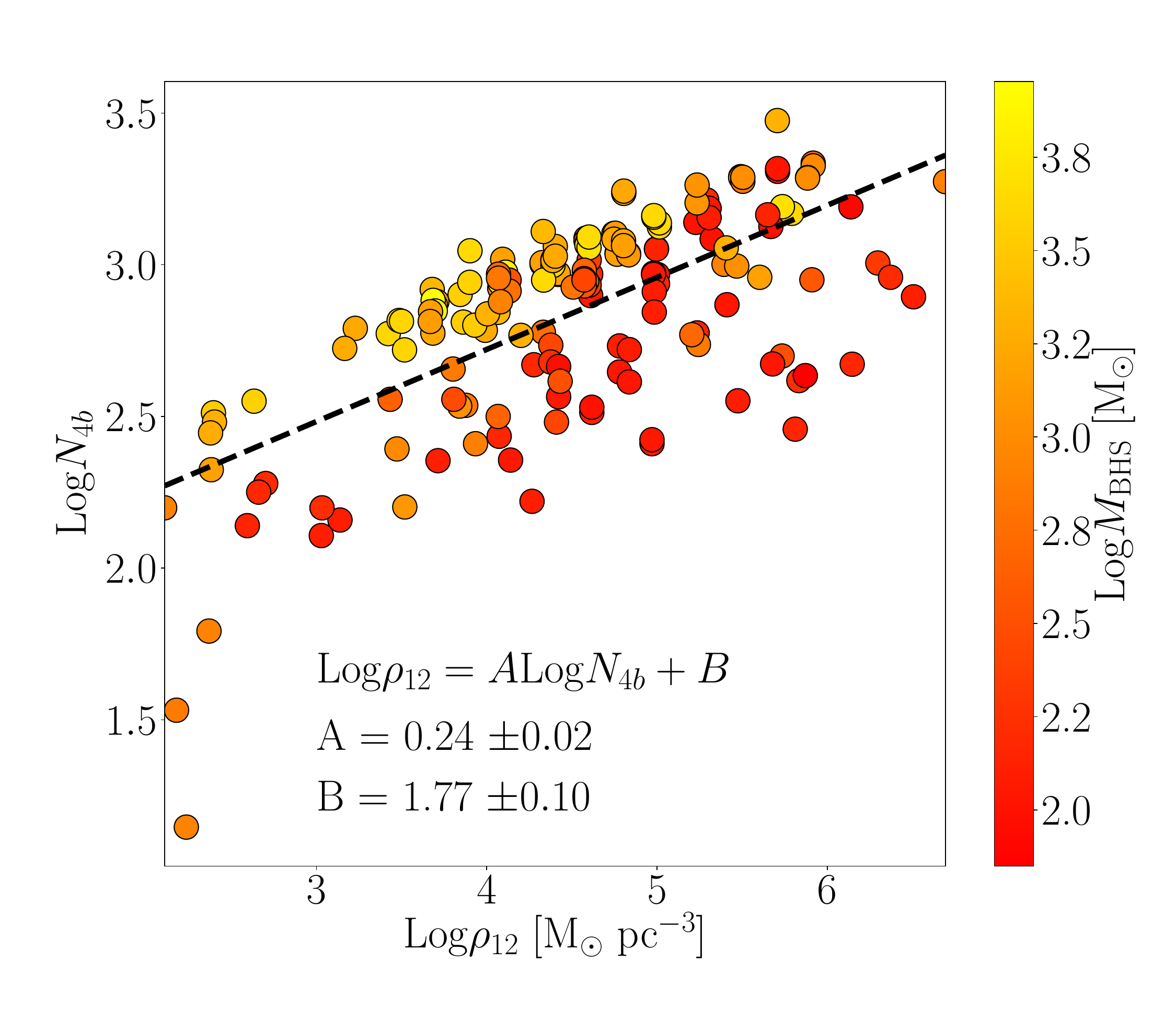}
\caption{Total number of BHB-BHB interactions as a function of the host GC central density at 12 Gyr in the MOCCA database. Colour coding shows the mass of the BH subsystem in the host globular cluster core.}
\label{den4B}
\end{figure}

Although Monte Carlo techniques implement a treatment to model binary-single and binary-binary encounters, they capture the complex outcomes of such interactions only partly. In fact, usually these methods artificially disrupt any triples or multiples formed out of gravitational encounters if their lifetime exceeds a fixed computational time \citep{pattabiraman13,hypki13}. To cope with this limitation, several works modelled the long-term evolution of multiples formed in binary-single and binary-binary scattering through a post-processing of Monte Carlo simulations \citep[e.g.][]{antonini16,samsing18b, Zevin18,fragione20}.
Utilizing the MOCCA SURVEY DATABASE I, a suite of over 2000 Monte Carlo models of GCs with different properties modelled over a 12 Gyr time span we reconstruct the history of all BHB-BHB scatterings in all GC models and find that the total number of such interactions correlates with the final GC central density via a simple power-law
\begin{equation}
\log N_{\rm 4BH} = (0.24\pm 0.02) \log \frac{\rho_{\rm 12}}{\Ms {\rm pc}^{-3}} + (1.77\pm 0.010).
\end{equation}
In general, this relation indicates that a typical GC witnesses, on average, 900 interactions in 12 Gyr. Figure \ref{den4B} shows this correlation for the MOCCA models containing a BH subsystem \citep{Arca18,Askar18}, highlighting that the denser the cluster and the BH subsystem the larger the number of BHB-BHB interactions. Note that the number of scatterings seem to depend on the BHS mass, i.e. on the number of BHs retained in the host cluster. In particular, we see that at a fixed value of $\rho_{12}$ a heavier BHS favour a larger number of scattering.

We explore the MOCCA database to determine the properties of the binary scatterings in terms of dynamical parameters, namely the semimajor axes and eccentricities of the two binaries, the masses of the BHs involved, the position of the binaries during the scattering, the velocity dispersion of the cluster at the time of the scattering. We found that all scatterings are hyperbolic, and $97\%$ of them take place in environments with a velocity dispersion in the range $\sigma_c = [1-20]$ km s$^{-1}$, with more than $81\%$ being characterised by $\sigma_c=[5-15]$ km s$^{-1}$. Around $\gtrsim 79\%$ of the scatterings take place in a regime of strong deflection, meaning that the angle $\delta$ between the inward and outward directions of the binaries centre of mass exceeds 90 degrees, and involve BHs with masses larger than $M_{\rm BH} > 10 \Ms$. These interactions are the most energetic and have the largest probability to impinge a substantial evolution in one, or both, binaries and, in some cases, to trigger the formation of a triple \citep{bacon96,fregeau04,Zevin18}.  

In order to assess how frequently a triple might form from a binary-binary scattering, we create a sample of 500 binary-binary scatterings with properties extracted directly from MOCCA models, namely from the general distribution of parameters that characterise the binary-binary interaction, the binary component masses, semimajor axis, and eccentricity. We name this simulation set MOCC.

Our main purpose is to demonstrate that binary-binary interactions can form a significant amount of triples, a fraction of which will be in a non-hierarchical configuration. A detailed study on the outcomes of binary-binary interactions in MOCCA models and their implications for BH mergers, instead, is left to a follow-up work.

It is important to stress that the MOCCA database contains models spanning a wide range of initial conditions like the cluster initial mass and metallicity, the fraction of primordial binaries, the stellar prescription adopted, the location in the Galaxy, the initial concentration and depth of the potential well. Such an heterogeneous mixture of variables can impact the properties of binary-binary scatterings in a non-trivial way, possibly impinging some bias in the choice of the binary-binary configuration parameters.
Therefore, we carry out three additional simulation sets modelling binary-binary hyperbolic encounters with well defined properties to compare with MOCCA models. For these models, we assign binary velocities assuming that the velocity dispersion at the time of the scattering is either $\sigma_c = 5,10,15$ km s$^{-1}$, thus bracketing the values that characterize MOCCA models, performing 500 simulations for each value.  
In the following, we refer to these additional models as BIN0, BIN1 and BIN2 (see Table~\ref{tab:4body}).

\begin{table*}
\centering{}
\caption{Main properties of 4-body models}
\begin{center}
\begin{tabular}{ccccccccc}
\hline
model & $\sigma_c$ & $f_{\rm tri,end}$ & $f_{\rm BBH}$ & $N_{\rm tri}$ & $f_{\rm hie}$& $f_{\rm uns}$ & $f_{\rm octu}$ & $f_{\rm stab+oct}$ \\ 
 & km s$^{-1}$& $\%$& $\%$ & & $\%$& $\%$& $\%$ & $\%$\\
\hline
\hline
MOCC & 6.5& 31.2 & 69  & 736 & 58 &  30 & 72 & 28\\
BIN0 & 5  & 22.2 & 76.6& 547 & 49 &  34 & 69 & 31\\
BIN1 & 10 & 20.4 & 80.6& 369 & 53 &  37 & 68 & 32\\
BIN2 & 15 & 21.4 & 83.8& 640 & 43 &  39 & 66 & 34\\
\hline
\end{tabular}
\end{center}
\begin{tablenotes}
\item 
Col 1: model name. The first model refers to the set of simulations tailored on the MOCCA binary-binary scattering database. 
Col 2: cluster velocity dispersion. 
Col 3: percentage of models in which the endstate is a triple (normalized to $N_{\rm sim}$). 
Col 4: percentage of models in which the end-state is at least one BBH with merger time below 14 Gyr, smaller than the evaporation time and than its initial value. 
Col 5: Number of triples (temporary and end-state) formed $N_{\rm tri}$. 
Col 6: percentage of triples with the ratio between outer binary pericentre and inner binary semiaxis larger than 10 (normalized to $N_{\rm tri}$).
Col 7: percentage of triples that are unstable -- \citet{mardling01} criterion not fulfilled. 
Col 8: percentage of systems for which secular effects can develop, i.e. having the octupole parameter $\epsilon<0.1$. 
Col 9: percentage of systems that both satisfy the stability criterion $a_3/a < K$ and have $\epsilon<0.1$. 
\end{tablenotes}
\label{tab:4body}
\end{table*}

Whilst for simulations set MOCC we extract all the parameters of the hyperbolic trajectories from MOCCA modes, to initialize the binary scattering in sets BIN0-1-2 we define the interaction semi-major axis
\begin{equation}
    a_{4b} = -\mu/v_\infty^2,
\end{equation}
where $\mu = G\sum_i m_i$ the {\it gravitational parameter} and $v_\infty = v_{12} - v_{34}$ the relative velocity between the binaries. 
The two binaries' velocities are then drawn from a Maxwellian distribution characterized by dispersion $\sigma_c$ and used to calculate $v_\infty$.
A hyperbolic encounter can be also characterized through two further parameters, namely the impact parameter and the deflection angle. 
The former is defined as $b_{4b} = -a_{4b}\sqrt{e_{4b}^2-1}$, while the latter represents the angle $\delta$ between the inward and outward direction of binaries centre of mass motion, and is connected to the eccentricity parameter via equation
\begin{equation}
    e_{4b} = \frac{1}{\sin(\delta/2)},
\end{equation}
note that the limit of large deflection angles, $\delta>\pi/2$, corresponds to the strong deflection regime. 

We draw the initial impact parameter assuming that it follows a linear distribution, as expected from simple geometrical arguments, for an isotropic velocity distribution $f(b){\rm d}b\propto 2\pi b{\rm d}b$. We limit the distribution between a minimum $b_{\rm min} = 0$, corresponding to head-on trajectories, whereas $b_{\rm max}$ is set in such a way that the maximum pericentre passage $r_{\rm p,4b,max}$ is limited to 5 times the largest semimajor axis of the two binaries, where $r_{p,4b}=b_{4b}\sqrt{(e_{4b}-1)/(e_{4b}+1)}$.

Binary components masses are extracted between $10\Ms$ and $50\Ms$, assuming a mass function $f(m)\propto m^{-2}$. Binary semi-major axes are drawn between 0.1 and 10 times the  {\it hard} binary critical value \citep{heggie75}, i.e. $a_{\rm ij,h} = GM_{\rm ij}/(2\sigma_c^2)$ according to a distribution flat in logarithmic values, whereas their eccentricities are selected from a thermal distribution \citep{jeans19}. All other orbital parameters for each binary --- longitude of periastron and ascending node, and mean anomaly --- are selected randomly between $0$ and $2\pi$.

We note that the initial conditions adopted for model sets BIN0, BIN1, and BIN2 are in broad agreement with the binary-binary scattering properties extracted directly from MOCCA models.

\begin{figure}
\centering
\includegraphics[width=8cm]{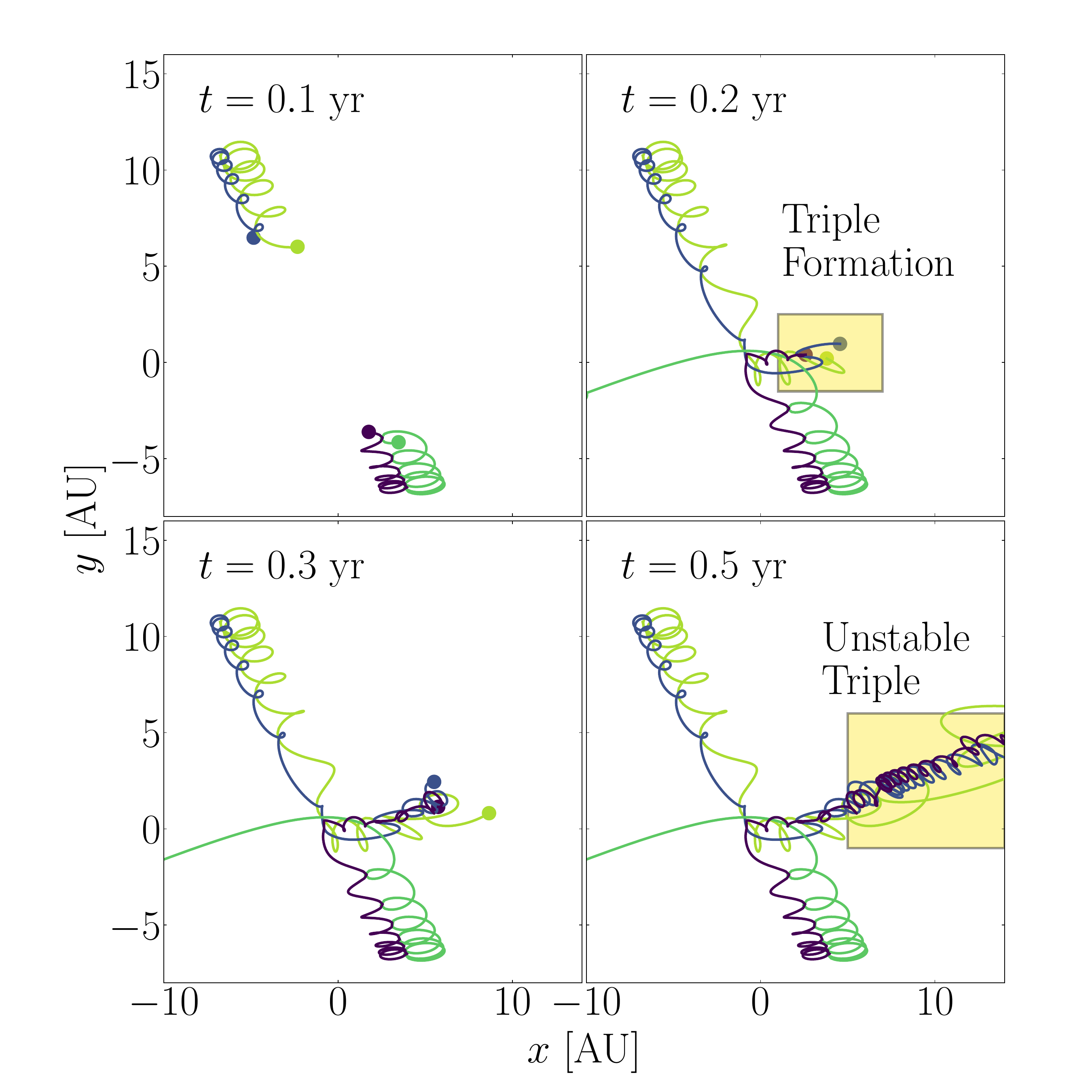}
\caption{Binary-binary interaction in one of the cases investigated. After the close encounter, one BH is ejected and the remaining form an unstable triple.}
\label{4bd}
\end{figure}

All binary-binary simulations are halted if one of the four BHs is ejected away or if the BHBs reach a final distance comparable to the initial one, i.e. the two BHBs are completely unbound. The evolution of one typical example is shown in Figure \ref{4bd}.

\subsection{Properties of triples formed from binary-binary scatterings}

The main end-states of a binary-binary scattering can be either one or two bound BHB, or a bound triple. In 
the latter case, we usually distinguish between an inner binary, i.e. the most bound BH pair, and an outer binary, composed of the inner binary centre of mass and the third bound object.

Whenever a BHB is left behind the binary scattering, it will undergo coalescence due to angular momentum loss via GW emission over a timescale \citep{peters64}
\begin{equation}
t_\gw =  \frac{5}{256}\frac{c^5 a_f^4 f(e_f)}{G^3M_1M_2(M_1+M_2)},
\label{peters}
\end{equation}
where
\begin{equation}
f(e) = \frac{(1-e^2)^{7/2}}{1+(73/24)e^2+(37/96)e^4},
\end{equation}
provided that either the BHB has no further companion or if the perturbation of the third BH on the evolution of the inner binary is negligible.

In the following we will focus mainly on the properties of triples. For clarity, we summarize the nomenclature used to identify the orbital parameters of the inner and outer orbits:
\begin{itemize}
 \item $a$,   semi-major axis;
 \item $e$,   eccentricity; 
 \item $R=a(1-e)$, binary pericentre;
 \item $\omega$, argument of pericentre;
 \item $\Omega$, longitude of the ascending node;
 \item $\varpi = \omega + \Omega$, longitude of pericentre;
 \item $t_{\gw}$, GW timescale;
 \item $i$, orbital inclination.
\end{itemize}
The subscript $3$ is used to label the quantities related to the outer binary.
The inclination is the angle between the angular momentum vectors of the inner and outer binaries.

\begin{figure}
    \centering
    \includegraphics[width=\columnwidth]{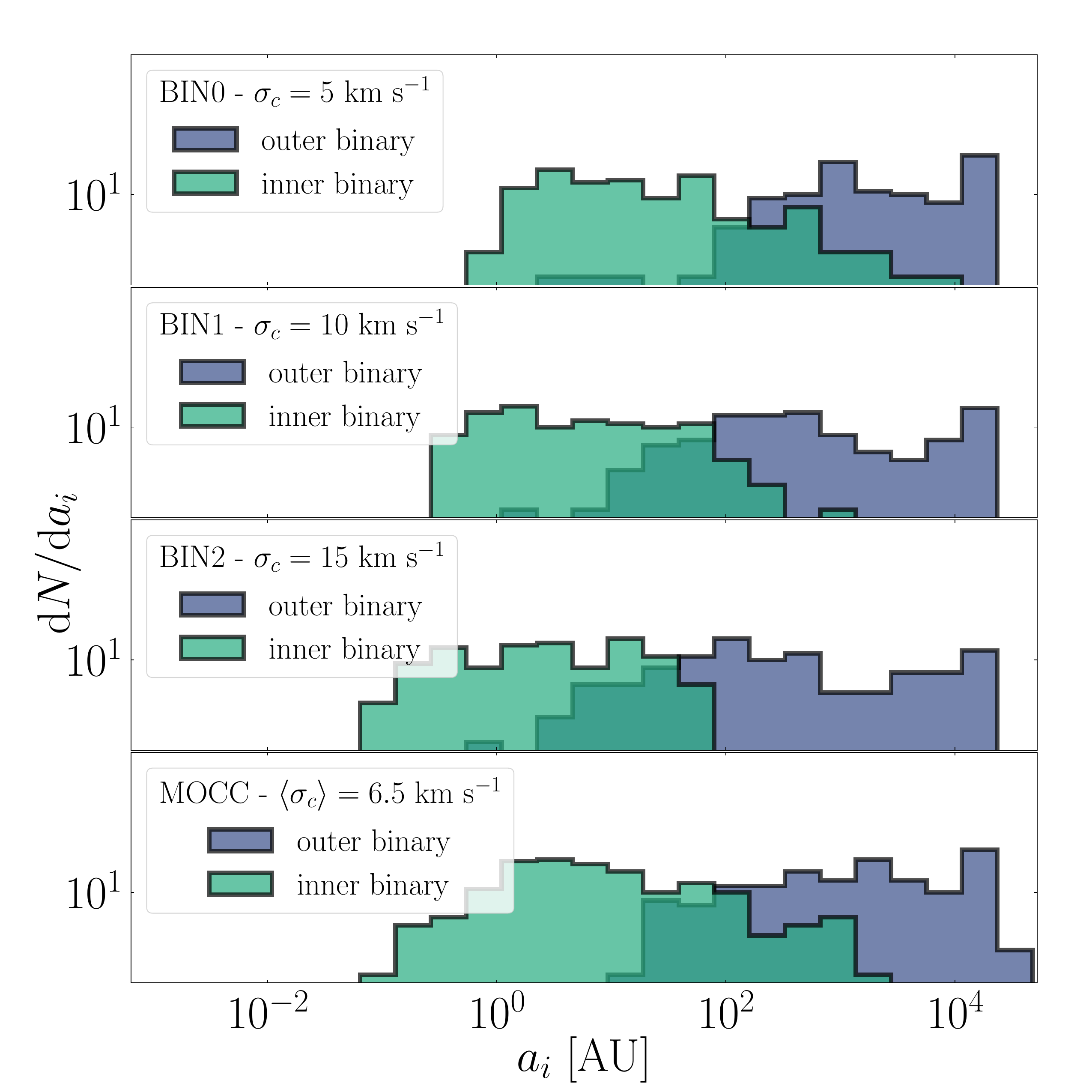}\\
    \includegraphics[width=\columnwidth]{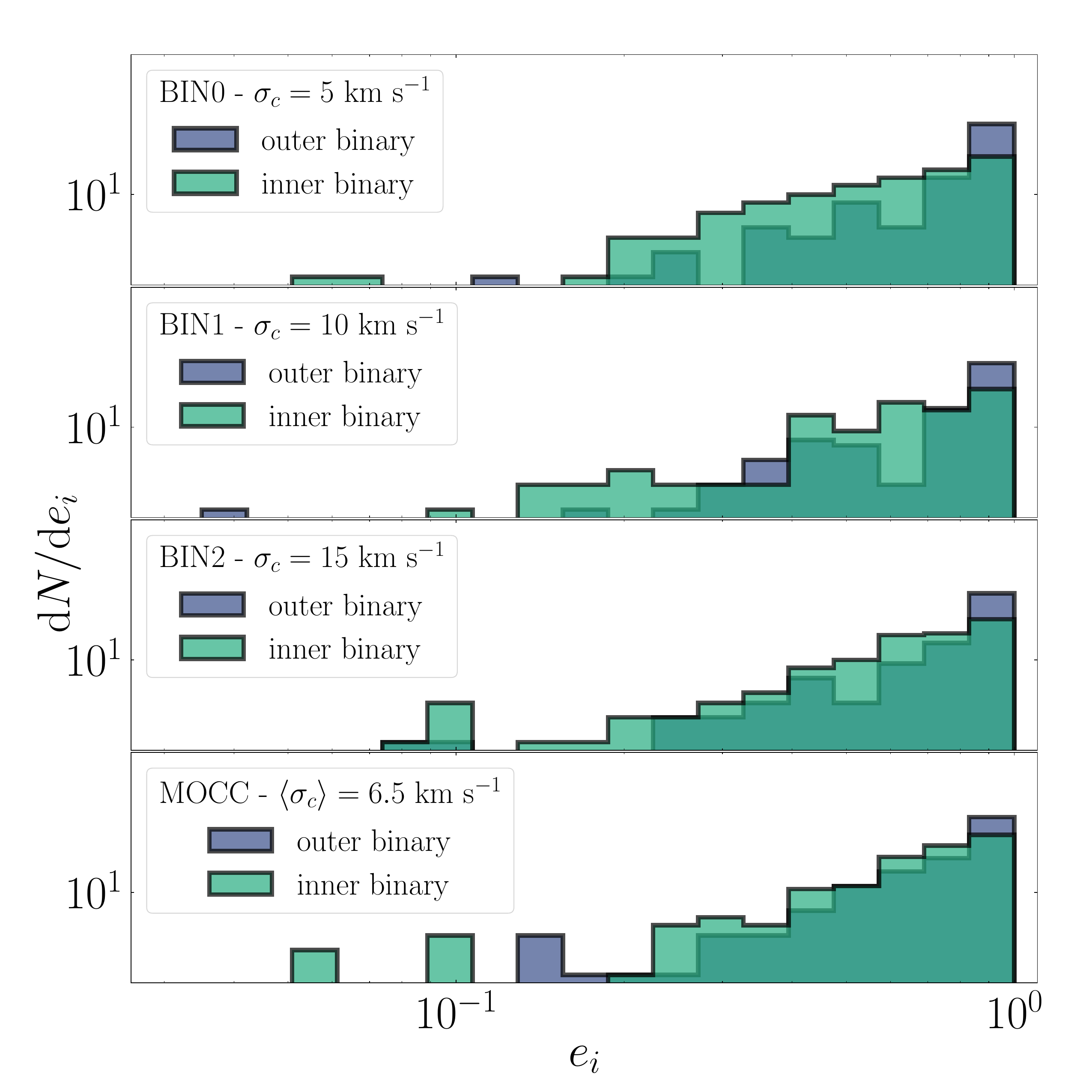}
    \caption{Top figure: semi-major axis distribution for the resulting outer (filled blue steps) and inner binaries (filled green steps) following the binary-binary scatterings. Bottom figure: eccentricity distribution of outer and inner binary, using the same color-coding as in the top figure. 
    In both figures, panels from top to bottom refer to cluster velocity dispersion values $\sigma_c = 5-~10-~15$ km s$^{-1}$, respectively.}
    \label{fig:fourtothree}
\end{figure}

\begin{figure}
    \includegraphics[width=\columnwidth]{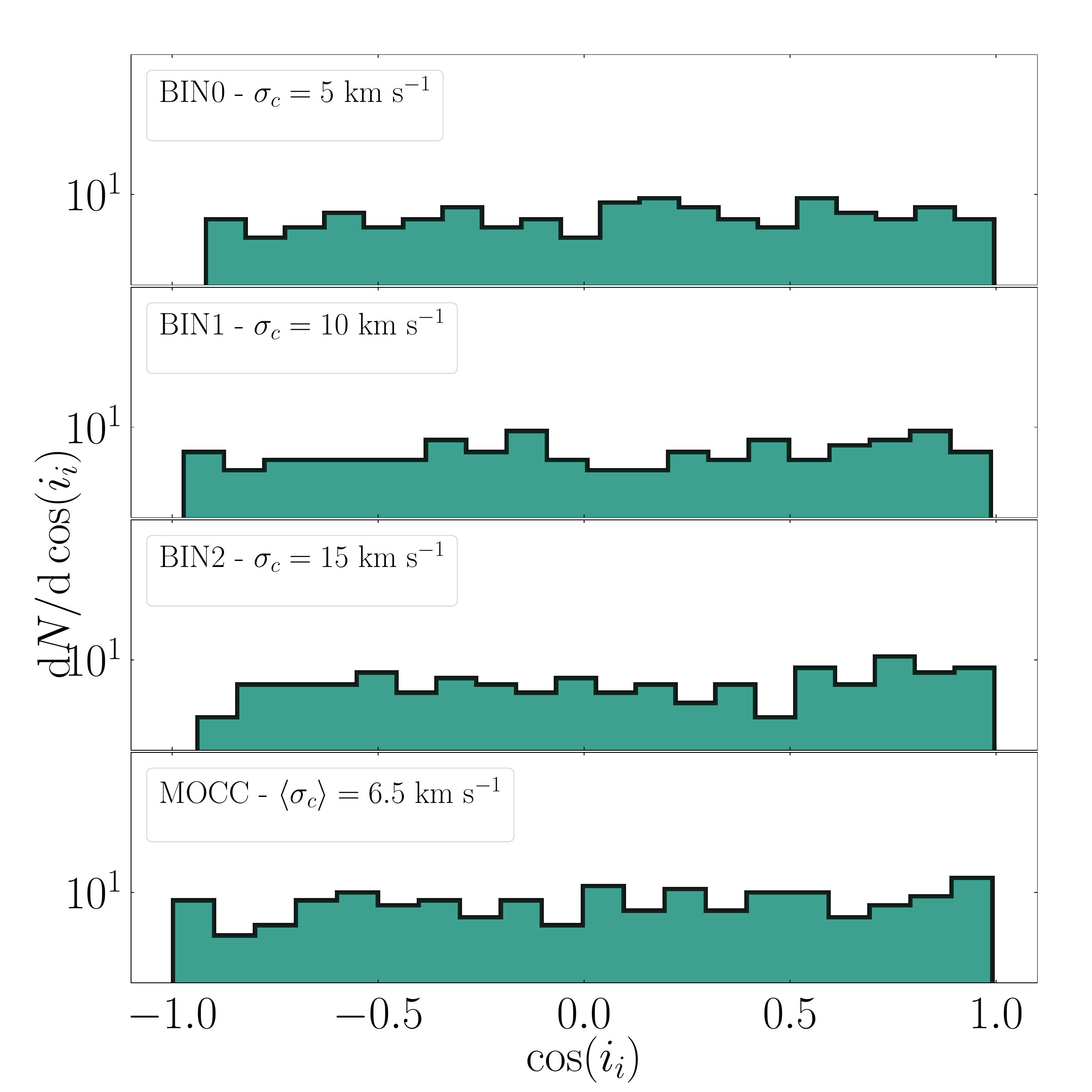}
    \caption{Mutual inclination distribution for the simulations shown in Figure \ref{fig:fourtothree}. }
    \label{fig:incli4}
\end{figure}

Figure \ref{fig:fourtothree} shows the distribution of semi-major axis and eccentricity for the inner and outer binaries of triples formed from BHB-BHB scatterings. It is apparent that the inner and outer semi-major axes' distributions depend on the cluster velocity dispersion, that favour the formation of triples with tighter inner binaries for higher velocity dispersion cluster $\sigma_c$ as shown in the bottom row panels of the two plots in the figure.  In particular, the inner and outer binaries are in the range $(5-3000)\au$.
Also, the overlap between inner and outer binary semi-major axis values implies the potential formation of triples whose evolution is chaotic. 
It is worth noting that triples formed in this way are equally distributed between prograde (inclination $\cos(i)>0$) and retrograde ($\cos(i)<0$) configurations, as shown in Figure \ref{fig:incli4}. As we discuss in the following sections, the relative orbital direction plays a critical role in determining the fate of the inner binary.

A triple system is stable if the semi-major axis of the inner binary remains constant secularly, while for unstable systems the chaotic nature of the system leads to energy transfer between the components. \citet{mardling01} determined a stability criterion through Newtonian $N$-body simulations, defined as \citep[see also][]{antonini14b,toonen16}
\begin{equation}
K_{\rm stab} < \frac{a_3}{a},
\label{eq1}
\end{equation}
where $a~(a_3)$ are the inner(outer) orbit semi-major axis, and
\begin{equation}
K_{\rm stab} = \frac{3.3}{1-e_3}\left[\frac{2}{3}\left(1+\frac{M_3}{M_\bhb}\right)\frac{1+e_3}{(1-e_3)^{1/2}}\right]^{2/5}\left(1-0.3 \frac{i}{\pi}\right),
\end{equation}
thus implying, in general, that larger inclinations reduce the critical value. Note that for a co-planar, prograde, and circular outer orbit and an equal mass triple, $M_3 = 0.5M_{\rm BHB}$, the stability criterion implies $K\simeq 0.3$, and reduces to $K\simeq 0.19$ for retrograde motion with $e_3=0.5$. 
Depending on the orbital configuration, after formation the triple can develop KL cycles \citep{lidov62,kozai62}, according to which the inner binary eccentricity can undergo a periodic oscillations depending on the mutual inclination. In a multiple expansion of the 
triple Hamiltonian, the leading order KL cycles are caused by the quadrupole term. In this case, KL oscillations can develop over a timescale
\begin{equation}
    t_{\rm KL} = \frac{8}{15\pi}\left(1+\frac{m_1+m_2}{m_1+m_2+m_3}\right)\left(\frac{P_3^2}{P_1^2}\right)(1-e_3^2)^{3/2},
\end{equation}
where $P_{1,3}$ is the orbital period of the inner(outer) binary, and drive the inner binary to reach a maximum eccentricity
\begin{equation}
    e_{\rm max} = \sqrt{1-5/3\cos^2 i_3}.
\end{equation}

We note that the onset of KL oscillations can drive away the system from stability, especially in the case of highly inclined and retrograde orbits \citep[see e.g.][]{Grishin2017}.
In cases where the octuple term is nonnegligible, the secular perturbation from the outer BH can drive the inner binary eccentricity to values close to unity. This is called eccentric Kozai-Lidov (EKL) effect \citep{blaes02,naoz11,lithwick11,naoz13,katz11,naoz16}. A general criterion to identify potential configurations that can be prone to the KL mechanisms is \citep[for a review, see][]{toonen16,naoz16}
\begin{equation}
\epsilon_{\rm KL} = \frac{|M_2-M_1|}{M_2+M_1} \frac{a}{a_3} \frac{e_3}{1-e_3^2} < 0.1.
\label{eq:epsilonKL}
\end{equation}

\begin{figure}
    \centering
    \includegraphics[width=\columnwidth]{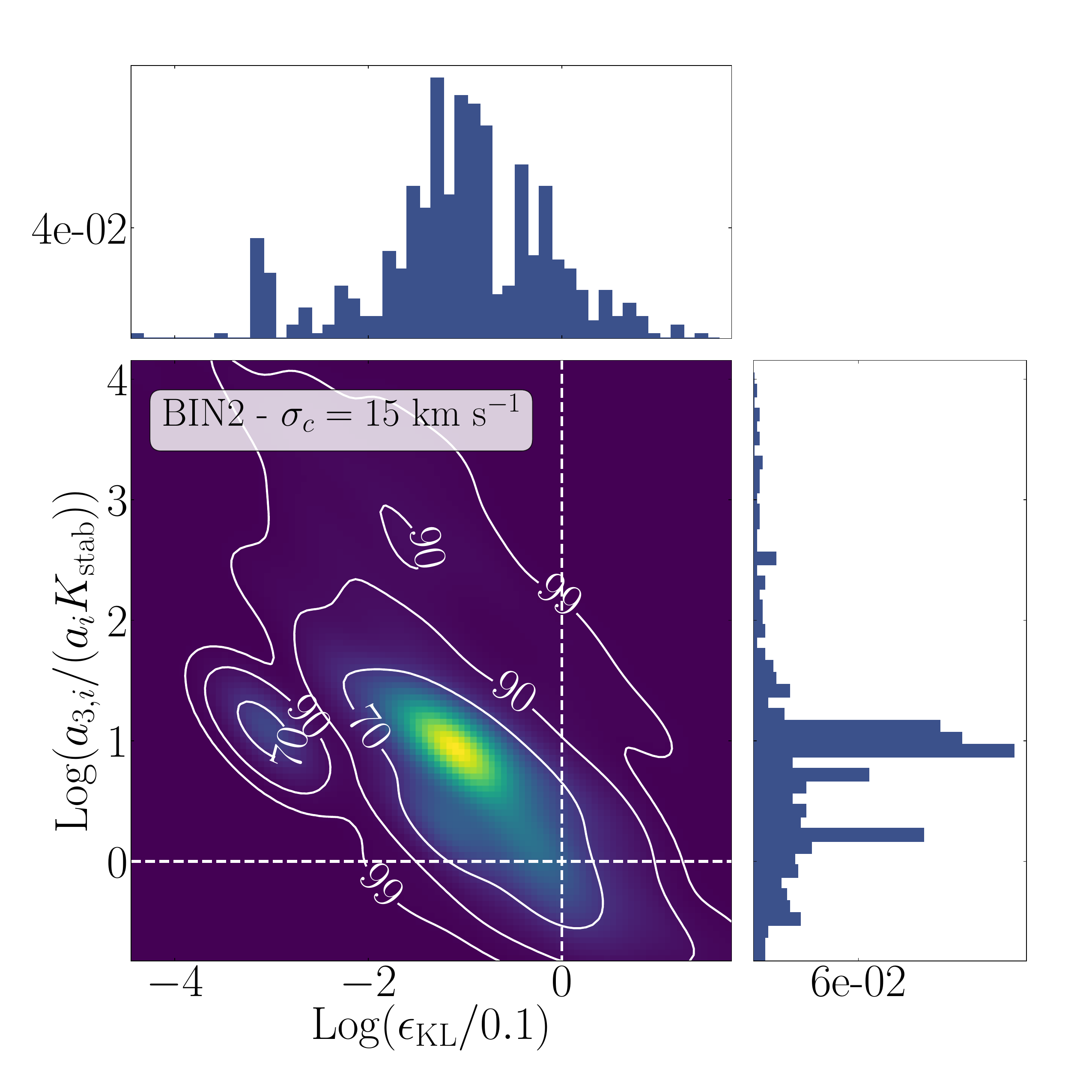}
    \caption{
    Surface number density for the stability and octupole-level approximation criteria for triples formed in model BIN2. The two dotted lines delimit regions where the effect of octupole level approximation can become important (left side of the vertical line) or the triple is stable according to \cite{mardling01} criterion (upper side of the horizontal line).
    }
   \label{fig:hiera}
\end{figure}

We show in Figure \ref{fig:hiera} the values of $K_{\rm stab}$ and $\epsilon_{\rm KL}$ normalised to their critical values for all triples formed in set BIN2. We find that around $18\%$ of models do not fulfill either of the two criteria, thus implying that their evolution can be substantially driven by instabilities and/or chaotic processes.

Since these triples or the binaries left after the binary-binary interaction will be embedded in the host cluster, their disruption will take place over an evaporation time \citep{bt}:
\begin{equation}
    t_{\rm evap} = \frac{\sqrt{3}\sigma_c}{32\sqrt{\pi}G\rho \ln \Lambda a}\frac{m_{\rm sys}}{m_*},
\end{equation}
where $\rho = M_{\rm GC}/R_h^3$ is the average density of a star cluster with mass $M_{\rm GC}$ and half-mass radius $R_h$, $\ln \Lambda = 6.5$ is the Coloumb logarithm, $m_* = 1\Ms$ is the average stellar mass, and $m_{\rm sys}$ is the mass of the system considered, i.e. either a triple or a binary. To calculate this quantity for our models we adopt a half-mass radius of $R_h = 3.2$ pc, i.e. the average value of observed GCs and use the relation between $\sigma_c$, $M_{\rm GC}$, and $R_h$ described in \citep{arca20} to calculate the cluster mass. In the following we identify the evaporation time for triples with pedix ${\rm evap, 3}$ and that for binaries with ${\rm evap, 2}$.

At the end of each simulation we check whether 
\begin{enumerate}[i)]
  \item \label{i:case1} at least one binary hardened in consequence of the scattering, or
  \item \label{i:case2} a bound triple is left behind following the interaction,
\end{enumerate}
and we derive the orbital parameters of any triple, short- or long-lived, formed during the binary scattering evolution. 
Note that the formation of a short-lived triple can fall in case \ref{i:case1}, as these systems tend to eject one of the components and leave a tight binary as final state. 

Figure 2 represents one of such cases: a triple clearly forms in consequence of the BHB-BHB scattering  but  eventually  breaks-up,  thus  its  end-state  is  composed of two single BHs and a bound BHB. We note that it is possible for the same BHB-BHB system to develop a temporary triple  more  than  once  thus  the  total  number  of  triples can  exceed the number of simulations ran.

In case \ref{i:case2}, instead, we distinguish three classes of triples depending on their orbital properties:
\begin{itemize}
    \item {\bf hierarchical triples}: bound triples with $a_3(1-e_3)/a > 10$ [$f_{\rm hie}$]; 
    \item {\bf unstable triples}: bound triples for which Eq. \eqref{eq1} is not satisfied [$f_{\rm uns}$]; 
    \item {\bf octupole triples}: bound triples that could undergo secular effects like KL cycles, which have $\epsilon_{\rm KL} < 0.1$ [$f_{\rm oct}$] (see Eq.~\ref{eq:epsilonKL}).
\end{itemize}

Note that these conditions are not mutually exclusive, as  one  system  can  undergo  several  of  the  phases above during its evolution. Here, we define a triple whenever three out of the four BHs are gravitationally bound, thus we include also short-lived systems. According to the classification above, in sets BIN0-BIN1-BIN2 we find that almost half ($f_{\rm hie} \sim 43-53\%$) of the temporary triples have $a_3(1-e_3)/a > 10$ thus being, potentially, hierarchical. Around $f_{\rm oct} = 68\%$ of triples have the octupole parameter $\epsilon <0.1$ and satisfy the \cite{mardling01} stability criterion. Around $f_{\rm uns} = 34-39\%$ of models do not fulfill the \cite{mardling01} criterion. Finally, a significant fraction ($f_{\rm stab+oct} = 25-33\%$) of triples satisfy both the stability criterion and have the octupole parameter $\epsilon <0.1$.

The binary-binary scatterings tailored to MOCCA initial conditions (set MOCC) return similar results, having around $58\%$ of the triples in a hierarchical configuration and $\sim 30\%$ in an unstable configuration. 

The vast majority of the temporary triples breaks-up and recombines into triples multiple times, thus suggesting that these type of triples can play a major role in dense clusters, where BHB-BHB interactions are frequent. 

In some cases, the perturbations induced by the outer BH or by one BHB can trigger the swift coalescence of the other BHB, provided that the timescales involved are compatible with the evaporation time of the system.

If the binary scattering leaves behind a stable triple, and the KL timescale is shorter than the triple evaporation time, then KL oscillations can develop before the triple is ionized by further stellar encounter. This mechanism drives the inner binary eccentricity up to $e_{\rm max}$ and causes a periodic burst of GWs that shorten the binary lifetime. Under such circumstances the inner binary undergoes merger over a timescale \citep{wen03,antonini12}
\begin{equation}
    t_{\gw ,f} = t_{\gw}(a,e_{\rm max})\frac{1}{f(e)\sqrt{1-e_{\rm max}^2}},
    \label{eq:an12}
\end{equation}
where $a,e$ represent the semimajor axis and eccentricity of the inner binary at the triple formation.

\begin{figure}
    \centering
    \includegraphics[width=\columnwidth]{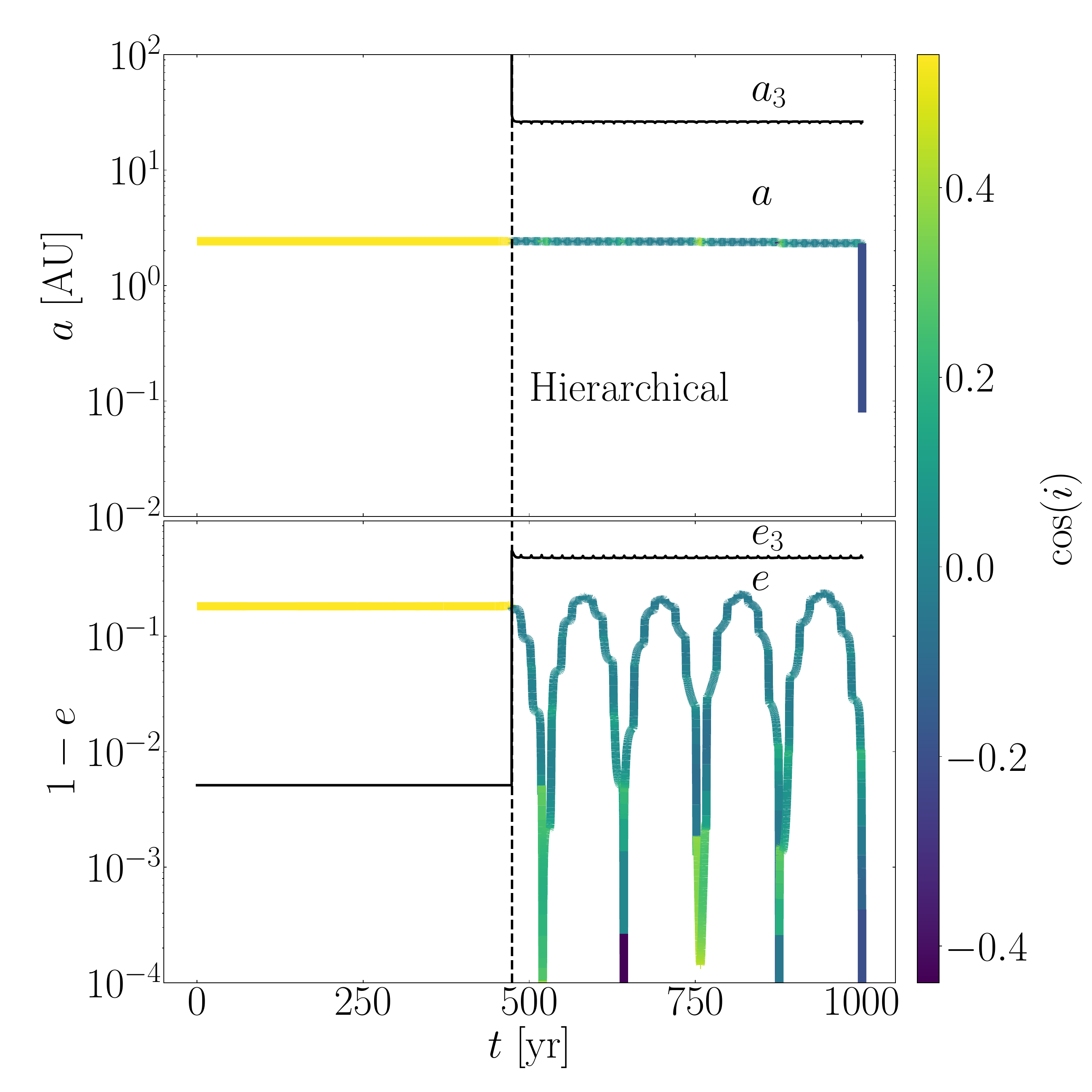}\\
    \includegraphics[width=\columnwidth]{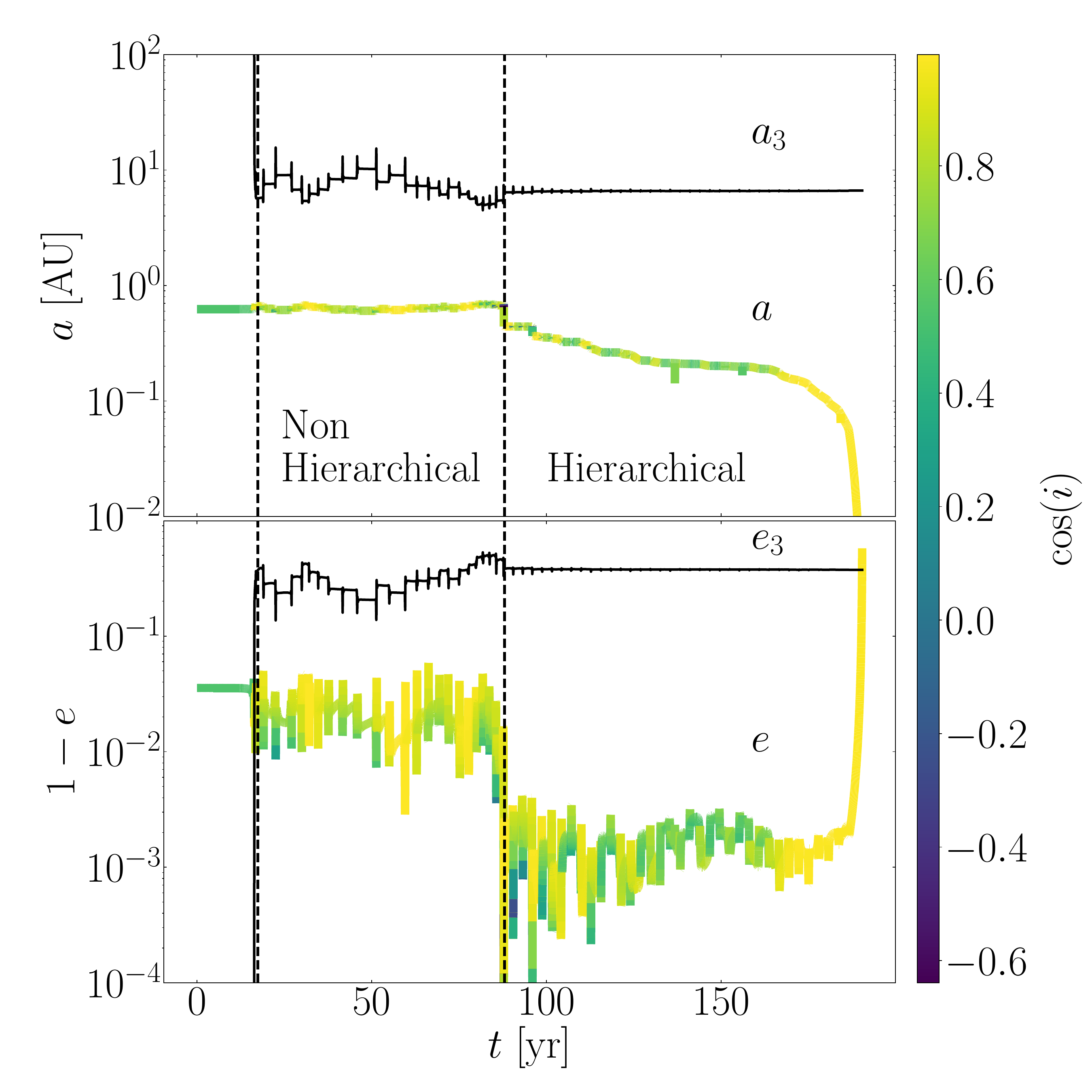}
    \caption{Top panel: time evolution of the inner binary eccentricity in the simulated triple BIN1-374. At $\sim 460$ yr, the binary-binary scattering drives the formation of a triple arranged in a hierarchical configuration, while the fourth object is ejected away. KL oscillations are apparent after the formation of the triple system. Bottom panel: same as above, but for model BIN2-152. In this case, the rapid variation are due to the chaotic interaction with the outer perturber, due to the formation of a non-hierarchical triple. In both cases, color-coding marks the mutual inclination. }
    \label{fig:hiermer}
\end{figure}

Two interesting examples of the evolutionary history of merging BHBs are displayed in Figure \ref{fig:hiermer}. In the first, taken from set BIN1, the binary scattering drives the formation of a hierarchical triple, having an inner binary with components $(M_1+M_2) = (23 + 18.4)\Ms$, semimajor axis $a = 2.4\au$ and eccentricity $e=0.8$, and outer binary with $(M_3,a_3,e_3) = (35.5\Ms,26.2\au,0.52)$. The inner binary undergo several KL oscillations that drive the eccentricity up to $e_{\rm max}=0.99995$, as shown in Figure \ref{fig:hiermer}, leading ultimately to the inner binary merger over a total timescale $10^3$ yr, with about $\sim 530$ yr spent in the hierarchical configuration. We note that the merger time calculated with Equation \ref{eq:an12} adopting the same $e_{\rm max}$ value returns a merger time $616$ yr, thus fully compatible with the full $N$-body model.

In the second example, taken from set BIN2, the scattering drives the formation of a non-hierarchical triple that eventually forms a hierarchical triple and triggers the inner BHB merger in a short time-scale, $\sim 10^2$ yr. In this case, the inner binary is characterized by BH masses $(M_1+M_2) = (11.6+10.4)\Ms$, semi-major axis $a=0.66\au$, and eccentricity $e=0.78$, while the outer orbit is characterized by $(M_3,a_3,e_3)=(13.9\Ms, 8.75\au, 0.72)$.

More in general, we can distinguish several pathways to BHB mergers depending whether the binary scattering:

\begin{enumerate}
    \item leaves behind a BHB with a final merging time smaller than 14 Gyr and the evaporation time $t_{\rm GW} < {\rm min}(14 ~{\rm Gyr},t_{\rm evap})$;
    \item induces the formation of a stable triple characterised by an inner binary with merging time $t_{\rm GW} < {\rm min}(14 ~{\rm Gyr},t_{\rm evap,2})$ and a KL time longer than the triple evaporation time ($t_{\rm KL} > t_{\rm evap,3}$). In such case, stellar interactions are likely to ionize the triple before KL oscillations take place;
    \item induces the formation of a stable triple having
    $t_{\rm KL} < t_{\rm evap,3}$. In this case KL oscillations develop before the triple evaporates and can drive the inner BHB to merger;
    \item induces the formation of an unstable triple for which the inner BHB merger time is shorter than the triple evaporation time ($t_{\rm GW} < t_{\rm evap,3}$) and $t_{\rm GW}<{\rm min}(14{\rm ~Gyr},t_{\rm evap,2})$, thus the binary likely merges before the triple is ionized;
    \item induces the formation of an unstable triple having $t_{\rm GW} > t_{\rm evap,3}$ and $t_{\rm GW}<{\rm min}(14{\rm ~Gyr},t_{\rm evap,2})$, thus the triple is ionized before the inner BHB merge.
\end{enumerate}

According to the merger classes defined above, we find that BHB-BHB scattering leads to at least one bound BHB with a merger time shorter than the evaporation time for $69-84\%$ of the cases while the end-state is a bound triple for around $20.4-31.2\%$ of the cases.

In the next section, we will model the evolution of triples with properties extracted from the binary scattering model set BIN2, i.e. the one with the largest chance to produce unstable  triples, which are the main focus of this work.

\begin{table*}
\caption{Main parameters of the numerical models}
\begin{center}
\begin{tabular}{ccccccccccc}
\hline
SET & $N_{\rm mod}$ & $M_{3,i}$ & $R_{3,i}/R_i$ & $e_{3,i}$ & $M_{1,i}$ & $M_{2,i}$ & $e_i$ & $\Delta \varpi_i$ & $i_i$ & $R_i$\\
 & & $\Ms$ & & & $\Ms$ & $\Ms$ &   & & $^\circ$ & AU \\
\hline
\multicolumn{11}{c}{\textbf{Realistic ICs}}\\
\hline
$0$ & $1000$ & $ 10-50 $ & $ 10^{-1}-10^{5} $ & $ 0-1 $ & $ 10-50 $ & $ 10-50 $ & $ 0-1 $ & $ 0-2\pi $ & $ 0-180 $  &$ 10^{-3}-2\times 10^{2}$ \\
\hline
\multicolumn{11}{c}{\textbf{Impact of the triple configuration}}\\
\hline
$1$  & $1000$ & $ 20 $    & $ 1-6 $    & $ 0-1 $ & $ 20 $    & $ 10 $    & $ 0 $   & $ 0-2\pi $ & $ 0 $      &$20$ \\
$2$  & $1000$ & $ 20 $    & $ 2-7 $    & $ 0-1 $ & $ 20 $    & $ 10 $    & $ 0 $   & $ 0 $      & $ 0-10 $   &$20$ \\
$3$  & $1000$ & $ 20 $    & $ 1-6 $    & $ 0-1 $ & $ 20 $    & $ 10 $    & $ 0-1 $ & $ 0 $      & $ 0-10 $   &$20$ \\
$4$  & $2000$ & $ 20 $    & $ 2-7 $    & $ 0-1 $ & $ 20 $    & $ 10 $    & $ 0-1 $ & $ 0 $      & $ 0-180 $  &$20$  \\
$5$  & $2000$ & $ 20 $    & $ 2-7 $    & $ 0-1 $ & $ 20 $    & $ 10 $    & $ 0.6 $ & $ 0-2\pi $ & $ 0 $      &$20$  \\
$6$  & $2000$ & $ 20 $    & $ 2-7 $    & $ 0-1 $ & $ 20 $    & $ 10 $    & $ 0.6 $ & $ 0-2\pi $ & $ 180 $    &$20$  \\
$7$  & $2000$ & $ 20 $    & $ 2-7 $    & $ 0-1 $ & $ 20 $    & $ 10 $    & $ 0-1 $ & $ \pi $    & $ 0-10 $   &$20$ \\
$8$  & $2000$ & $ 20 $    & $ 2-7 $    & $ 0-1 $ & $ 20 $    & $ 10 $    & $ 0-1 $ & $ \pi $    & $ 170-180 $&$20$ \\
\hline
\multicolumn{11}{c}{\textbf{Impact of component masses and orbital properties}}\\
\hline
$9$  & $8000$ & $ 10-50 $ & $ 2-7 $    & $ 0-1 $ & $ 10-50 $ & $ 10-50 $ & $ 0-1 $ & $ 0-2\pi $ & $ 0-180 $  &$20$ \\
$10$ & $8000$ & $ 10-50 $ & $ 2-7 $    & $ 0-1 $ & $ 10-50 $ & $ 10-50 $ & $ 0-1 $ & $ 0-2\pi $ & $ 0-180 $  &$1$ \\
\hline
\end{tabular}
\end{center}
\begin{tablenotes}
\item Column 1: Model name. Column 2: Number of runs. Column 3: mass of the outer BHs. Column 4-5: outer binary initial separation, normalized to the value of the inner binary, and eccentricity.
Column 6-7: Masses of the inner binary components. Column 8: inner binary eccentricity. Column 9: difference between the inner and outer longitude of pericentre. Column 10: mutual inclination between the inner and outer orbits. 
\end{tablenotes}
\label{tab:models}
\end{table*}

\subsection{Black hole triples evolution in astrophysical systems: implications for binary mergers}
\label{sec:BBHfull}

In this section we model the evolution of triples whose initial conditions are drawn from the results of our BHB-BHB scattering experiments described in Section \ref{sec:GC}. On the one hand, this enable us to focus only on the formation of triples, rather than on the much larger parameter space that characterises binary-binary scatterings, whilst on the other hand it permits us to obtain general properties of potential BHB mergers originating from this channel, and to compare them with present-day GW detections. We run 1000 simulations drawing the initial inner/outer semimajor axis and eccentricity and mutual inclination from the distribution obtained for 
BIN2 simulations set, which assumes a host cluster with velocity dispersion $\sigma = 15$ km s$^{-1}$ (see Figures \ref{fig:fourtothree} and \ref{fig:incli4}).

In the following we refer to this triple models as SET0, as summarized in Table \ref{tab:models}. The main objective of these simulations is to determine the role played by both hierarchical and non-hierarchical triple evolution in the formation of BHB mergers. 

To identify potential mergers we calculate the merger time in two different ways, depending on the level of hierarchy of the triple as discussed in the previous section. For each triple we calculate the \cite{mardling01} stability criterion, $a_{3,f}/a_f < K$ (see Equation \ref{eq1}), and the octupole-level approximation criterion $\epsilon_{\rm oct} < 0.1$ (see Equation \ref{eq:epsilonKL}). In the case in which the triple does not fulfill these criteria, or it is disrupted, we calculate the merger time using Equation \ref{peters}.
If, instead, the triple satisfies both criteria above, the merger time is calculated following Equation \ref{eq:an12}.

We identify this way 216 mergers out of 961 models, corresponding to a merger probability of $22.7\%$. At the end of the simulation, around $\sim 92.5\%$ of merger candidates are the inner BHB in a triple that satisfy both stability and octupole criteria ($a_{3,f}/a_f < K$; $\epsilon_{\rm oct} > 0.01$) thus we classify them as {\it hierarchical} mergers, whereas the remaining $7.5\%$ are BHBs originated from disrupted triples.
It must be noted that around two third of the triples are initially in a hierarchical configuration due to the constraints imposed by the binary scatterings, thus this partly explains the much larger amount of ``hierarchical'' mergers compared to ``non-hierarchical'' one. 

\begin{figure}
    \centering
    \includegraphics[width=\columnwidth]{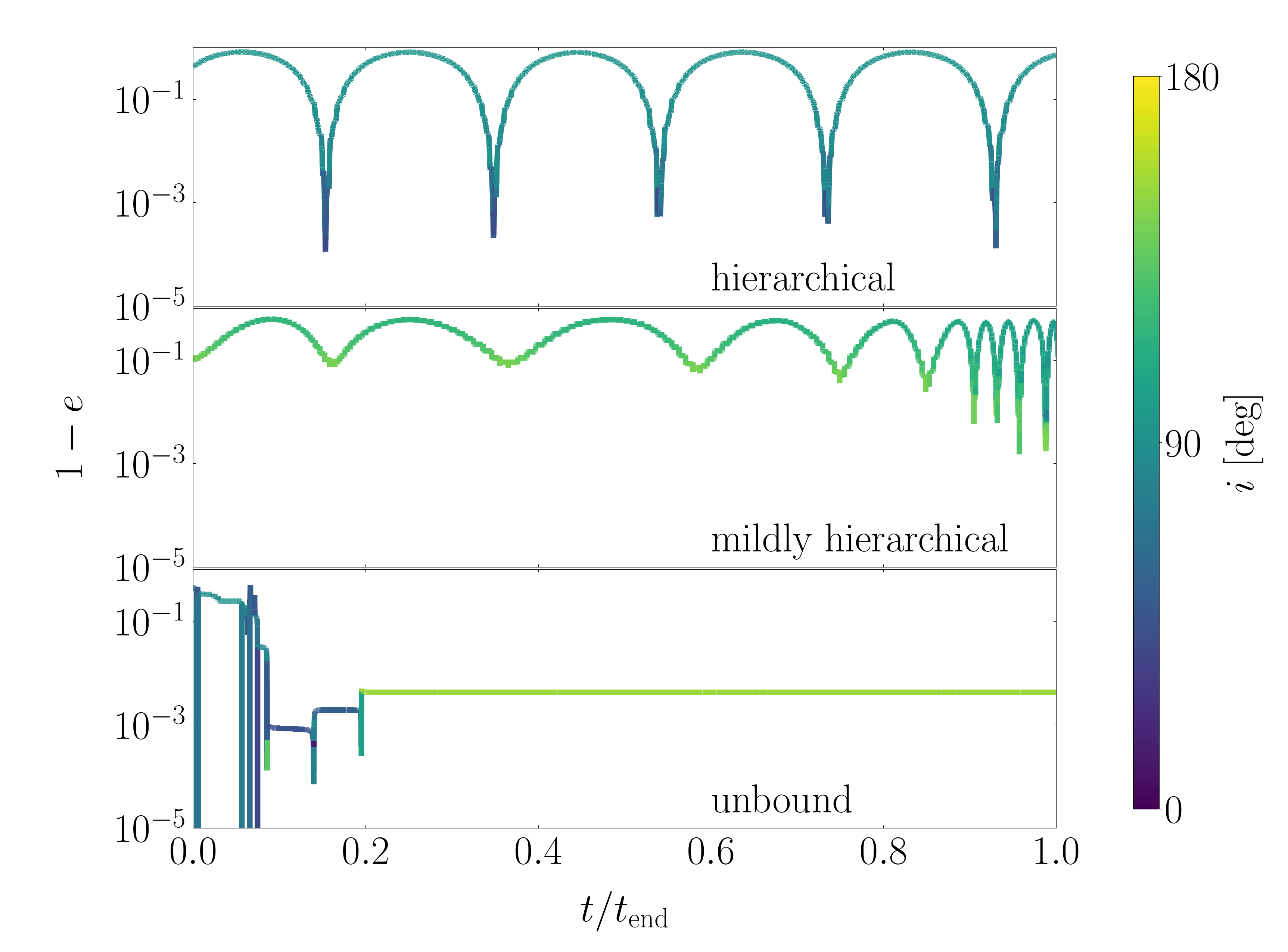}
    \caption{Eccentricity variation as a function of time in three systems that lead the inner binary to merge. Top panel shows a triple in which the inner binary undergoes typical KL oscillations. Central panel shows a case where KL oscillations vary in phase and amplitude over time. Bottom panel shows the case of disrupted triple, with the inner binary evolution driven solely by GW emission.}
    \label{fig:samples}
\end{figure}

We also find a subclass of systems characterised by a peculiar behaviour in the evolution of the eccentricity, which differ substantially from the standard KL predictions. Figure \ref{fig:samples} shows three systems in this sample, one showing typical KL eccentricity variation, one showing the peculiar KL-like cycles mentioned above, and the latter showing the case of a disrupted triple. 

\begin{table}
    \centering
    \caption{Mergers in SET0}
	\begin{center}	    
    		\begin{tabular}{ccccc}
        		\hline
        		\hline
	        SET &  $N_{\rm sim}$ & $N_{\rm mer}$ & $N_{\rm mer,hie}$ & $N_{\rm mer,dis}$ \\
        		\hline
	        0 & 1000 & 218 & 200 & 18 \\
    		    \hline
	    \end{tabular}
	\end{center}    
    \begin{tablenotes}
    \item Col. 1: model ID. Col. 2: total number of simulations. Col. 3-4: number of mergers, of those developing in hiearchical configurations, and in disrupted triples, respectively.
    \end{tablenotes}
    \label{tab:my_label}
\end{table}

\begin{figure}
    \centering
    \includegraphics[width=\columnwidth]{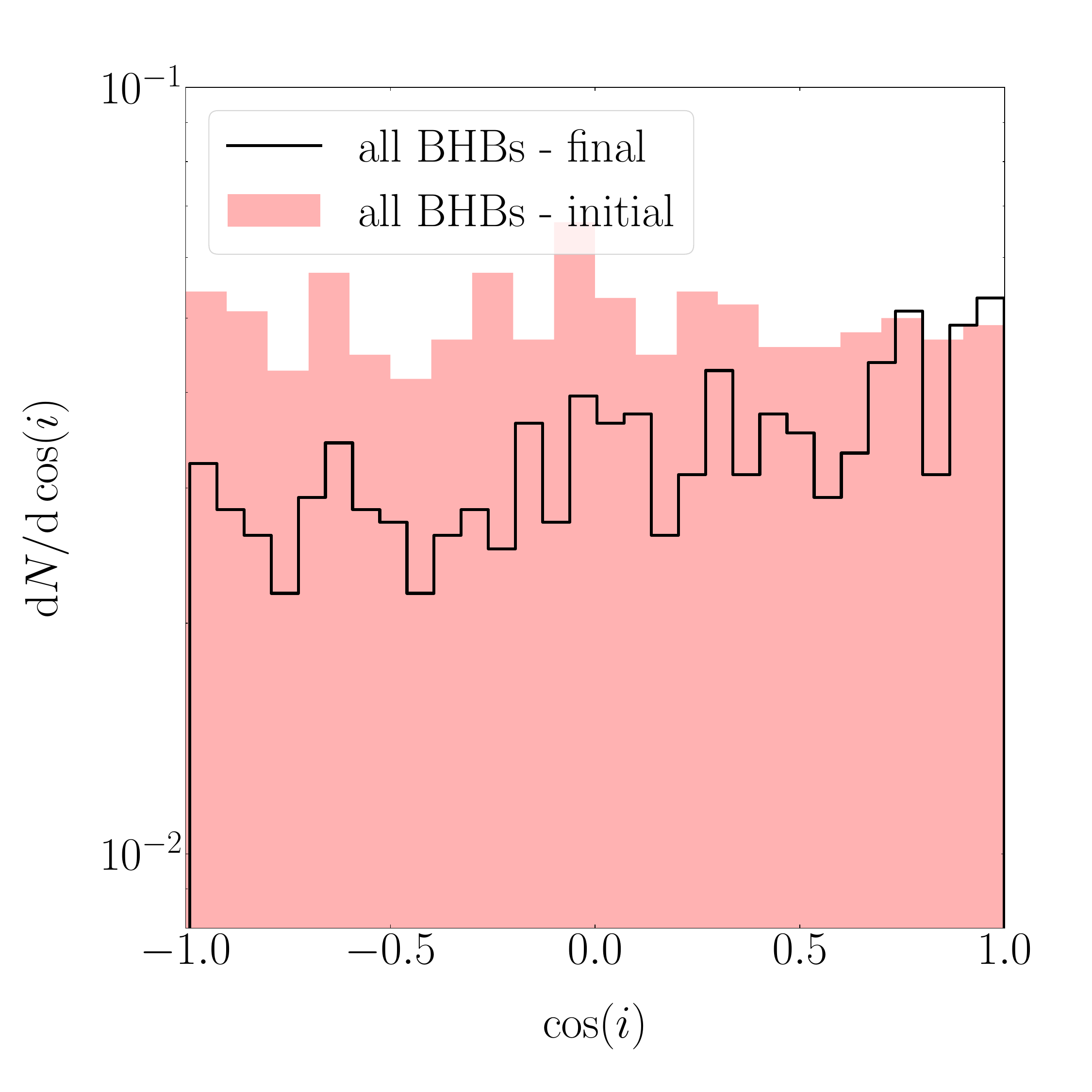}
    \includegraphics[width=\columnwidth]{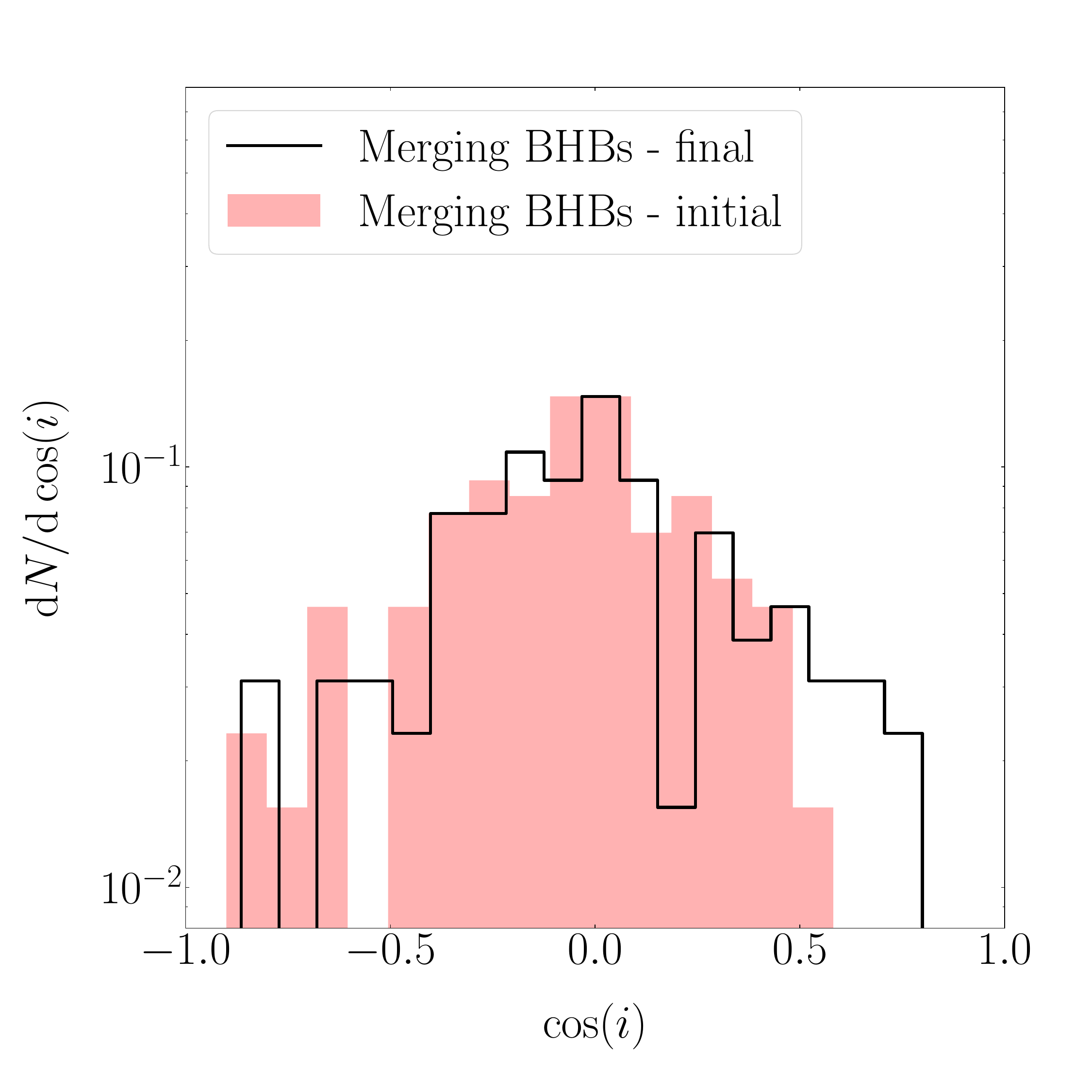}
    \caption{Top panel: initial (red filled boxes) and final (black open steps) inclination distribution for all BHBs in SET0. Bottom panel: same as above, but only for merging BHBs in SET0.}
    \label{fig:inc11}
\end{figure}

The top panel of Figure \ref{fig:inc11} shows the distribution of initial and final values of the inclination for all BHBs in SET0. We find that the interaction with the perturber tends to decrease the fraction of retrograde systems: the initial models are equally distributed between prograde and retrograde configuration, whereas at the end prograde configurations outnumber retrograde ones by a factor $1.34$. This suggests that retrograde systems have the tendency to reduce the inclination, and flip the orbit in some cases, whereas prograde tends to maintain their original configuration. Limiting the analysis to mergers only, we find that the population concentrates around nearly perpendicular configurations, i.e. with a $\cos(i)\simeq 0$, as shown in Figure \ref{fig:incli4}.

\begin{figure}
    \centering
    \includegraphics[width=\columnwidth]{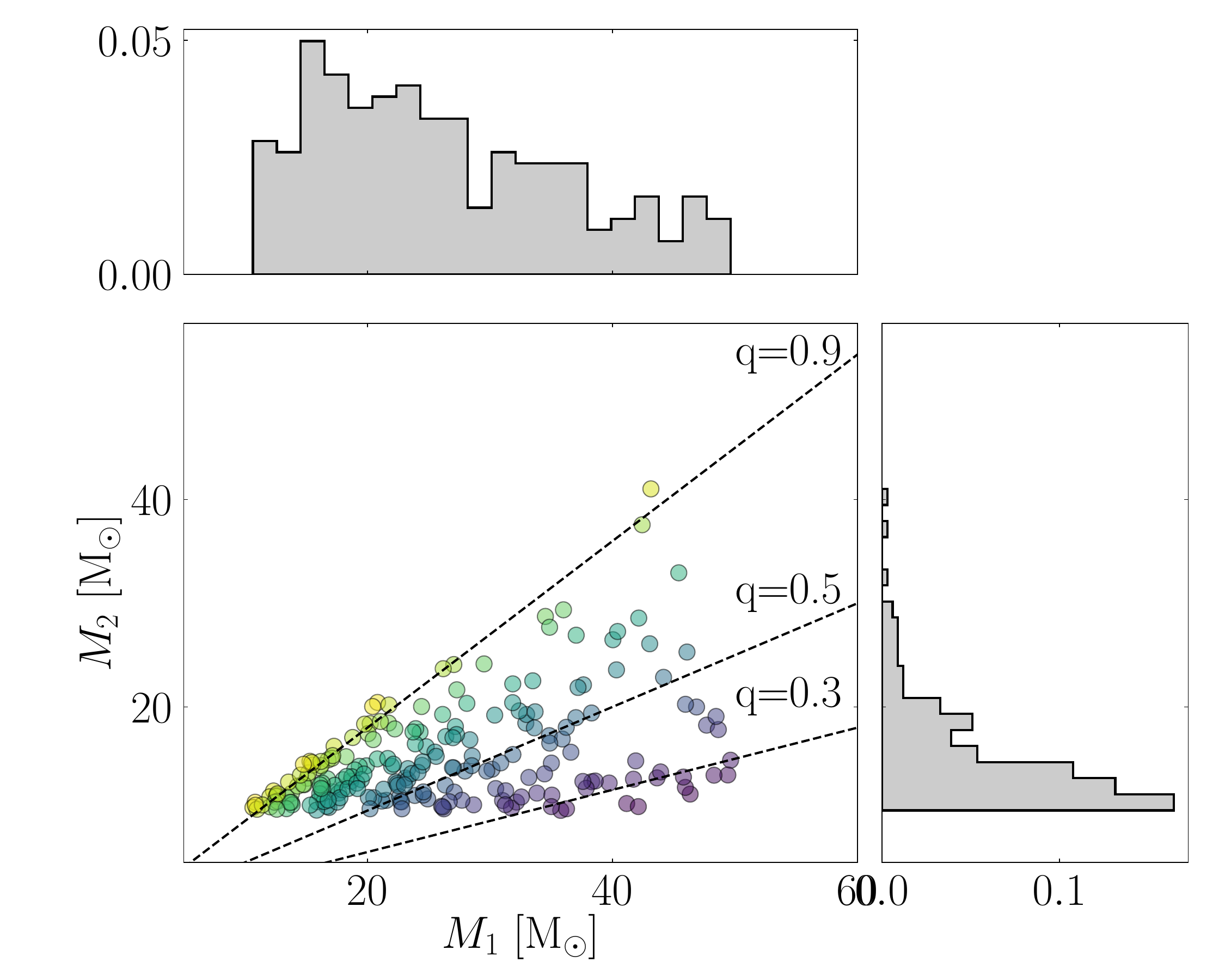}
    \includegraphics[width=\columnwidth]{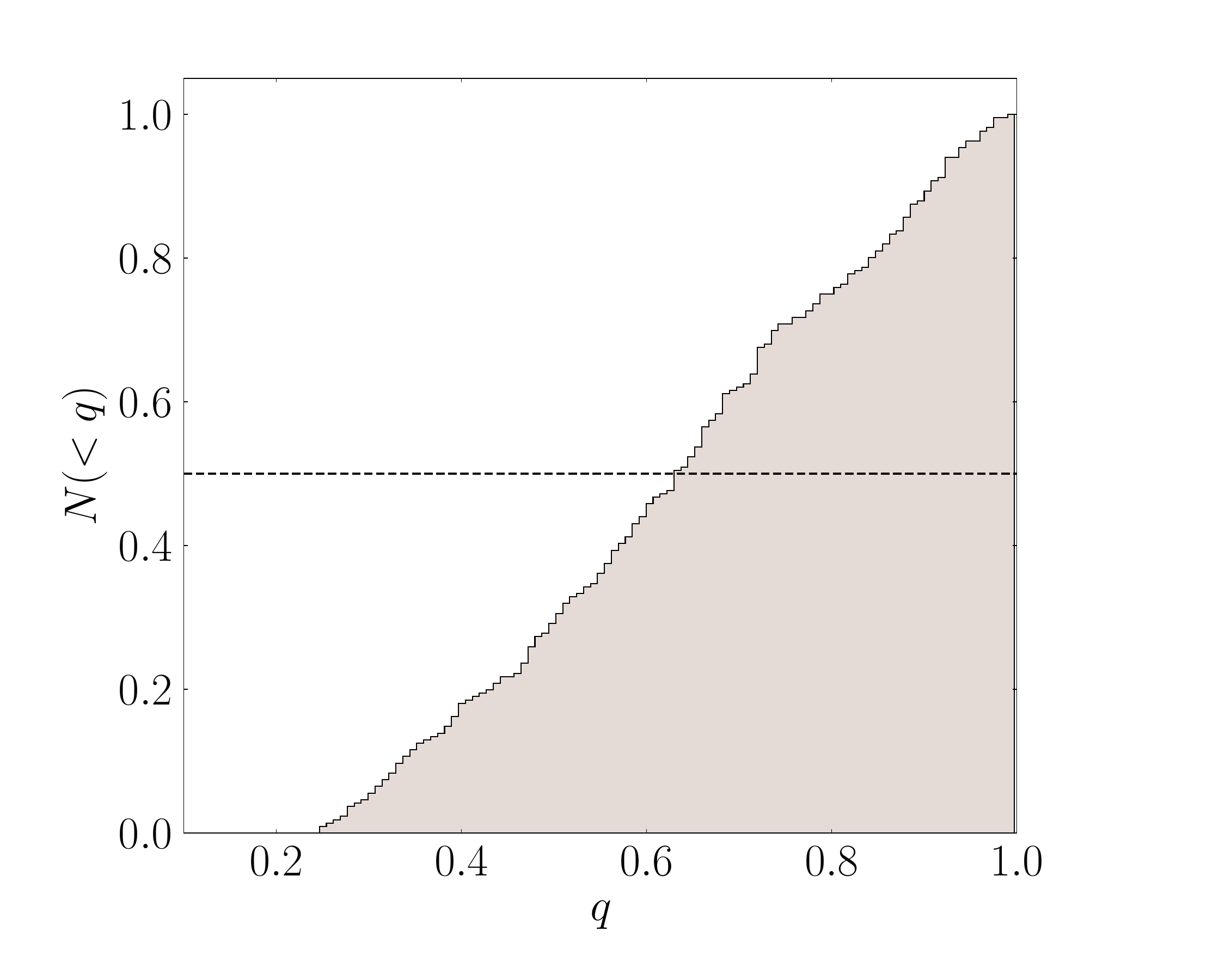}
    \caption{Top panel: combined mass distribution of merging BHBs components for SET0. Dashed line identifies loci in the plane at fixed value of the mass ratio. Darker(lighter) colors identify larger(smaller) mass ratios. Bottom panel: mass ratio distribution for merging BHBs in SET0.}
    \label{fig:massdistri}
\end{figure}

As shown in Figure \ref{fig:massdistri}, the mass distribution of the primary in merging BHBs of SET0 appears to be almost flat, with a weak peak at values $M_1 = 15-25\Ms$. The mass distribution of secondary component, instead, shows a strong peak at values $M_2 = 10\Ms$ and a sharp decline at values $M_2>20\Ms$. As a result, $50\%$
of mergers in SET0 have a mass ratio in the range $q = 0.2-0.6$. 

Around $18\%$ of models undergo component swap. Among them, in a few cases ($3\%$) the outer BH mass initially exceeds the whole BHB mass, $M_\bhb < M_3$, thus the component swap determines the formation of a final BHB much heavier than the initial. The majority of component swaps ($15.2\%$) leads to a final BHB in which the companion is heavier than the original thus confirming that component exchanges tend to promote an increase in the binary binding energy. However, component exchanges affect only $4.6\%$ of all mergers (10 mergers out of 216). In 8 cases the BHB mass change by less than $20\%$ whereas in the remaining 2 cases the new companion increases the BHB mass by a factor 1.3 and 1.8 respectively, acquiring a net positive gain of $10\Ms$ and $30\Ms$ respectively.

In section \ref{sec:an} we will explore the role of the triple configuration in determining the formation of merging BHBs.

\subsection{GW emission frequency and merging BHB eccentricity}

The GWs signal emitted during the BHB evolution is characterized by a broad frequency spectrum, whose peak frequency can be calculated as \citep{wen03,antonini14b}
\begin{equation}
f_\gw = 0.35 ~{\rm mHz} 
\left(\frac{M_{\bhb,f}}{30 \Ms}\right)^{1/2}
\left(\frac{a_f}{0.01 \au}\right)^{-3/2}
\frac{\left(1+e_f\right)^{1.1954}}{\left(1-e_f^2\right)^{1.5}}.
\end{equation}
As noted by \cite{antonini14b}, the maximum eccentricity attainable by the inner binary can be expressed as 
\begin{equation}
\sqrt{1-e_{\rm max}} = 5\pi \frac{M_{2,f}}{M_{\bhb,f}}\left[\frac{a_f}{a_{3,f}(1-e_{3,f})}\right]^3.
\end{equation}
The Equation above can be used to constrain the region of parameter space in which the BHB can attain 
an eccentricity above 0.1 at frequencies audible to ground-based detector, a condition that can be expressed as $3<a_{3,f}(1-e_{3,f})/a_{1,f}\lesssim 10$ in the case of BHBs with components having comparable masses. It must be noted that our models are wider systems compared to \cite{antonini14b}. As shown in the following, this reduces significantly the probability to observe eccentric binaries in the Hz band.

\begin{figure}
\centering
\includegraphics[width=8cm]{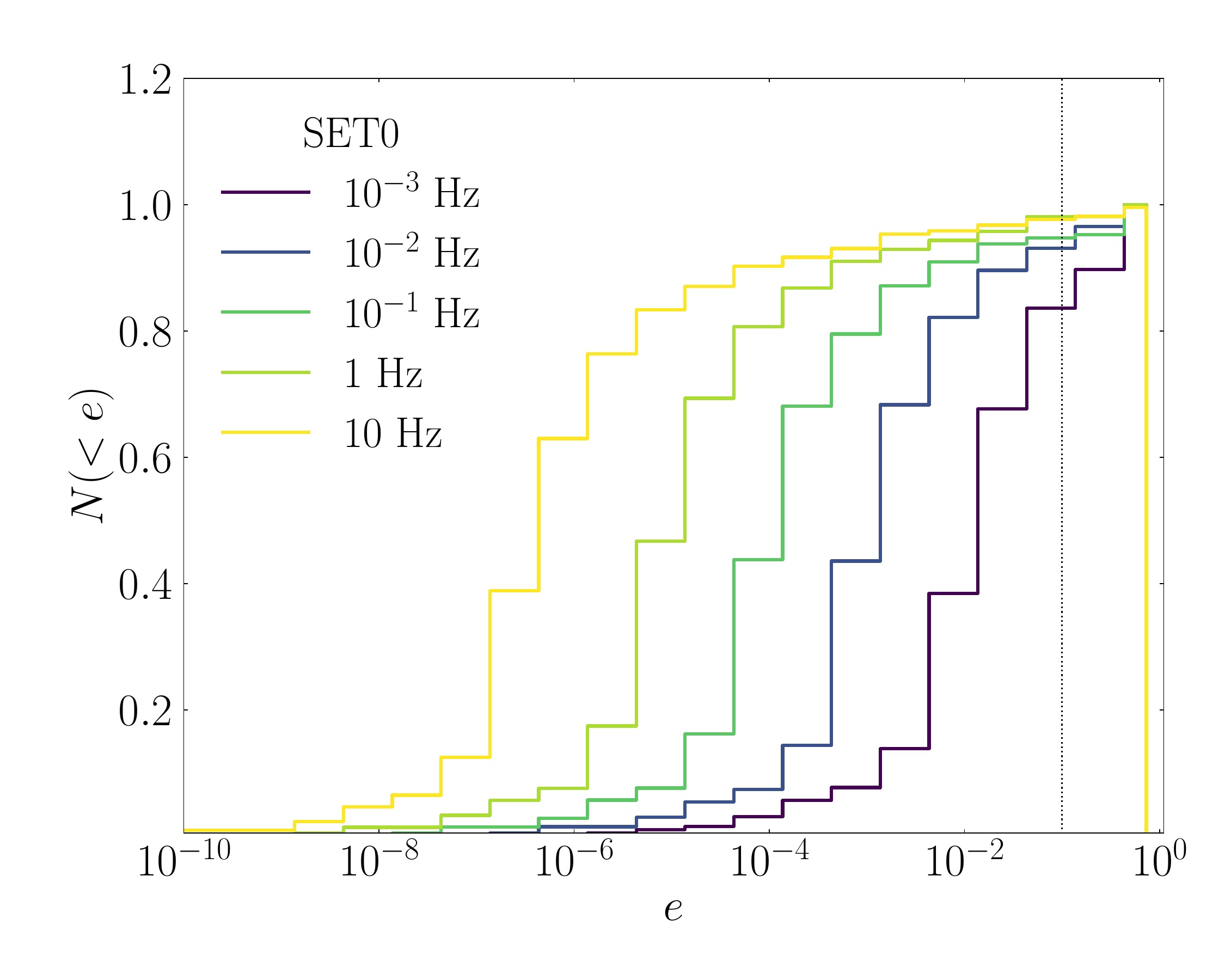}\\
\includegraphics[width=8cm]{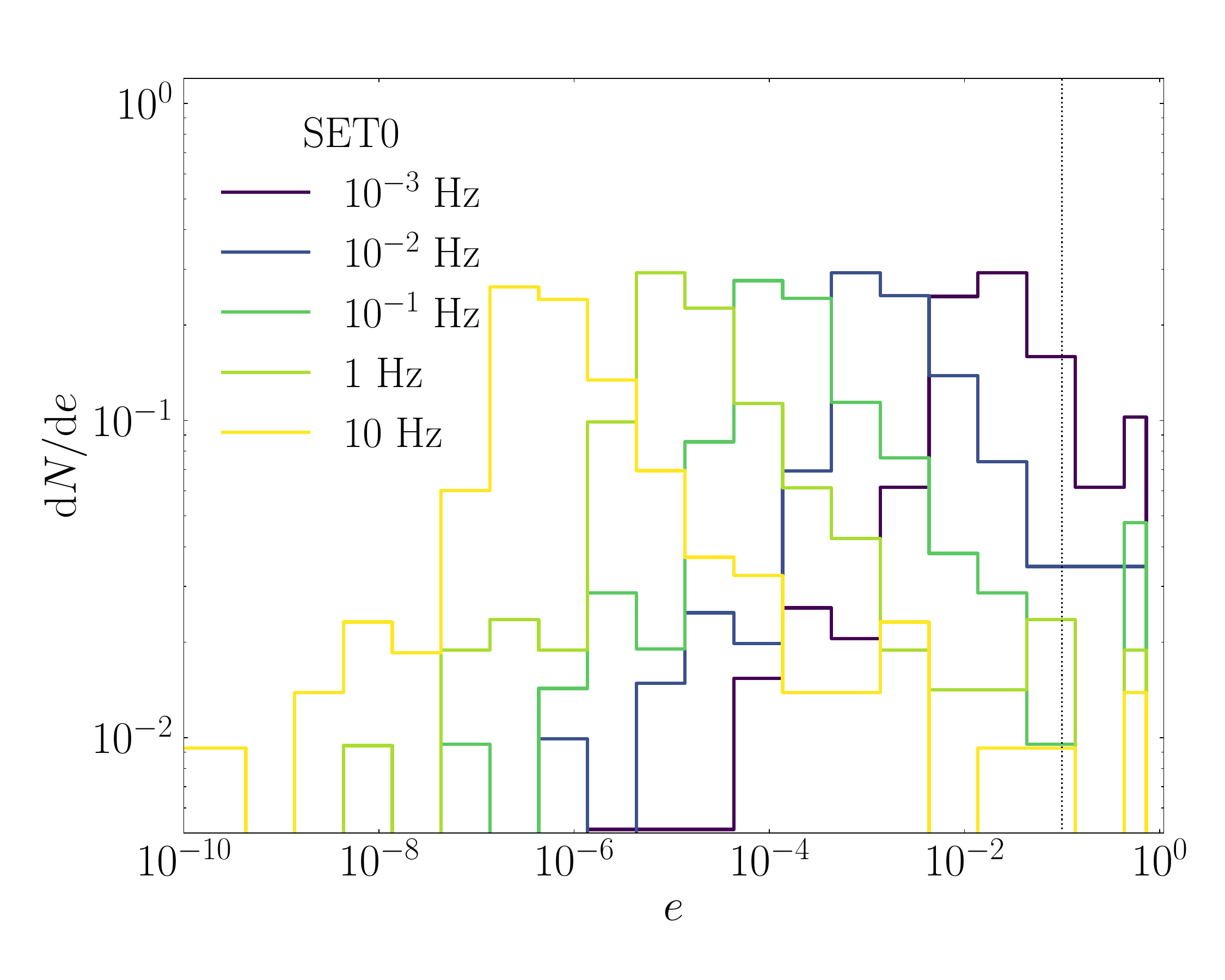}\\
\caption{Eccentricity cumulative distribution (top panel) and distribution (bottom panel) for all the BHB mergers in model 11. The eccentricity is calculated as the BHB crosses a given frequency, as indicated in the legend. Dotted line indicates a value of eccentricity $e=0.1$.}
\label{histoGW0}
\end{figure}

In \argdf, the last phases preceding the binary merger are extremely time-consuming, since the code needs extremely small steps to integrate precisely the components equation of motion. In order to find a good balance between computational load and modelling reliability, for each binary we integrated numerically \cite{peters64} equations:
\begin{eqnarray}
\frac{\der a_f}{\der t} &=& -\frac{64}{5}\beta(M_{1,f},M_{2,f})\frac{F(e_f)}{a_f^3},\\
\frac{\der e_f}{\der t} &=& -\frac{304}{15}\beta(M_{1,f},M_{2,f})\frac{e_f G(e_f)}{a_f^4},
\label{peters2}
\end{eqnarray}
with
\begin{eqnarray}
F(e_f)          &=& (1 - e_f^2)^{-7/2}\left(1 +\frac{73}{24} e_f^2 + \frac{37}{96}e_f^4\right);\\
\beta(M_{1,f},M_{2,f})  &=& (G^3/c^5) M_{1,f}M_{2,f}(M_{1,f}+M_{2,f});\\
G(e_f)          &=& (1-e_f^2)^{-5/2}\left(1+\frac{121}{304}e_f^2\right).
\end{eqnarray}
These equations have analytical solutions for $a_f(e_f)$ \citep{peters64} and time evolution \citep{Mikoczi2012}.
For each simulation, we select the ($a_f,~e_f,~M_{\bhb,f}$) values available at the last snapshot and we use them to integrate the equation above. This choice implies three possibilities: a) if the triple breaks up before the simulation ends the values at the last snapshot represent the BHB endstate, b) if the inner BHB merges before the simulation ends, the values at the last snapshot represent the BHB status if the outer BH does not have any effect on the inner BHB evolution, c) if the triple is still bound but the inner BHB merger time is smaller than 14 Gyr, we take the maximum eccentricity if the binary is classified as ``hierarchical'' or the value at the last snapshot otherwise. Figure \ref{histoGW0} shows the cumulative distribution of merging BHBs eccentricity calculated in different frequency windows, namely $f = 10^{-3},~10^{-2},~10^{-1},~1,~10$ Hz. 

We found that around $60\%$ of mergers have a noticeable eccentricity $e>0.1$ in the LISA frequency band \citep{LISA17}, i.e. $f = [10^{-4} - 10^{-2}]$ Hz, whereas this fraction drops to $3\%$ for eccentric mergers falling in the LIGO-Virgo sensitive band, $f>1$ Hz.

Interestingly, around one third of the mergers that are eccentric in the LISA band form from non-hierarchical triples. More specifically, $\gtrsim 75\%$ of non-hierarchical mergers are potential eccentric LISA sources, while this percentages decreases to $60\%$ for hierarchical mergers. This suggests that the chaotic evolution of non-hierarchical triples might have an higher efficiency in forming eccentric sources at least at low frequencies and for environments typical of GCs. 

\section{Numerical simulations of non-hierarchical triples}
\label{sec:an}

In the previous section, we have shown that non-hierarchical triples can constitute around two third of triples formed in binary scatterings in dense clusters. Our results suggest that mergers from non-hierarchical triples can constitute around $10\%$ of all mergers from triples and that around $80\%$ of them will have a significant eccentricity in the LISA band. Moreover, our preliminary analysis discussed in the previous section suggests that the configuration of the triple might play a role in determining the final fate of the inner BHB. In the following, we thus focus on the evolution of non-hierarchical and mildly hierarchical triples to assess the importance of the orbital configuration in determining the fate of the triple.

In an attempt to discern what parameters are most effective in determining the states emerging from the triple evolution, in the following we present and discuss  30,000 simulations gathered in 10 groups, performed using \argdf \citep{ASCD17b}. The main features and initial values for each set are summarized in Table \ref{tab:models}. Since in the case of co-planar orbits the argument of node $\Omega$ is ill-defined, in the table we refer to $\Delta \varpi$ as the difference between the inner and outer binary longitude of pericentre. In non-coplanar cases, this quantity reduces to $\Delta \varpi=\omega$, as we fixed all the other terms initially to 0.

Different from simulation SET0 (discussed in section \ref{sec:GC}), which includes initial conditions drawn from BHB-BHB scattering, we study a large series of non-hierarchical triple simulations (SET1-10) characterized by an inner binary with predefined propertie in this section. These sets of simulations allow us to carry out a systematic analysis on how the evolutionary outcome depends on the initial conditions for non-hierarchical triples. Simulation sets from SET1 to SET4 are designed to explore the role played by the inner binary eccentricity and the outer binary inclination, SET5 to SET8 allow us to quantify the dependence on the orbital orientation, while SET9 and SET10 are devoted to investigate the role of the tertiary component's mass and, more generally, the distribution of the outcome of the triple evolution if changing all the orbital parameters.
As shown by Table~\ref{tab:5}, in all the cases but SET9 and SET10, we set the inner binary pericentre to $R_i=20\au$ and masses $(20\Msun,10\Msun,20\Msun)$ implying $t_{\gw,i} = 10^{18}$ Gyr for a circular binary. 
In SET10, instead, we choose $R_i=1\au$, a value typical of binaries forming in the densest regions of globular and nuclear star clusters \citep{heggie75,miller09,antonini16} for which $t_{\gw,i} \simeq 3000$ Gyr for $e_i=0$.

We stop our simulations if one of three conditions is satisfied: 1) the triple  breaks, 2) two components merge, 3) the integration time exceeds 5 Myr. The latter conditions allows us to identify stable or meta-stable triples on this timescale. The initial value of the inner binary orbital period is in between $\sim 1-100$ yr in all the sets.

\subsection{Dynamics of bound triples: an interesting reference case}
\label{sec:num}

The behaviour of an unstable bound triple cannot be described through a secular approximation formalism, as the triple lifetime is usually comparable to a few orbital periods of the outer binary (see Figure \ref{sec:GC}). We look for the distinct statistical features in the evolution of these systems using direct $3$-body simulations. For the sake of simplicity, in the following the subscript $f$($i$) denotes final(initial) quantities.
First, let us assume the case of a BHB, comprised of BHs with masses $M_{1,f}=20\Ms$ and $M_{2,f}=10\Ms$, characterised by an initial eccentricity $e_i=0$ and semi-major axis $a_i=20\au$.
The BHB is orbited by a third BH, with mass $M_{3,i}=M_{1,i}$, initially placed at a pericentral distance $R_{3,i} = 200\au$ from the inner binary centre of mass, with eccentricity $e_{3,i} = 0.7$. In the following, we will refer to the BHB as the \textit{inner binary} and to the system composed by the outer BH and the inner BHB centre of mass as the \textit{outer binary}.

\begin{figure}
\centering
\includegraphics[width=8cm]{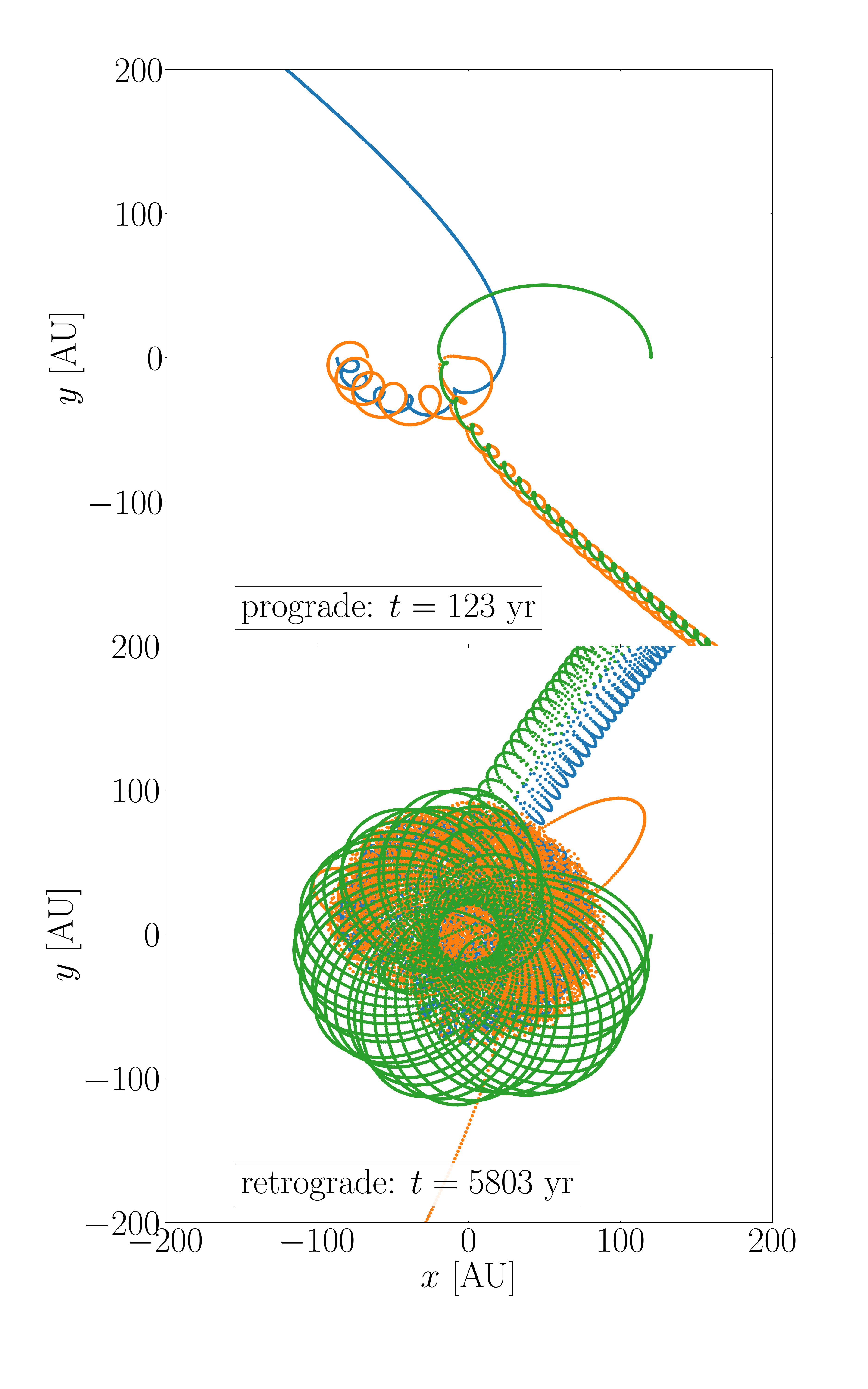}
\caption{BHB-BH trajectory for a non-hierarchical triple initially in a prograde (top panel) or retrograde (bottom panel) configuration. The time indicated in the figure marks the moment at which the triple breaks up.}
\label{snap}
\end{figure}

Figure \ref{snap} shows how the two systems evolve in time. In the case of an initial prograde configuration (top panel in Figure \ref{snap}), the triple breaks immediately after the first passage at pericentre ($\sim 100$ yr), the outer BH swaps with the lighter component of the initial inner BHB. The lighter BH is ejected and the new BHB is composed of two equal mass BHB with eccentricity $e_i\simeq 0.15$ and semi-major axis $a_i=24.3\au$.

\begin{figure}
\centering
\includegraphics[width=8cm]{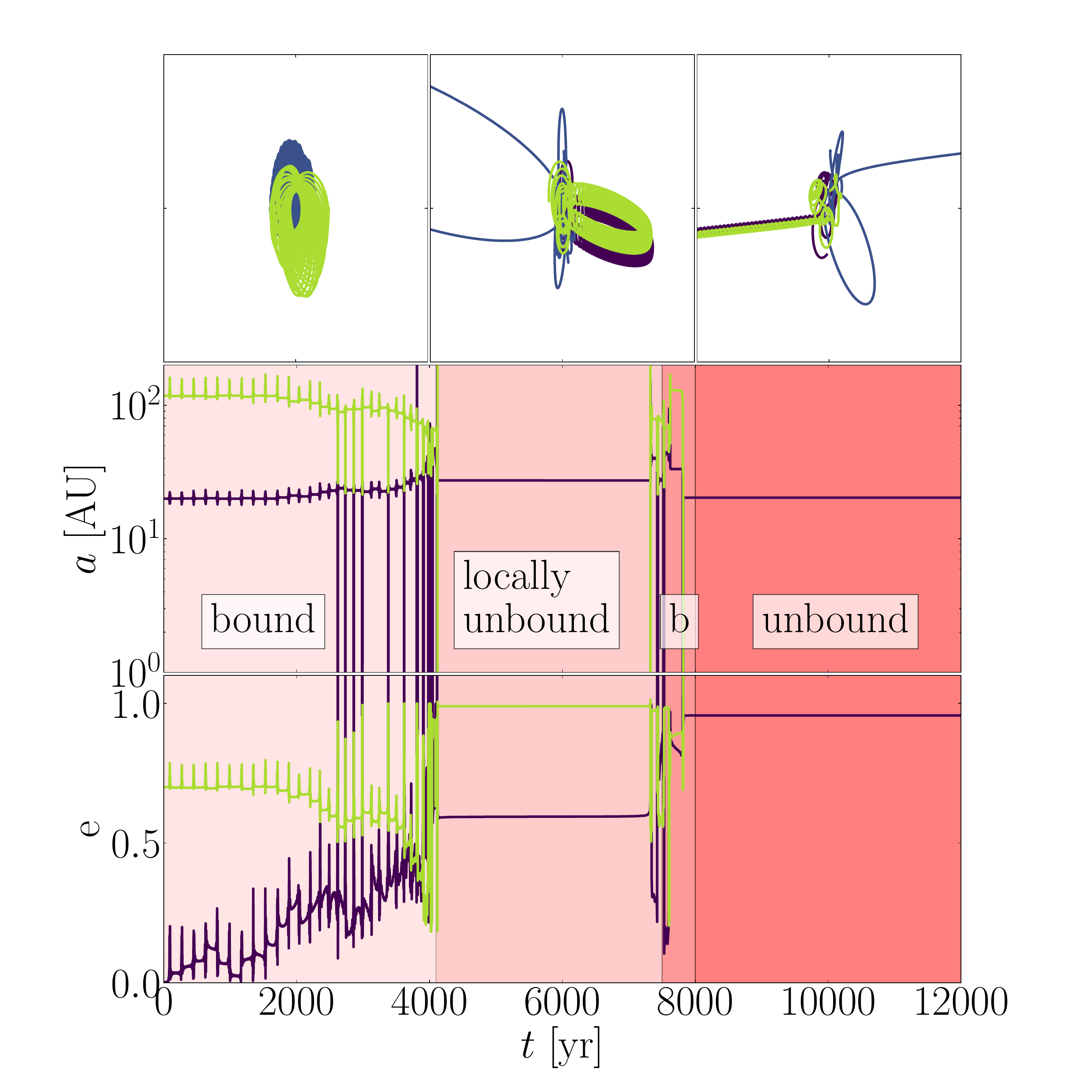}\\
\caption{From top to bottom: the semi-major axis and eccentricity of the inner (red solid line) and outer binary (black dotted line) for the triple shown in the bottom panel of Figure~\ref{snap}. The system shows a bound triple in the first $\sim$4000 yr, then one object marked with light blue is ejected on a weakly bound eccentric orbit, and ultimately it is ejected completely ejected on an unbound trajectory.}
\label{results}
\end{figure}

In the case of an initially retrograde configuration, instead, the overall picture is much more complex. The inner and outer binary orbits undergo mutual precession for $21$ orbital periods of the outer binary. During this phase, the eccentricities and semi-major axes develop several oscillations, shown in Figure \ref{results}, that culminate after $\sim 6\times 10^3$ yr in a series of strong and repeated scatterings. During this phase, which lasts for $\gtrsim 600$ yr, the inner binary changes its components frequently until the triple breaks up. The lighter object is again ejected and the resulting binary is now composed of two equal mass objects with mass $20\Ms$, a separation slightly larger than its initial value, $a_f= 25\au$, and a high eccentricity $e_f=0.999275$. The corresponding coalescence time-scale for such a system is $t_{\gw ,f} \simeq 1.68$ Gyr, thus suggesting that counter-rotation can facilitate coalescence in non-hierarchical triples. 

The overall evolution of the initially counter-rotating triple, sketched in Figure \ref{results}, can be divided in four main stages. In stage I, the inner and outer binary orbits undergo precession, the outer semi-major axis $a_3$ slowly decreases while $a$ increases. Moreover, the outer binary decreases its average eccentricity, while the inner binary eccentricity increases up to 0.6.
At some point (stage II) the three objects come close enough that the triple loses any kind of hierarchy. A short phase of scatterings drive the formation of a binary composed of the outer BH and the heavier component of the inner binary (red and blue lines in Figure \ref{results}) while the original secondary of the inner binary is pushed toward a wider, nearly parabolic but bound orbit. At the next close approach of the third object leads to a new series of strong interactions (stage III), which results in the ultimate disruption of the triple. This complex series of interactions leaves behind a highly eccentric binary composed of the two most massive members of the triple (stage IV).

The evolutionary differences between the initially prograde and retrograde cases could be related to the different level of stability in the two configurations, as made evident in the \cite{mardling01} stability criterion. In our reference model, $a_3/a \simeq 5.9$, while the critical value for stability (see Equation \ref{eq1} above) is $K = 12.7$ for a retrograde configuration and $K=18$ for a prograde configuration, thus $a_3/a<K$ in both cases. 
The triple is initially unstable and the evolution is chaotic \citep[see][for results in the non-coplanar case]{Grishin17,Grishin18}.

The simple model presented above shows at a glance the importance of the triple orbital configuration, which can lead to enormous differences in the final status of the system. Some of these results may be explained by energy and angular momentum transfer during the repeated close interactions. 
The initial total energy of the system is given by the sum of the binding energies of the inner and outer binary, namely
\begin{equation}
E_i = \frac{GM_{1,i}M_{2,i}}{2a_i} + \frac{G(M_{1,i}+M_{2,i})M_{3,i}}{2a_{3,i}} + H_{\rm int},
\end{equation}
where $H_{\rm int}$ the interaction energy term, which is usually small in the  hierarchical limit \citep{harrington68}.  
Assuming that the binary breaks up after several interactions, ultimately producing a binary and an unbound star, and that the energy taken away is much larger than the interaction energy, the final energy budget will be dominated by the binary binding energy
\begin{equation}
E_f = \frac{GM_{1,f}M_{2,f}}{2a_f}.
\end{equation}
The difference between the final and initial energy must equal the kinetic energy of the escaping third object and the centre of mass of the final binary, which is non-negative $(E_f - E_0) \geq 0$. Combining and re-arranging the equations above, yields a constraint on the final binary hardness 
\begin{equation}
\frac{M_{1,f}M_{2,f}}{a_f} \geq 
\left[1+\left(\frac{M_{3,i}}{M_{1,i}}+\frac{M_{3,i}}{M_{2,i}}\right)\frac{a_i}{a_{3,i}}\right]\frac{M_{1,i}M_{2,i}}{a_i}
> \frac{M_{1,i}M_{2,i}}{a_i}.
\label{energytr}
\end{equation}
Thus, the disruption of the triple inevitably leads to the formation of a harder binary, either by shrinking ($a_f<a_i$) or acquiring a heavier component ($M_{2,f}>M_{2,i}$) via component swap. In the representative example presented above, the final binary mass is $3/2$ larger than its initial value due to the swap of the outer BH and the secondary component of the inner binary, while its semi-major axis remains almost constant.

Let's now consider the triple angular momentum $\vec{L} = \vec{L}_\bhb + \vec{L}_3$. In the simplest approximation that the triple is initially coplanar, the norm of the angular momentum vector is simply $L_i = L_{\bhb,i} \pm L_{3,i}$, where the sign $+$($-$) identifies an initially prograde(retrograde) configuration. Let's now consider the case in which the triple flips configuration, given the angular momentum conservation we can write the final BHB angular momentum as
\begin{equation}
L_{\bhb,f} = 
\begin{cases}
L_{\bhb,i} + L_{3,i} + L_{3,f}     &~~  {\rm pro- ~to ~retrograde},\\
L_{\bhb,i} - L_{3,i} - L_{3,f}     &~~  {\rm retro- ~to ~prograde},
\end{cases}
\end{equation}
where the two equations hold for a flip from a prograde to a retrograde configuration and viceversa.
The two equations above tells us that in a pro-to-retrograde flipping the BHB acquires angular momentum ($L_{\bhb,f} > L_{\bhb,i}$), thus implying that its eccentricity decreases, whereas for a retro-to-prograde flipping $L_{\bhb,f}<L_{\bhb,i}$, thus the final eccentricity rises up.
Thus from energy and angular momentum conservation we find that
\begin{itemize}
\item the final binary is harder, i.e. either by becoming tighter or more massive, than the initial inner binary;
\item orbital flips can determine the increase (from retro- to prograde orbits) or decrease (from pro- to retrograde orbits) of the BHB eccentricity.
\end{itemize}

In the example presented above, note that the initial merger time for the inner BHB is $t_{\gw ,i} = 8.7\times 10^{18}$ yr, thus much larger than a Hubble time. However, in consequence of the triple evolution, the GW timescale of the outcoming BHB reduces in both the configurations, due to the efficient energy transfer that drives the component swap. In the prograde case the final BHB GW time is still larger than a Hubble time, but in the retrograde configuration the GW timescale drops by 9 order of magnitudes, i.e. $t_{\gw ,f} = 1.68$ Gyr. Hence, this process facilitates the formation of merging binaries.
In the next section, we explore in detail how GW times change in non-hierarchical and unstable triples, and what orbital configurations favour a more efficient shrinkage of the inner BHB. 

\subsection{Distribution of GW merger times}

\begin{figure}
\centering
\includegraphics[width=8cm]{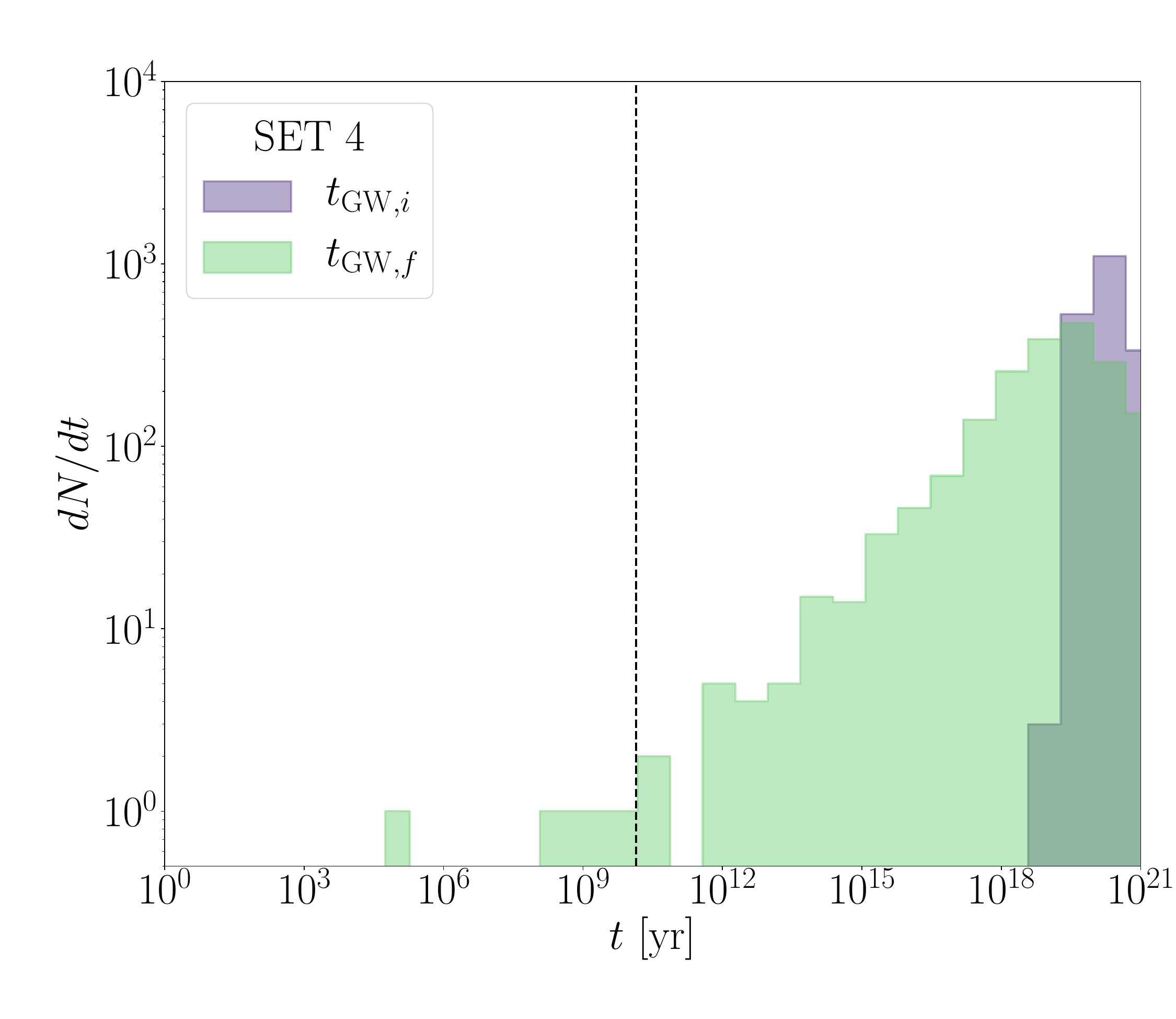}\\
\includegraphics[width=8cm]{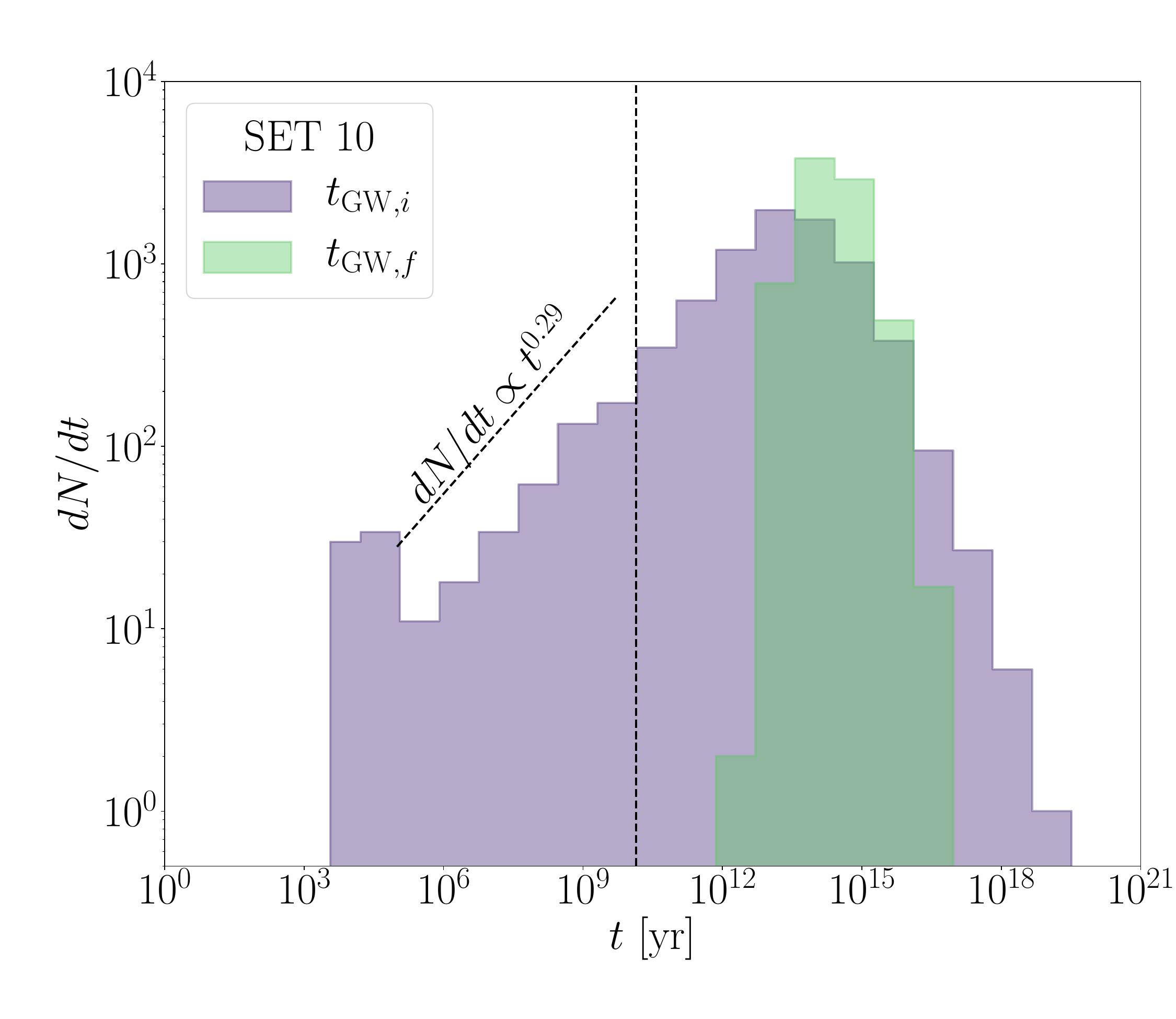}
\caption{Initial (filled red boxes) and final (blue steps) GW merger time distribution for all BHBs in SET4 (top panel) and SET10 (bottom panel) (see Table~\ref{tab:models} for parameters). The dotted vertical line marks $t_\gw = 14$ Gyr. A broad tail extending to low GW timescales are produced with $dN/d\ln t_{\rm GW} \propto t_{\rm GW}^{0.32}$ for SET4 and $t_{\rm GW}^{0.29}$ for SET10, as expected (see text).   
}
\label{tgwdis}
\end{figure}

As a first step, we exploit SET4 and SET10 to explore how the evolution of a non-hierarchical triple affect the BHB merger time. In these simulation sets (see Table~\ref{tab:models}): a) the initial pericentre is fixed at $20\au$ and $1\au$, respectively; b) the initial eccentricity and inclination are widely distributed; and c) the BH masses are either fixed or allowed to vary, respectively.
Figure \ref{tgwdis} compares the initial and final distribution of the merger times for all the models in SET4 and in SET10. The broadening of the merger time distribution is apparent in both cases, falling below 14 Gyr in 5 cases in SET4 ($0.25\%$ of the total sample) and up to 677 cases in SET10 ($8.5\%$). 

\begin{figure}
\centering
\includegraphics[width=\columnwidth]{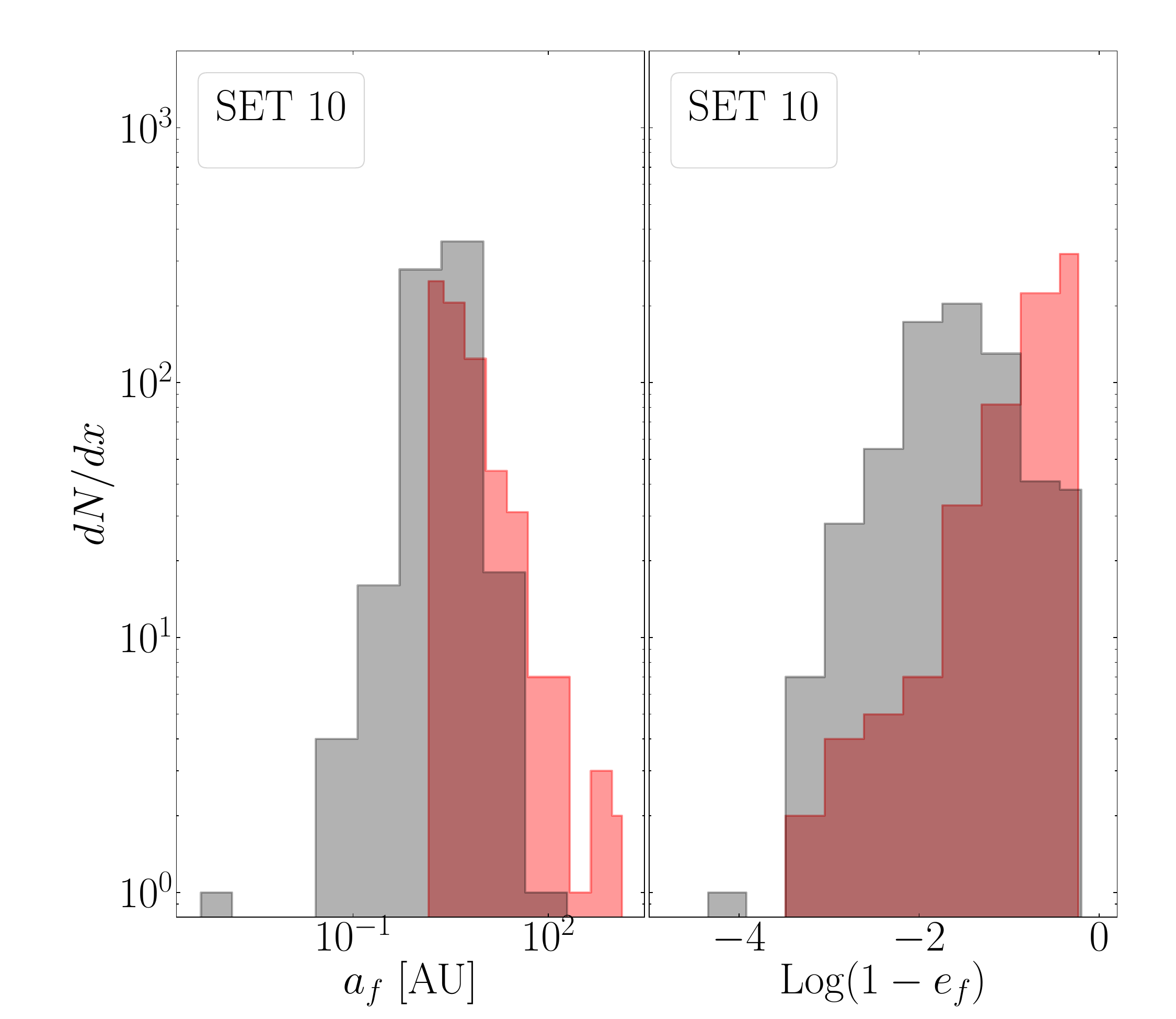}
\caption{Semi-major axis (left panel) and eccentricity (right panel) distribution for the initial inner BHB population (red filled boxes) and for mergers only (black steps) in SET10.}
\label{sGW}
\end{figure}

The results show that the GW merger time distribution follows approximately a power-law 
\begin{equation}
\frac{dN}{d \ln t_{\gw ,f}} \propto t_{\gw ,f}^{0.3}
\end{equation}
or equivalently 
\begin{equation}
\frac{dN}{d t_{\gw ,f}} \propto t_{\gw ,f}^{-0.7}
\end{equation}
for both SET4 and SET10 for GW time-scales less than its initial value. This outcome is mostly due to the fact that $t_{\gw ,f}$ given by equation (\ref{peters}) is most sensitive to the eccentricity, approximately as $t_{\gw ,f}\propto (1-e_f)^{7/2}$. This implies that $(1-e_f)\propto t_{\gw ,f}^{2/7}$, and so for all other parameters fixed, we get
\begin{equation}
\frac{dN}{d t_{\gw ,f}} = \frac{dN}{d (1-e_f)} \frac{d (1-e_f)}{d t_{\gw ,f}} \approx  \frac{dN}{d (1-e_f)} t_{\gw ,f}^{-5/7}\,.
\end{equation}
Thus, if $1-e_f$ is approximately uniformly distributed after the encounters, the distribution of GW time-scale follows $t_{\gw ,f}^{-5/7}$, i.e. $t_{\gw ,f}^{-0.71}$. A similar argument shows that for fixed $e_f$, $dN / d t_{\gw ,f} \propto  (dN/da_f)\times t_{\gw ,f}^{-3/4}$. However, as shown in Figure \ref{sGW}, we find that the final semimajor axis distribution spans typically only a decade around the initial semimajor axis distribution and many of the merging binaries have an eccentricity very close to 1 following the three body dynamics. Thus the eccentricity distribution after the three body dynamics is crucial for the GW merger time distribution.

\subsection{Effects of initial eccentricity, inclination, and argument of pericentre}

In this section we focus on the role of the combined effects arising from the inner binary eccentricity and the outer binary inclination. The values allowed for the outer pericentre $R_{3,i}$ are, depending on the simulations set, between 1 and 7 times the initial inner binary pericentre which typically represents a non-hierarchical systems. Fixing the ratio between the initial pericentre of the inner and outer orbit and allowing the eccentricities to vary implies the possibility in some cases that $R_i/(1-e_i) > R_3/(1-e_{i,3})$, thus implying that the semimajor axis of what we identify as inner BHB is actually larger than the outer. Therefore, in all the analysis presented below we get rid of all models that initially have $R_i/(1-e_i) > R_3/(1-e_{i,3})$, corresponding to around $10\%$ of all models. Due to the assumptions made, the ratio $a_{3,i}/a_i$ ranges between $1$ and $10^3$. Note that in set BIN2 (BHB-BHB scatterings) we find $94\%$ of triples having semimajor axis ratio in the same range, thus this kind of configuration describe the majority of triples formed in BHB-BHB strong encounters. The eccentricity is selected according to a thermal distribution \citep{jeans19} both for $e_{3,i}$ and $e_i$, although in some sets the value of the inner BHB eccentricity is kept fixed, as listed in Table \ref{tab:models}. 
We selected a co-planar, prograde configuration with $e_i=0$ as a reference model for SET1 simulations.
In order to shed light on the importance of the orbital parameters, we then assumed $0\leq i_i <10^\circ$ in SET2 and, along with this choice, we vary also $e_i$ in SET3. In SET4 we varied the outer inclination in between $0\leq i_i<180^\circ$, thus investigating both prograde and retrograde orbits, and $e_i<1$, thus investigating the evolution of both circular and nearly radial BHBs.
This sample of 5000 simulations gathered in 4 groups represents a quick way to highlight the differences between prograde and retrograde tight triple systems. These differences are further investigated in SET5 and SET6, where we examine exactly co-planar orbits in a co-rotating and counter-rotating configuration, respectively, with fixed inner binary eccentricity at $e_i=0.6$ and random $\Delta \varpi_i$. SET7 and SET8 are nearly coplanar configurations where the eccentricity is drawn with $0\leq i_i<10^\circ$ and $170^{\circ}\leq i_i <180^\circ$, respectively and unlike in SET3 we set $\Delta \varpi_i=\pi$. 

\begin{figure}
\includegraphics[width=8cm]{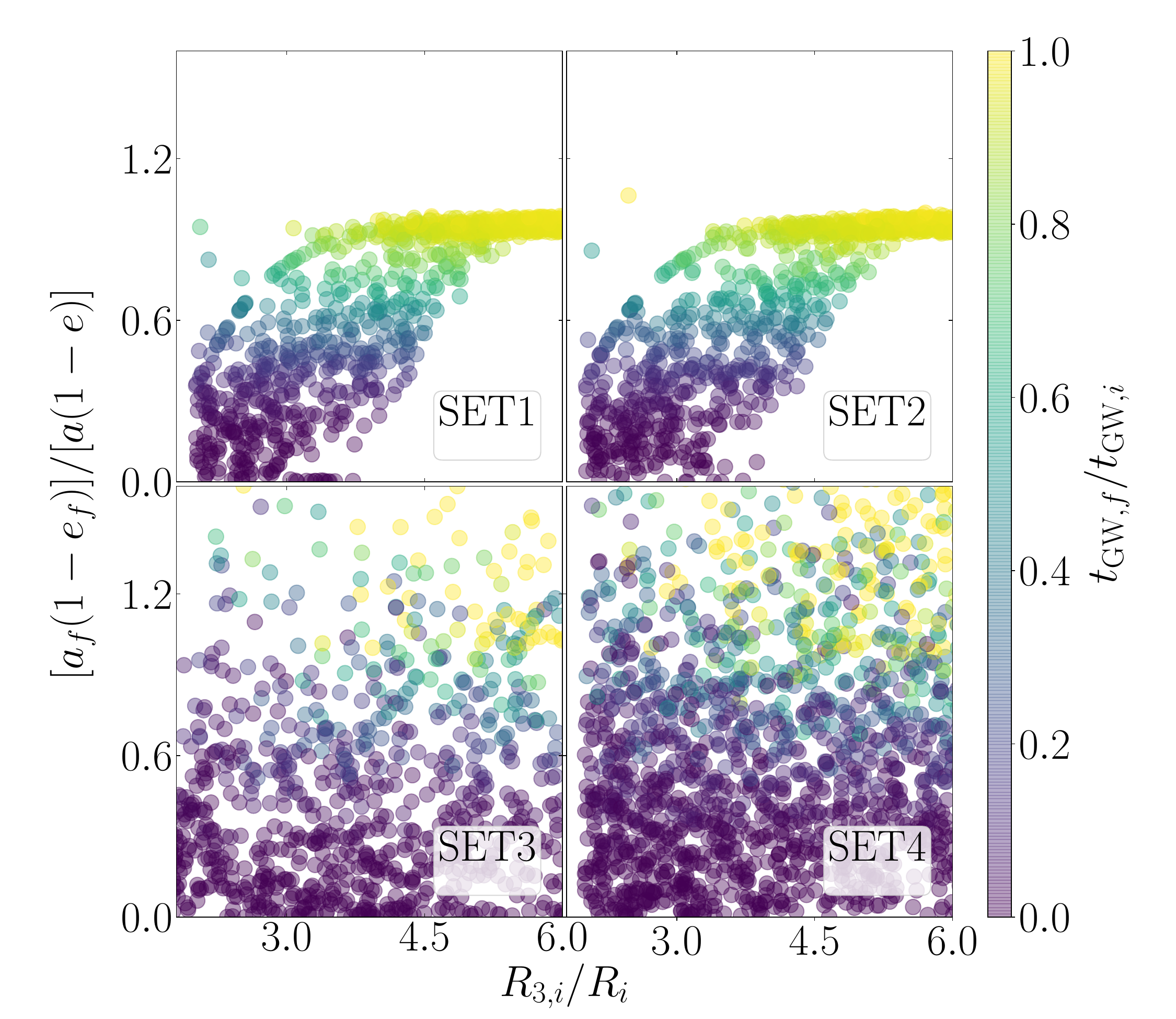}\\
\includegraphics[width=8cm]{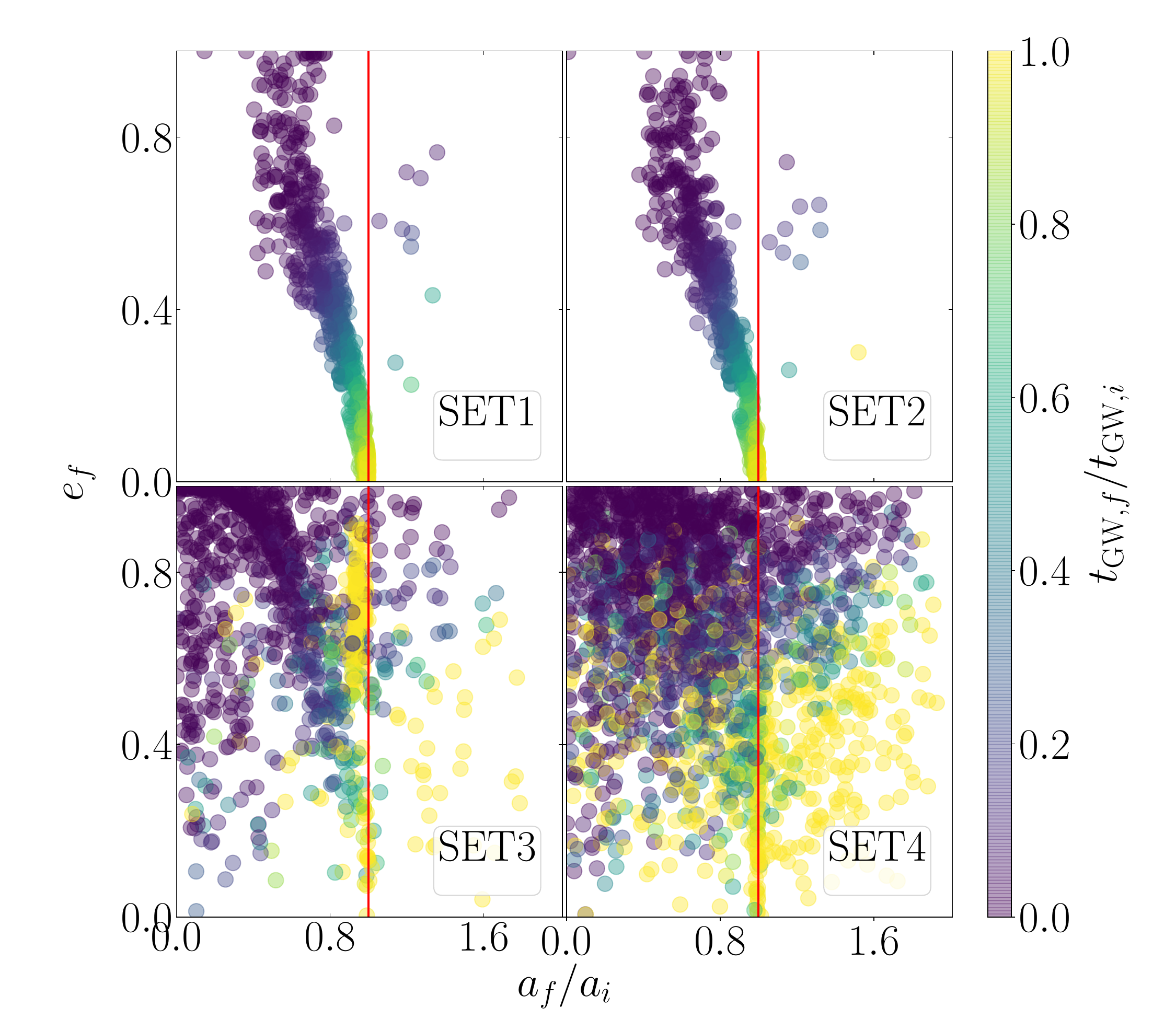}
\caption{Ratio between the initial and final value of the inner binary pericentre with respect to the outer binary initial separation, normalized to the inner BHB initial semi-major axis, $a_i=20\au$. From left to right and from top to bottom panels refers to SET1 ($e_i=0$, $\Delta\varpi_i = 0-2\pi$, $i_i = 0^\circ$), SET2 ($e_i=0$, $\Delta\varpi_i = 0$, $i_i = 0-10^\circ$), SET3 ($e_i=0-1$, $\Delta\varpi_i = 0$, $i_i = 0-10^\circ$) and SET4 ($e_i=0-1$, $\Delta\varpi_i = 0$, $i_i = 0-180^\circ$), respectively, see Table~\ref{tab:models}. The colours represent the ratio between the final and initial value of the GW time-scale for the inner binary. Bottom panels: same as in top 4 panels, but here we represented the inner binary final eccentricity vs. the semi-major axis ratio.
}
\label{evsR3}
\end{figure}

The top 4 panels in Figure \ref{evsR3} shows the ratio between the final and initial value of the inner binary pericentre, $r_{p,f}/r_{p,i}$, plotted against the ratio of the outer binary's to the inner binary's apocentre initially, denoted by $R_{3,i}/R_i$, in the case of SET1 to SET4. 
The figure clearly shows that $R_{3,i}\lesssim 5R_i$ is necessary for a significant decrease of the pericentre distance and the GWs time-scale for a nearly coplanar triple. 

On the other hand, allowing both $i_i$ and $e_i$ to vary (SET3 and SET4) makes the distribution of $e_f$ and $a_f$ much broader for all $R_{3,i}$ values. Indeed, in this case the third object can lead to a significant reduction of the inner binary pericentre and $t_{\gw,i}$ even if it initially moves far from the inner binary. The bottom 4 panels of Figure \ref{evsR3} show the inner binaries' final eccentricity and the factor by which the semi-major axis changes. In SET1 and SET2 the semimajor axis is typically conserved to within a factor of 2 but the eccentricity is broadly distributed up to almost unity, also showing a correlation with $a_f/a_i$. SET3 shows that when the inner binary initial eccentricity is allowed to vary then its semimajor axis can shrink by a larger factor and there is a larger fraction of highly eccentric sources. There is no prominent clustering/correlation in the case of SET4.
We conclude that all orbital elements play a crucial role in determining the outcome of the triple evolution. Next we discuss the effect of the outer and inner eccentricity for nearly coplanar, prograde systems (SET1, SET2 and SET3) in more detail, and discuss the outer binary inclination in the next section.

\subsection{Pro- or retrograde: does it matter?}

\begin{figure}
\centering
\includegraphics[width=8cm]{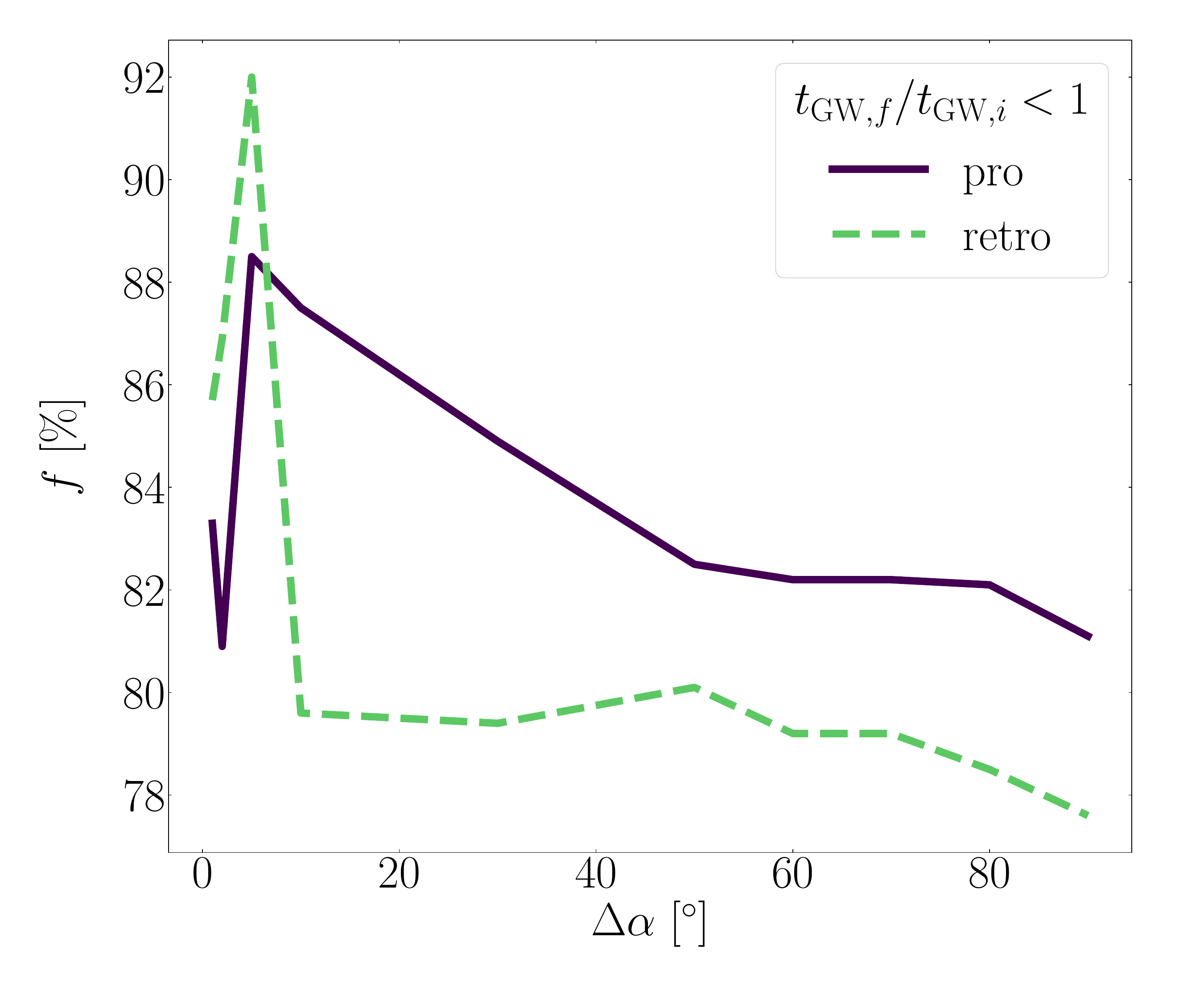}\\
\includegraphics[width=8cm]{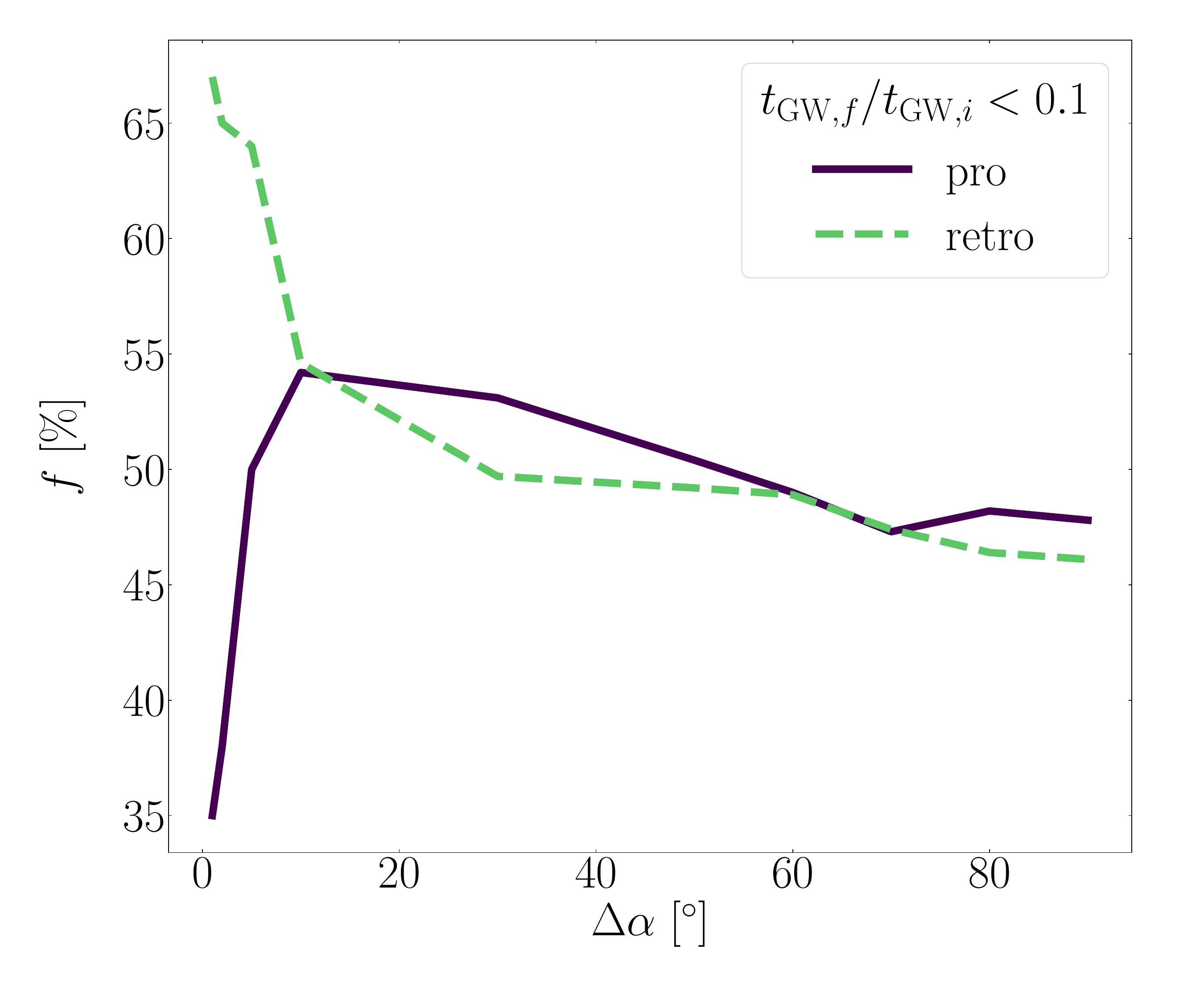}\\
\includegraphics[width=8cm]{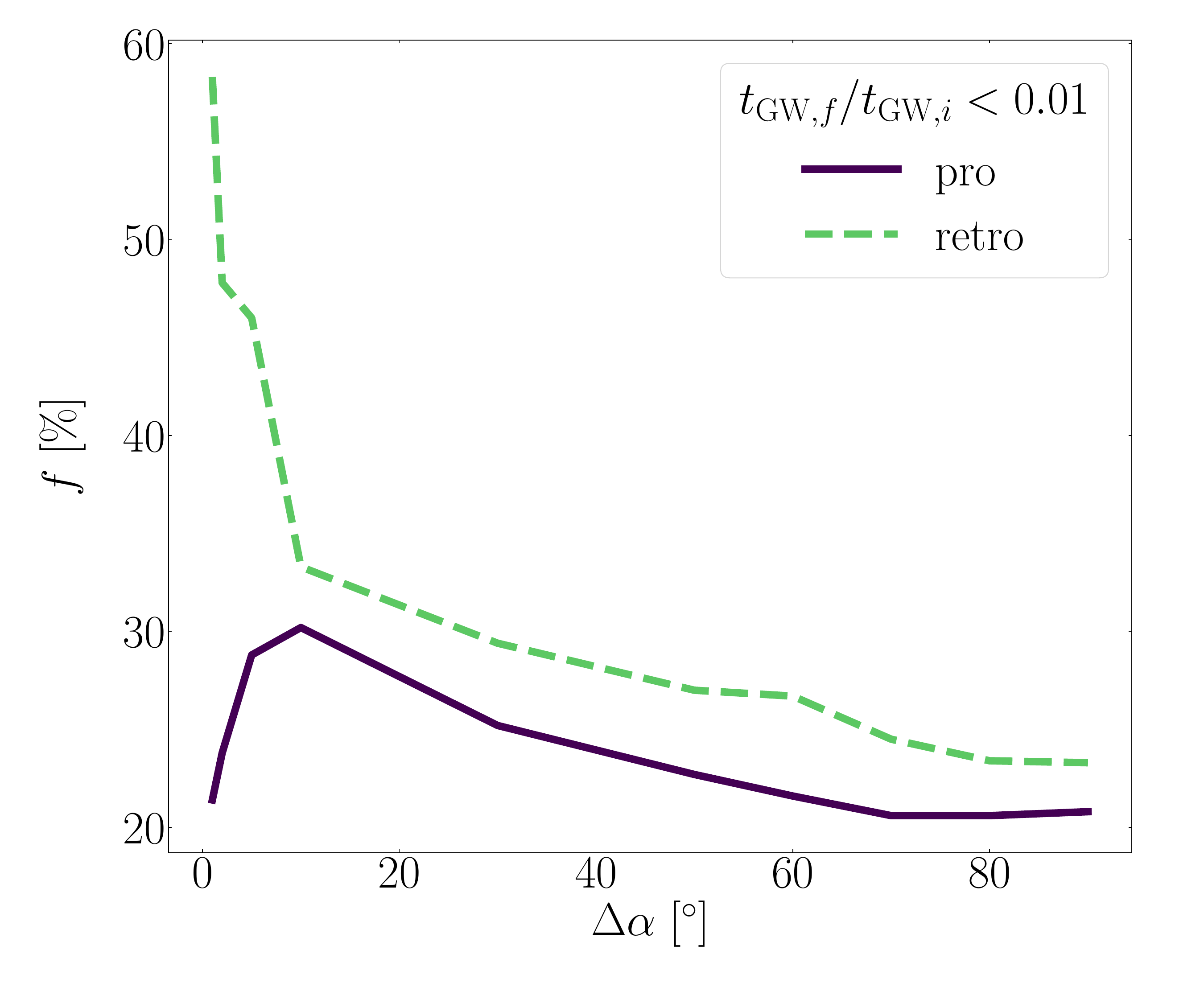}\\
\caption{Percentage of models with $i_i<\Delta\alpha$ (red solid line) or $i_i>\Delta\alpha$ (black dotted line) and $t_{\gw,f}/t_{\gw,i}<1, ~0.1$ and $0.01$ (from top to bottom) in SET4. Nearly coplanar retrograde orbits have the highest probability to greatly reduce the GW timescale. 
}
\label{new1}
\end{figure}

In order to shed light on the evolution of triple systems, we calculated for SET4 the fraction of models that satisfy a particular constraint among all orbits that have an initial inclination smaller than a given value $\Delta \alpha$ and larger than $\pi - \Delta \alpha$. In particular we examine the fraction of simulations for which the merger time is decreased by a factor 1, 0.1 and 0.01, as shown in Figure \ref{new1}.
For instance, setting this fraction at $\Delta \alpha = 5^{\circ}$ means that we consider only models for which $i_i < 5^\circ$ (almost coplanar prograde) or $i_i > 175^\circ$ (almost coplanar retrograde)
and calculate the percentage of cases in which the merger time decreases by at least the given factor. We verified that in both the prograde and retrograde subsamples selected at varying $\Delta \alpha$ the number of objects is similar, as so the implications in the two cases have similar statistical significance.

In the top panel of Fig. \ref{new1} we show the fraction of binaries for which the GW timescale decreases as a function of $\Delta \alpha$ in the prograde and retrograde cases, respectively. The middle and bottom panels show cases when the GW timescale decreases by at least a factor 10 or 100, respectively.
We find that with the exception of nearly coplanar orbits $\Delta \alpha<10^{\circ}$, prograde and retrograde orbits generate similar fractions at fixed $\delta \alpha$ to within $\sim 10\%$, with retrograde orbits producing slightly higher probability to decrease the GW timescale by a factor 100. The striking result of the figure is that the probability of decreasing the GW timescale by a factor of 100 is highly peaked at coplanar retrograde configurations.

More interestingly, the percentage of cases for which $t_{\gw ,f}/t_{\gw,i} < 100$ is much higher for retrograde orbits, as shown in Table \ref{tab2}. This implies that retrograde configurations drive the formation of much tighter BHBs than prograde, thus being characterised by much shorter GW timescales.

After a few passage at pericentre, the triple can either break up or remain in a meta-stable or stable configuration, depending on the initial conditions. 
For instance, the triple disruption occurs in $93.3\%$ of the models in SET4, thus outlining the chaotic nature of such configurations. In the following, if the triple disrupt and leaves behind a BHB and a BH, we calculate the inclination promptly after the disruption, thus measuring the angle between the angular momenta directions immediately after the BH ejection.

Under this assumption for disrupted models, we calculate the percentage of models that are in a prograde,$\eta_{\rm pro}$, or retrograde, $\eta_{\rm ret}$, configuration by the end of the simulation.

This quantity allows us to quantify the occurrence of orbital flipping in our models. 
As long as the initial conditions assume an almost co-planar prograde configuration and zero eccentricity for the inner orbit ($i_i<10^\circ$, SET1 and SET2) flipped systems are less than $0.02\%$. An eccentric inner binary (SET3) leads this percentage to slightly increase up to $3.4\%$. In the most reliable case in which the inclination is initially equally distributed among prograde and retrograde configurations (SET4) we find a clear excess of prograde systems ($80.5\%$) compared to retrograde ones ($19.5\%$). This likely implies that there is a preferential ``direction'' toward which this kind of system evolves. In the next section we will deepen this investigation.

\begin{table}
\centering{}
\caption{Percentage of tightened BHBs.}
\begin{center}
\begin{tabular}{cccccc}
\hline
SET & $f_{1.0}$ & $f_{0.01}$& $\eta_{\rm pro}$ & $\eta_{\rm ret}$ \\
    & ($\%$) & ($\%$) & ($\%$) & ($\%$)	\\
\hline
1 & 100 &8.2   & 100  & 0    \\
2 &99.9 &8.6   & 99.8 & 0.02 \\
3 &90.0 &35.8  & 98.2 & 1.8  \\
4 &78.4 &20.9  & 81.0 & 19.0 \\
\hline
\end{tabular}
\end{center}
\begin{tablenotes}
\item Column 1: Model name. Column 2(3): percentage of models in which the GW time-scale of the inner binary reduces by a factor 1(100). Column 4(5): percentage of models ending up in a prograde (retrograde) configuration. 
\end{tablenotes}
\label{tab2}
\end{table}

\begin{figure}
\centering
\includegraphics[width=8cm]{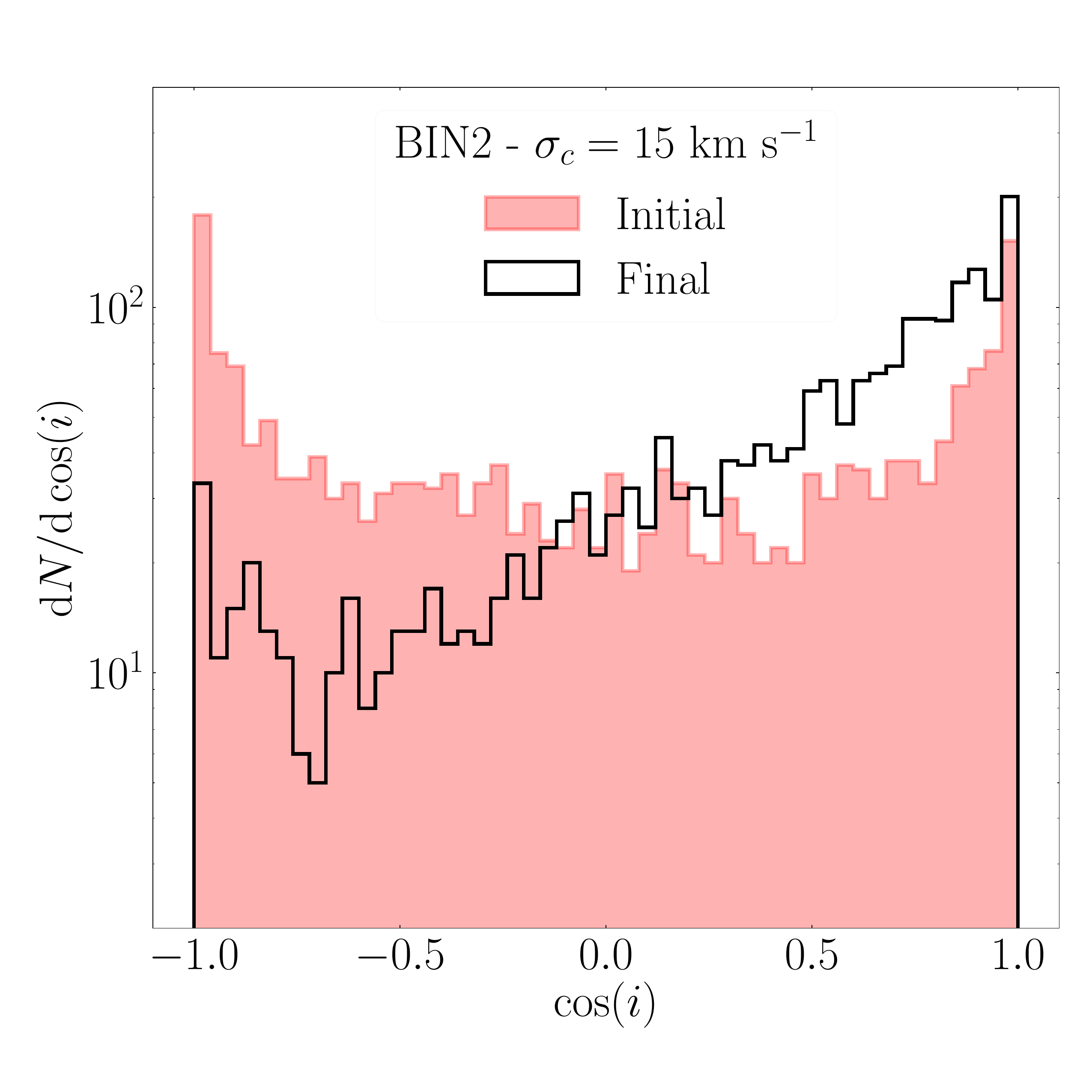}
\caption{Initial (filled red steps) and final (empty black steps) inclination distribution in SET4.}
\label{fig:inc}
\end{figure}

This trend is outlined by Figure \ref{fig:inc}, which shows a comparison between the initial and final distribution of the inclination cosine for all the models in SET4. Despite being initially nearly flat, the distribution is evidently unbalanced after the triple evolution, due to the orbital flipping more efficient for retrograde systems. At values $\cos(i_f) > -0.8$ the distribution seems well described by a power-law. At smaller values, corresponding to inclinations $i_f>140^\circ$, the power-law changes slope and maximizes for co-planar retrograde systems.

\subsection{The impact of coplanarity, inclination, $e$ and $\varpi$: orbital flipping and BHB merger}

SET5 to SET8 are designed to better understand nearly coplanar systems and to further constrain the differences between prograde and retrograde configurations.

In SET5 to SET8, the triple has a large probability to produce a binary and an unbound single. In coplanar prograde models (SET5), we find the maximum number of disruptions, $N_{\rm dis} = 1908$, corresponding to $\sim 95.4\%$ of the models investigated. Nearly coplanar retrograde models (SET7) have a somewhat smaller number of disrupted triples, $N_{\rm dis} = 1753$, at around $87.7\%$. The BHB emerging from the triple break-up has a final merger time at least 10 times smaller than its initial value in over $45.7 - 67.2\%$ of the cases, depending on the initial inclination, and maximizes in the case of co-planar retrograde orbits with a mildly eccentric outer binary (SET6). A comparison between all these quantities, summarized in Table \ref{tab:5}, suggests that the most effective configuration in producing merging BHBs is fully co-planar and retrograde. 

\begin{table*}
\centering
\caption{Main results for SET4 to 8}
\begin{center} 
\begin{tabular}{ccccc|cccccc}
\hline
SET & $\langle i_i\rangle$ & $f_{\rm mer}$ & $f_{0.1}$ & $f_{\rm dis}$ & $\eta_{\rm pro}$ & $f_{p1}$ & $f_{p0.01}$ & $\eta_{\rm ret}$&  $f_{r1}$ & $f_{r0.01}$  \\
& $\circ$ & ($\%$)& ($\%$)& ($\%$)& ($\%$)& ($\%$) & ($\%$) & ($\%$) & ($\%$) & ($\%$)\\
\hline
$4$ & $ 90$ & $ 0.25$ & $45.7$ & $93.4$  &$81.0$ & $77.6$ & $19.3$ & $19.0$ & $81.7$ & $27.3$ \\
$5$ & $  0$ & $ 0.7$  & $44.8$ & $95.4$  &$99.3$ & $80.2$ & $20.4$ & $0.7$ & $100.0$ & $100.0$ \\ 
$6$ & $180$ & $ 3.7$  & $67.2$ & $90.8$  &$77.6$ & $90.5$ & $49.5$ & $22.4$ & $68.8$ & $60.9$ \\
$7$ & $  5$ & $ 0.85$ & $57.6$ & $94.5$  &$98.1$ & $86.9$ & $29.3$ & $1.9$ & $96.9$ & $96.9$ \\
$8$ & $175$ & $ 0.7$  & $51.5$ & $87.7$  &$66.0$ & $80.3$ & $32.9$ & $34.0$ & $70.4$ & $26.1$ \\
\hline
\end{tabular}
\end{center}
\begin{tablenotes}
\item Col. 1: model name. 
Col. 2: average inclination. 
Col. 3-5: percentage of mergers, models with reduced GW time and disrupted models, respectively.
Col. 6: final percentage of prograde systems. 
Col. 7-8: percentage of prograde cases in which $t_{\gw,f}<t_{\gw,i}$ or $t_{\gw,f}<10^{-2}t_{\gw,i}$. 
Col. 9: final percentage of retrograde systems. 
Col. 10-11: percentage of retrograde cases in which $t_{\gw,f}<t_{\gw,i}$ or $t_{\gw,f}<10^{-2}t_{\gw,i}$.
\end{tablenotes}
\label{tab:5}
\end{table*}

Moreover, Table \ref{tab:5} shows the final number of models in a prograde and retrograde configuration, respectively, for each simulations set, along with the percentage of those models having a GW time ratio decreased below 1 or $10^{-2}$ of its original value.
In SET5, where we assumed an initially prograde configuration, we found that only $0.7\%$ of models evolve toward a retrograde configuration. In all these few cases, $t_\gw$ decreases by a factor 100 ($f_{r0.01}=100\%$), while among those that preserve the initial prograde configuration the merger time decrease is less effective, being $f_{p0.01}=20.4\%$.
On the other hand, when the initial configuration is retrograde (SET6), the triple tends to change configuration to prograde in $2/3$ of the cases. Note that those triples which flip into a co-rotating configuration shrink as well, thus having a reduction in the GW time-scale, but the probability for the merger time to decrease by a factor 100 is around $f_{p0.01}\lesssim50\%$, while it is higher for simulations in which the initial counter-rotating property did not change, being $f_{r0.01}\simeq 61\%$. Allowing small deviations from coplanarity and a distribution in the initial eccentricity does not change these results qualitatively. At initially prograde systems at low inclinations (SET7, $i<10^\circ$), only a handful triples change to retrograde, with $96.9\%$ of them achieving a factor 100 reduction in $t_{\rm GW}$. For initially retrograde systems at inclinations above $i>170^\circ$ (SET8), instead, $66\%$ of the models flip but only $30\%$ of them have $t_{\gw,f}/t_{\gw,i}<0.01$. Therefore, as a general result we note that the criterion $t_{\gw,f}/t_{\gw,i}<0.01$ is more common in retrograde cases which flip. From this comparison, we identify the following general properties of triple systems:
\begin{itemize}
\item triple systems tend to evolve toward co-rotation;
\item the orbital flip causes, on average, the formation of a tighter inner binary system;
\item \textit{nearly}\footnote{i.e. $i_i<10^{\circ}$ or $i_i>170^{\circ}$} coplanar initially prograde and retrograde systems have a similar probability to decrease the GW timescale by a factor 100, but \textit{exactly} coplanar retrograde systems are $2.5$ times as likely to decrease the GW timescale by a factor 100 than \textit{exactly} coplanar prograde systems.
\end{itemize}

\begin{figure}
\centering
\includegraphics[width=8cm]{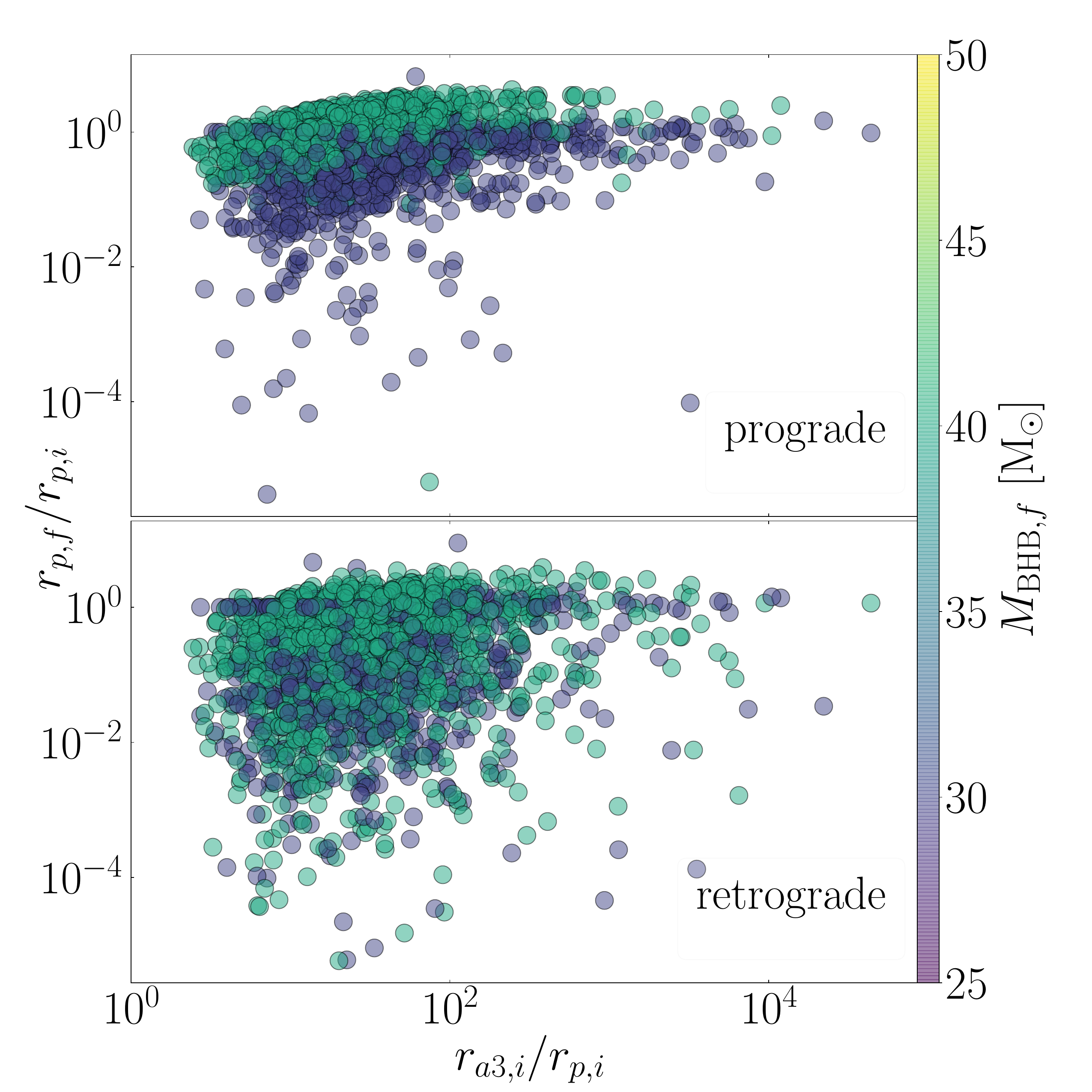}\\
\caption{Final inner binary pericentre as a function of the outer binary initial semi-major axis, both normalized to the initial pericentre of the inner binary for prograde (SET5, top panel) and retrograde (SET6, bottom panel) systems. Binaries maintaining their original components ($10\Msun+20\Msun$) are identified with blue dots, whereas those exchanging the lighter component are green dots and identify final binaries with component masses $20\Msun+20\Msun$. The final binary total mass is identified by the coloured map.}
\label{swap}
\end{figure}

The final BHB components show clearly that inclination also affects the capacity of the outer BH to take the place of the lighter BHB components. 
Figure \ref{swap} shows the ratio final-to-initial pericentre ratio $r_{p,f}/r_{p,i}$ as a function of the ratio between the outer apocentre and inner pericentre, $r_{a3,i}/r_{p,i}$, for models that either preserve or exchange one of the inner BHB components in SET5 (prograde co-planar) and SET6 (retrograde co-planar). In initially prograde configurations, we see that swapped binaries tend to preserve their original orbits, being $r_{p,f}/r_{p,i}>0.05 - 0.1$, whereas for "original" binaries this ratio can fall below $10^{-4}$ in a few cases, especially if $r_{a3,i}/r_{p,i} < 100$. For retrograde models, instead, the pericentre ratio broadly distribute in the range $10^{-5}-1$ regardless of the component swap and the mass of the secondary companion, thus suggesting that in retrograde systems the exchange of a component does not impact the overall shrinkage of the binary.

\section{Unravelling mergers' host clusters with gravitational waves}
\label{sec:gw}
As we discussed in Section \ref{sec:BBHfull}, BHB mergers developing in triples formed in clusters with $\sigma = 5-15$ km s$^{-1}$ can appear as eccentric sources in the LISA and, less frequently, in the LIGO-Virgo observational window. We have seen that the properties of the merging BHBs from non-hierarchical triples depend on the triple configuration and orbital parameters. In this section we exploit SET9 and SET10 to investigate whether is possible to place constrain on the environment in which a BHB merger develops on the basis of the associated GW signal. To do this, we first note that all the models presented here have inner BHBs with a fixed initial pericentre of either 20 or 1 AU. However, as shown above, dense stellar systems can contain even harder BHBs. To extrapolate our results for the sub-AU regime, we rescale the length scale in SET10 by a factor $1/l$ where $l>1$. For instance, circular binaries in the rescaled set have $a_i = 1/l\au$. Note that the eccentricity distribution, as well as the distribution of all the other parameters, remain unaltered.
Assuming Newtonian dynamics, the rescaling is equivalent to changing the time unit and the gravitational constant for a given set of $N$-body simulations. We discuss in the following the impact of unstable triples evolution on GW astronomy. In the following, we analyze the properties of the GW signal emitted by merging binaries in SET 10, scaling conveniently the results under the assumption of either $l=1$ or $l=100$. Note that the former, which implies $a_i=1\au$ for circular binaries, leads to values typical of low-density clusters, like open or young massive clusters, the latter, i.e. $a_i=0.01\au$, is more typical in GCs and galactic nuclei. Therefore, uncovering potential differences between these two kinds of merging BHBs can provide useful clues to infer the origin of observed GW sources.

In SET9, where the triple configuration is similar to SET4 but the BH masses are allowed to vary, we find a merger probability of $P_{\rm mer,SET9} = 0.25\%$ (i.e. 15 mergers out of 8000 simulations), pretty similar to SET4. This similarity is due to the fact that the GW time decreases most when the initial inner BHB mass is similar to that of the outer BH ($M_{\bhb,i} / M_{3,i} \sim 1-3$). Out of 15 mergers, we find 6 with an eccentricity above 0.1 when the transiting in the mHz frequency window, and only 1 eccentric source in the $0.1-1$ Hz range. 
In SET 10, instead, we find 677 merger events, corresponding to a $P_{\rm mer, SET10} = 8.5\%$. This is clearly related to the choice of an initially tighter system. Among all the mergers, 273 occur in a prograde configuration, $P_{\rm mer, pro} = 6.8\%$ of all the prograde models, while the remaining 404 come from retrograde configurations, $P_{\rm mer, ret} = 13.6\%$. This further confirms our earlier finding that, on average, models tend to evolve toward co-rotation, while the minority evolving from a co- to counter-rotation configuration shrinks more efficiently and merges with a higher probability.
As shown already in Figure \ref{tgwdis}, the final GW time distribution exhibits two distinct features: a small peak corresponding to $t_{\gw,f} = 10^4$ yr, and increasing trend in the time range $10^6-10^{10}$ yr, well described by a power-law ${\rm d}N/{\rm d} \ln t_{\gw ,f}\propto t_{\gw ,f}^{0.30\pm 0.02}$. Compared to the initial distribution, the final GW time distribution is much broader. Such broadening is caused by the tertiary BH perturbations, which increases the final eccentricity to values near unity.

\begin{figure}
\centering
\includegraphics[width=8cm]{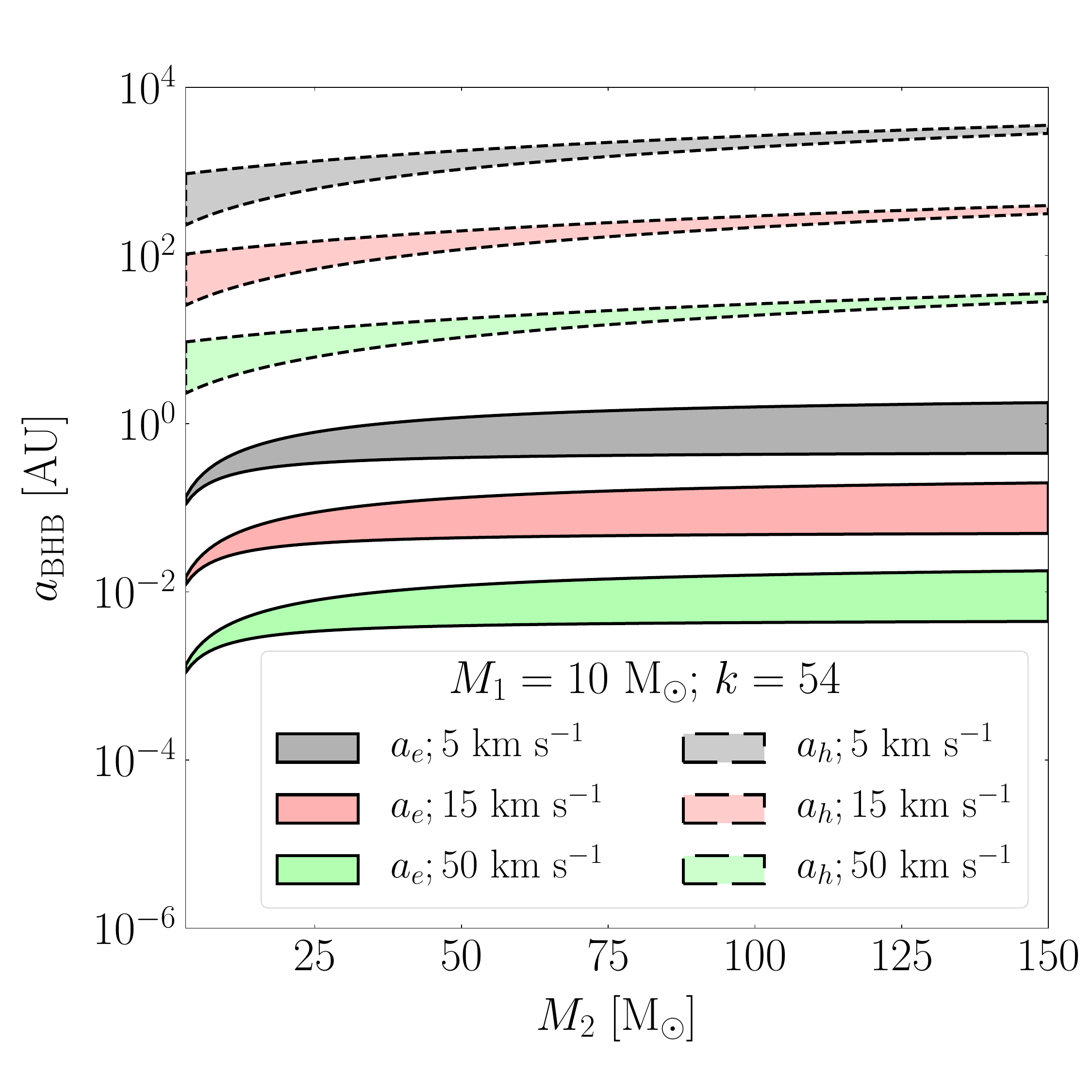}
\caption{Hard binary criterion $\Gamma_h$ as a function of the BHB semi-major axis and for different value of the host system velocity dispersion, with the upper(lower) limit corresponding to a BHB mass $M_\bhb = 30$($10$)$\Ms$. The horizontal black line marks the boundary between hard and soft binary configurations.}
\label{hard1}
\end{figure}

All the models presented here have inner BHBs with a fixed initial pericentre of either 1 or 20 AU. However, dense stellar systems can contain much harder BHBs depending on the environments. A BHB is defined ``hard'' if its binding energy $E_b$ exceeds the average field kinetic energy $\langle E_K\rangle$, i.e. if the semimajor axis is smaller than a limiting value $a_h$
\begin{equation}
a_h = \frac{GM_{\bhb,f}}{2\sigma^2} 
\end{equation}
As pioneered by \cite{heggie75}, hard binaries tend to become harder and harder as they interact with other stars, thus they are the most likely to survive in star clusters. Figure \ref{hard1} shows how $\Gamma_h$ varies with the BHB semi-major axis assuming a BHB mass of either $10$ (lower bound) or $30 \Ms$ (upper bound). Monte Carlo models of globular clusters suggested that the population of BHBs being processed via strong encounters and ultimately ejected from the cluster have semimajor axes that are linked to the cluster mass and half-mass radius \citep{rodriguez16}
\begin{equation}
a_e = \frac{\mu_{12}}{M_c}\frac{R_c}{k},
\label{eq:rod}
\end{equation}
where $k=54$ \citep{rodriguez16}, and $\mu_{12}=M_1M_2/(M_1+M_2)$ is the BHB reduced mass. Following \cite{arca20}, we can connect the cluster mass and half-mass radius to its velocity dispersion via the relation
\begin{equation}
{\rm Log} \frac{GM_c}{R_c} = \alpha + 2{\rm Log}\sigma,
\label{eq:arc}
\end{equation}
with $\alpha =1.14\pm0.03$, thus we can express the typical values of ejected BHBs semimajor axis in terms of the cluster velocity dispersion. Figure \ref{hard1} shows how $a_h$ and $a_e$ varies assuming a primary BH with mass $M_1 = 10\Ms$ and a cluster velocity dispersion $\sigma = 5,~15,~50$ km s$^{-1}$. If we consider $(a_e,a_h)$ as the limiting values of long-lived BHBs in dense clusters, we see that in dense nuclear clusters the tightest BHBs can have $a$ as low as $10^{-3}$ AU, while BHBs in ordinary globular clusters are expected to have $a$ in the range $10^{-2}-1$ AU.

To extrapolate our results for the sub-AU regime and extend our results to nuclear clusters, we rescale the length scale in SET10 by a factor $1/l$ where $l>1$. For instance, circular binaries in the rescaled set have $a_i = 1/l\au$. Note that the eccentricity distribution, as well as the distribution of all the other parameters, remain unaltered.
Assuming Newtonian dynamics, the rescaling is equivalent to changing the time unit and the gravitational constant for a given set of $N$-body simulations. We discuss in the following the impact of unstable triples evolution on GW astronomy.
In the following, we analyze the properties of the GW signal emitted by merging binaries in SET 10, scaling conveniently the results under the assumption of either $l=1$ or $l=100$. 
Note that the former, which implies $a_i=1\au$ for circular binaries, leads to values typical of low-density clusters, like open or young massive clusters, the latter, i.e. $a_i=0.01\au$, is more typical in GCs and galactic nuclei. Therefore, uncovering potential differences between these two kinds of merging BHBs can provide useful clues to infer the origin of observed GW sources.

The results of for both scaled and unscaled mergers in SET10 are summarized in Table \ref{eccen}.

\begin{table*}
\centering{}
\caption{Fraction of mergers with eccentricity > 0.1 in difference frequency bands for SET10}
\begin{center}
	\begin{tabular}{ccccccc}
	\hline
	\hline
	scaling factor $l$ & stellar environment & $10^{-3}$ Hz & $10^{-2}$ & $10^{-1}$ Hz & $1$ Hz & $10$ Hz \\
    		               &                     & LISA         &\multicolumn{2}{c}{DECIGO}& ET     & LIGO/Virgo\\
	\hline	
		 1             & loose star cluster  & $17 \%$      & $6\%$ & $4 \%$           &$0.2\%$ & $ - $ \\
		 100           & dense star cluster  & $92 \%$      & $30\%$ & $7\%$           &$2  \%$ & $ 0.5\%$ \\
	\hline
	\end{tabular}
\end{center}
\label{eccen}
\end{table*}

\subsection{GW from mergers in low-mass star clusters}

\begin{figure}
\includegraphics[width=\columnwidth]{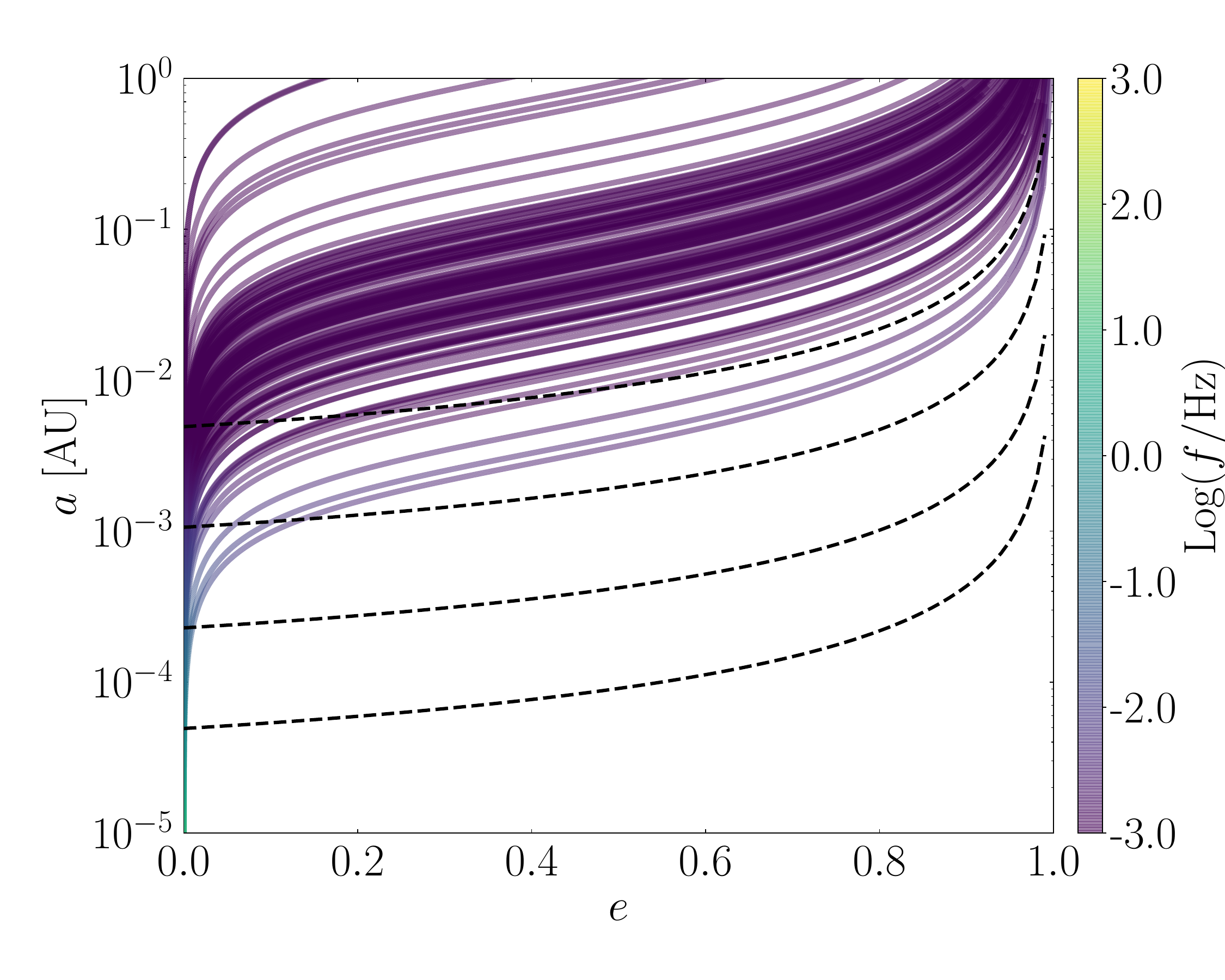}\\
\caption{Orbital evolution for merging BHBs in SET10. Each coloured track represents one model, with tracks moving from right to left and from top to bottom. The coloured map labels the BHB frequency.}
\label{fGW}
\end{figure}

\begin{figure}
\centering
\includegraphics[width=8cm]{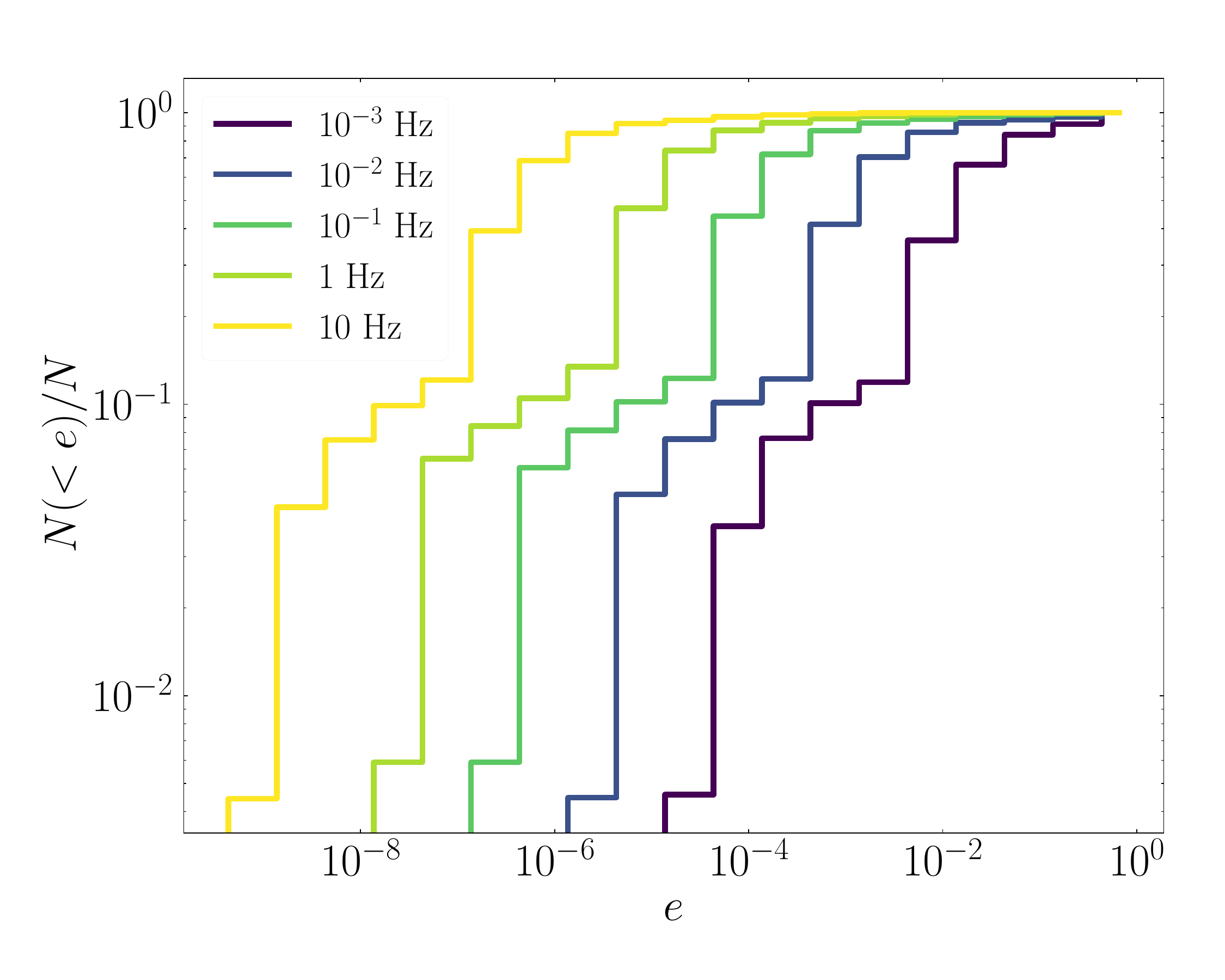}\\
\caption{Eccentricity cumulative distribution for all the merging BHBs in model 10. The $e_f$ values are calculated as the BHB crosses a given frequency, as indicated in the legend.}
\label{histoGW}
\end{figure}

In the following we focus on BHBs typical of low-mass star clusters, using results from SET10 and assuming $l=1$. This correspond to BHBs with an initial semimajor axis $a_i = 1$ AU, value typical for low-mass globular clusters and young massive clusters ($\sigma \simeq 5-10$ km s$^{-1}$). For all BHB mergers in SET10, we show the GW emission-driven evolution of eccentricity, semi-major axis and frequency in Figure \ref{fGW}. Each point in this plot provides us with the BHB orbital properties as it shifts toward different frequency bands. In order to understand how the distribution of eccentricities changes in different frequency bins and what are the implications for different GW detectors, for each BHB we record the eccentricity as the frequency crosses the $10^{-3} ~-~ 10^{-2} ~-~ 10^{-1} ~-~ 1 ~-~ 10$ Hz frequency tresholds. The distribution and cumulative distribution of eccentricities in different frquency bands is shown in Figure \ref{histoGW}. We find that none of our merging BHBs enter the $1 < f_\gw \lesssim 10$ Hz band with an eccentricity larger than $0.1$, thus suggesting that low-density star clusters might not be able to produce eccentric binaries visible in the LIGO \citep{ligo14}, Virgo \citep{virgo14}, and KAGRA \citep{akutsu18} observable window ($f_{\rm det} \gtrsim 10$ Hz). However, the smaller the frequency considered the larger the fraction $f_e$ of sources having eccentricity $e > 0.1$: we find $f_e \sim 0.2\%$ for sources with $f_{\rm det}\sim 1-10$ Hz, where the Einstein Telescope \citep{punturo10} will be sensitive, $f_e\sim 4-6\%$ at $f_{\rm det} \gtrsim 10^{-2}-{-1}$ Hz, the realm of decihertz observatories \citep{arca19c} like DECIGO \citep{Kawamura11}, and $f_e\sim 17\%$ in the range of $f_{\rm det} \gtrsim 10^{-3}$ Hz, the LISA observable window \cite{LISA17}. Table \ref{eccen} shows how eccentric mergers distribute across different frequency ranges.

\begin{figure}
\includegraphics[width=\columnwidth]{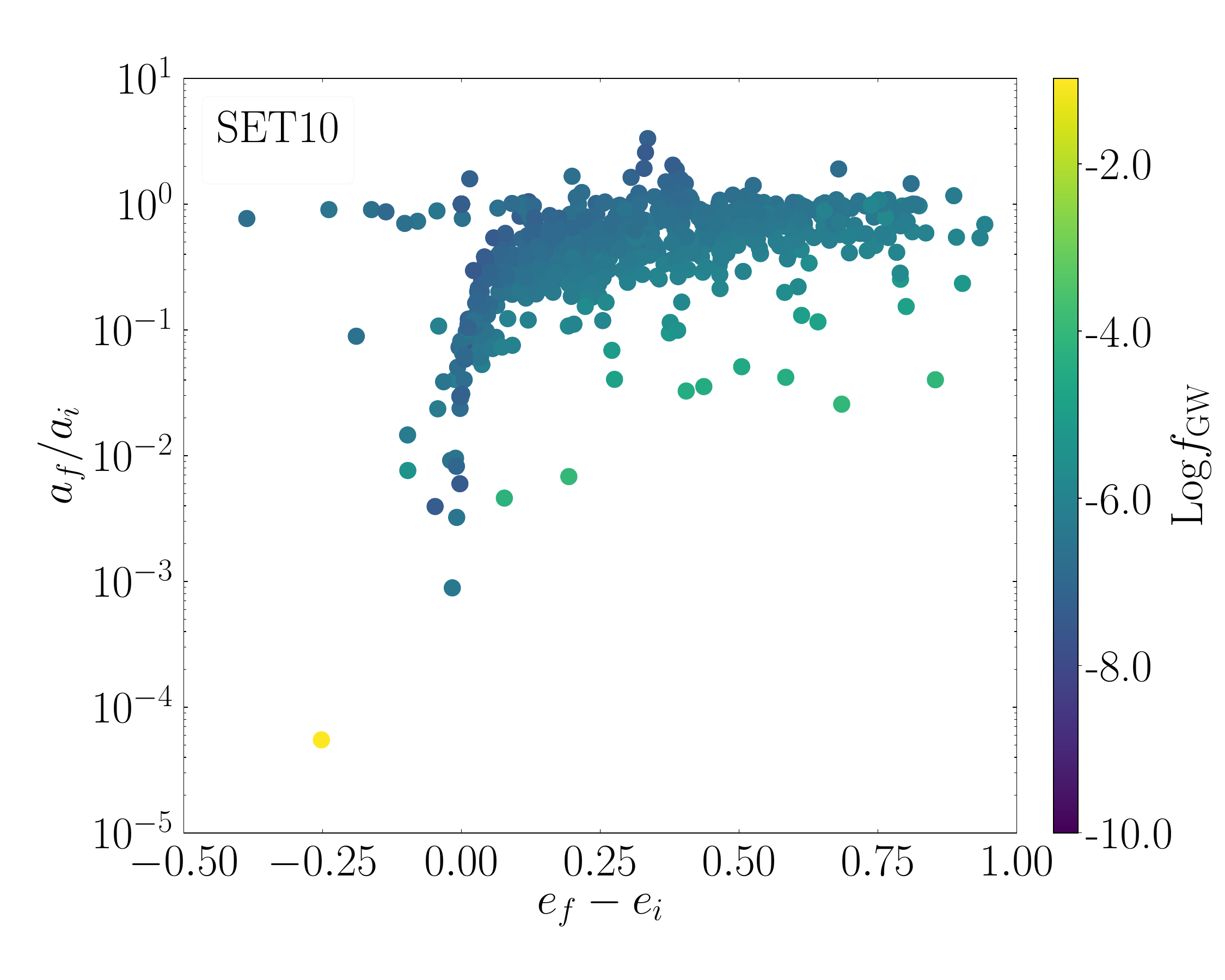}
\caption{Final to initial inner BHB semi-major axis ratio as a function of the difference between the final and initial BHB eccentricity. All the data points represent binaries that coalesce within 14 Gyr. The colour coded maps identify the binary frequency by the end of our simulations.}
\label{aredux}
\end{figure}

The absence of eccentric sources emitting in the 1-10 Hz band is essentially due to the triples initial orbital properties. This is clear from Figure \ref{aredux}, which shows, for the inner BHB, the ratio between the final and initial semi-major axis values and the difference between the final and initial eccentricity. Indeed, although the final BHBs tend to have significantly increased $e_i$ values, their semi-major axis decreases only by less than a order of magnitude in most of the cases, where $a_f \gtrsim 10^{-3}\au$ and $f_\gw \sim 10^{-3}-10^{-1}$ Hz. Interestingly, merging binaries draw a quite clear pattern in the $a_f/a$-$(e_f-e)$ plane, showing that the BHBs that shrink more ($a_f/a_i\lesssim 10^{-2}$) tend to maintain their initial eccentricity or circularize ($e_f-e_i<0$). Note that these quantities are extracted from the last snapshot of our simulations, thus they provide information on how stellar dynamics shape the BHB final properties.

\subsection{GW from mergers in dense star clusters}

\begin{figure}
\includegraphics[width=\columnwidth]{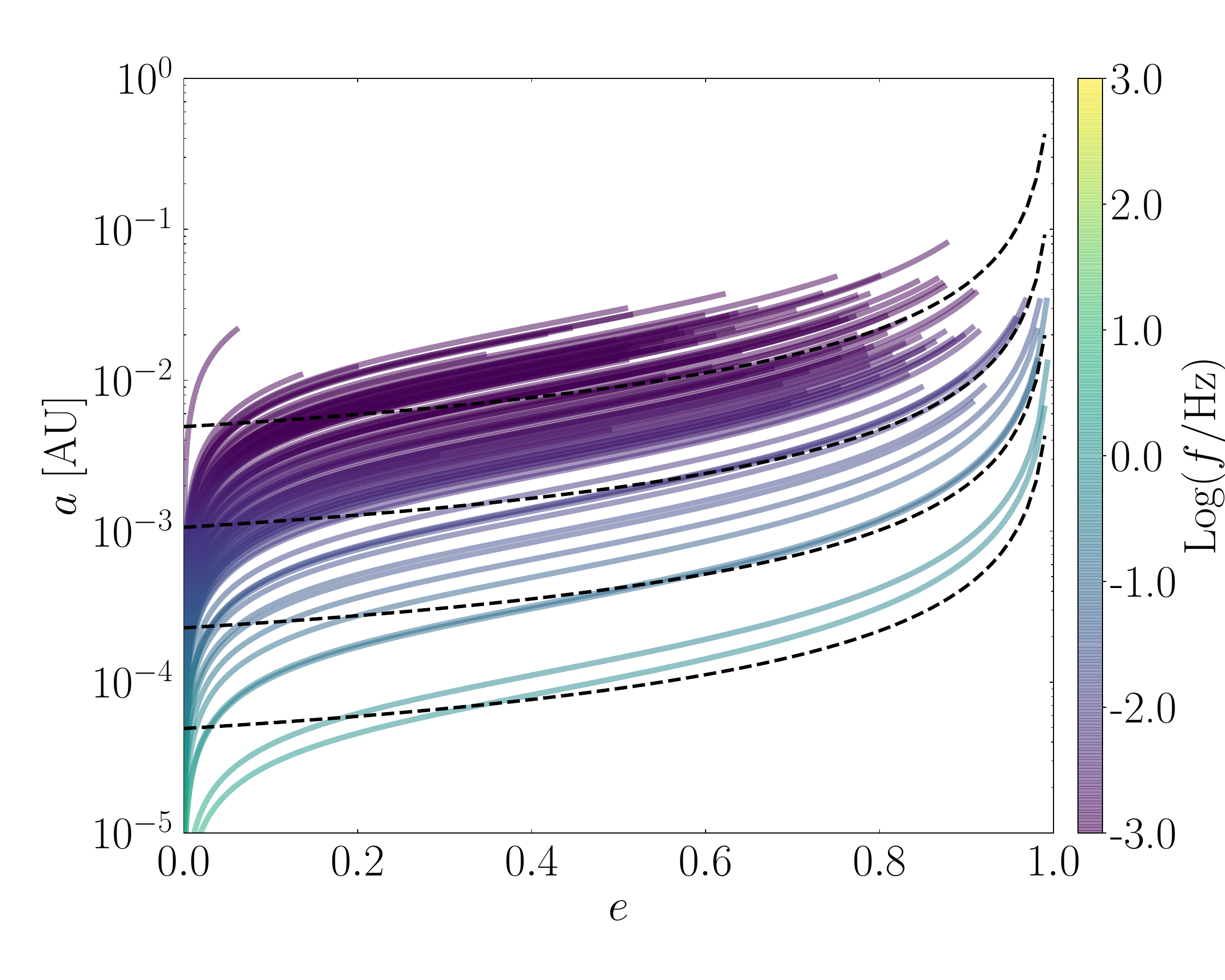}\\
\caption{As in Fig. \ref{fGW}, but assuming an initial BHB semi-major axis $a=0.01\au$ at zero-eccentricity.}
\label{fGWres}
\end{figure}

To explore whether harder initial BHBs can emit GWs in the Hz regime while having a large eccentricity, we assume in the following $l=100$.
This implies, for instance, that circular binaries in the rescaled set have $a_i = 0.01\au$, thus representing typical values of dense GCs and NCs. Note that the eccentricity distribution, as well as the distribution of all the other parameters, remain unaltered. This can be easily done in N-body simulations, provided that to the length scale rescaling is associated a corresponding rescaling of the time unit and the gravitational constant. 
In this case, as shown in Figure \ref{fGWres}, a sizeable number of binaries crosses the Hz band while having still a non-negligible eccentricity. 

\begin{figure}
\centering
\includegraphics[width=8cm]{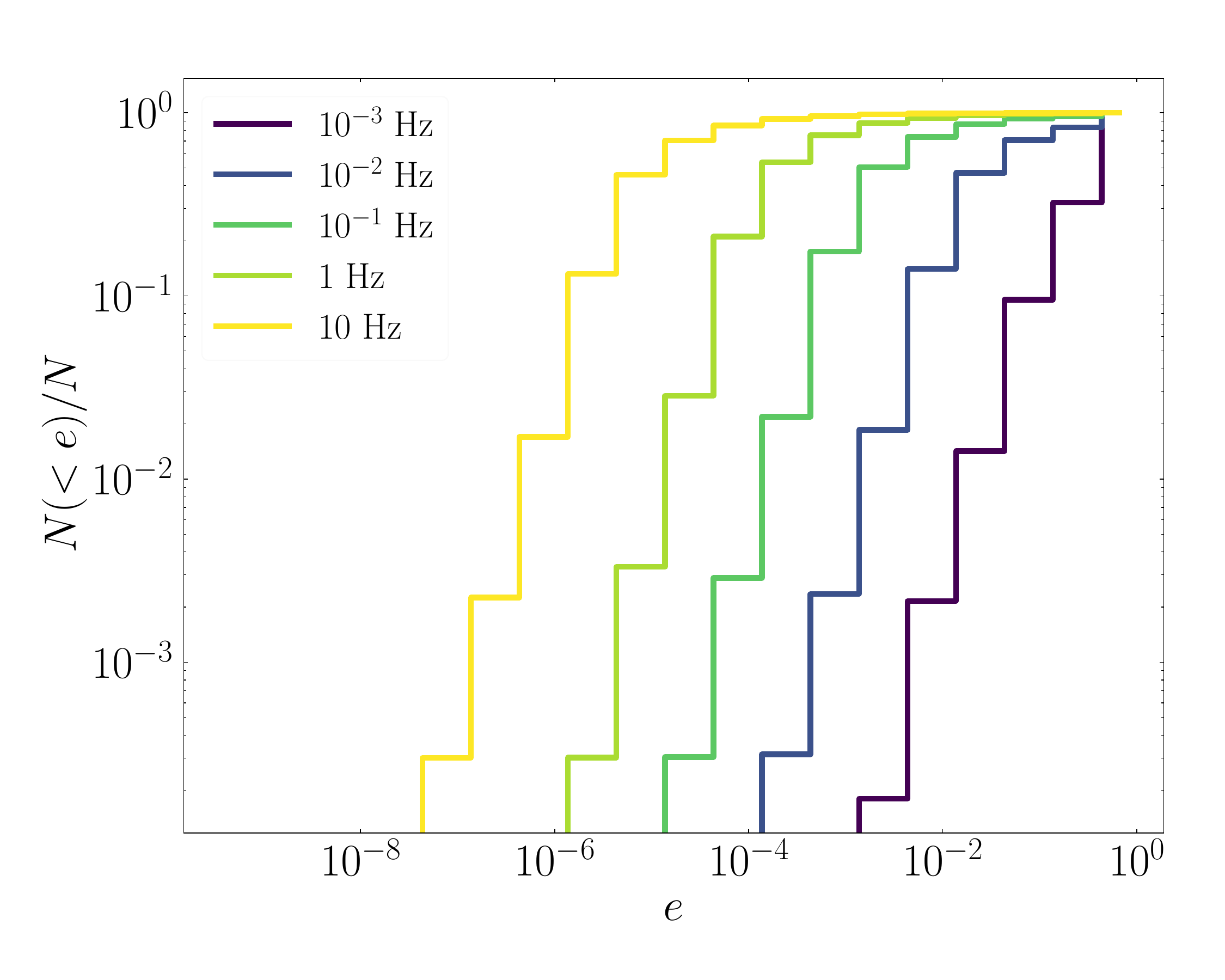}\\
\caption{As in Fig. \ref{histoGW}, but assuming an initial BHB semi-major axis $a=0.01\au$ at zero-eccentricity.}
\label{histoGWres}
\end{figure}

The eccentricity distribution and the corresponding cumulative distribution for our rescaled systems are shown in Figure \ref{histoGWres}. As summarized in Table \ref{eccen}, the percentage of eccentric sources increases from $0.5 - 2 - 7 - 30 - 92 \%$ at frequencies $10-1-10^{-1}-10^{-2}-10^{-3}$ Hz, thus almost all mergers appear to be eccentric in the LISA observational band. Our results suggest that hard BHBs ejected in non-hierarchical triples maintain a quite large eccentricity while passing through the $10^{-3}-10^{-1}$ Hz frequency range. Hence, these eccentric sources could be observed with DECIGO and LISA, but they become nearly circular by the time they enter the LIGO frequency band due to circularization due to GWs \citep{samsing18c}. Note that here we consider only mergers with GW time below 14 Gyr. On the other hand, BHBs with initially smaller semi-major axes ($a_i\lesssim 10^{-2}$) yield a large probability for mergers to enter the $f_\gw\sim 1$ Hz regime with large eccentricities, $e>0.7$, possibly being observable by the Einstein Telescope in the coming years. Approximately a third of the BHBs shift from 0.1 Hz toward the LIGO/VIRGO band at $\sim 10$ Hz, $70\%$ of them having eccentricities above 0.1 as shown in Figure \ref{fGWres}.
Note that this plot also suggests that sources possibly drifting toward the LIGO band form with initial eccentricities above $\sim 5\times 10^{-2}$, thus implying that they do not cross the LISA observational frequency range, as also suggested by \cite{samsing18b}.

\subsection{GW detectability}

Following \cite{kocsis12}, we calculated the characteristic amplitude $h_c$ for all the merging BHBs in SET10 in order to determine whether they are visible in low- or high-frequency detectors.
Differently from \cite{sesana16} assumptions, however, our BHBs have eccentric orbits. For sources which do not inspiral significantly during the observation time $T$, i.e. $T < f/\dot{f}_{p}$ the GW strain is composed of discrete harmonics with frequency
\begin{equation}
f_n = n \nu =  \frac{n}{2\pi} \left[\frac{G M}{a^{3}}\right]^{1/2}
\end{equation}
of width $\Delta f \sim 1/T$ each where $\nu = a^{3/2}/( G M )^{1/2}$
\citep{peters63}
\begin{equation}
h(a,e,t) = \Sigma_{n=1}^{\infty} h_n(a,e;f_n)\exp(2\pi i f_n t),
\end{equation}
where 
\begin{equation}
h_n(a,e;f_n) = \frac{2}{n}\sqrt{g(n,e)}h_0(a).
\label{amp}
\end{equation}
The condition on the inspiral timescale $T>f_{\rm p}/\dot{f}_{\rm p}$ can be calculated using for frequency and its derivative the relations \citep{peters64}, 
\begin{align}
f_{\rm p} &= \frac{ (G M)^{1/2} (1+e)^{1/2}}{2\pi a^{3/2}(1-e)^{3/2}} \label{pp}\\
\dot{f}_{\rm p} &=
\left(-\frac{3}{2}\frac{\dot{a}}{a} - k(e) \dot{e}\right)f_{\rm p}\\
k(e) &= \frac{1}{(1-e)^{1/2}(1+e)^{3/2}} - \frac{3}{2}\frac{(1+e)^{1/2}}{(1-e)^{2/3}},\\
\end{align}
where $\dot{a}$ and $\dot{e}$ are defined in Eq. \ref{peters2}. In the equations above, $h_0$ is the characteristic strain for a circular source \citep{oleary09,kocsis12,sesana16}
\begin{equation}
h_0(a) = \frac{\sqrt{32}}{5} \frac{G^2}{c^4} \frac{M_z\mu_z}{D a},
\end{equation}
$D$ is the distance of the source, $a$ the binary semi-major axis, $M_z=(1+z)(M_1+M_2)$ and $\mu_z=M_1M_2/(M_1+M_2)$ are the total and reduced binary masses, respectively, where $z$ is the cosmological redshift.

The function $g(e,n)$ in Equation \eqref{amp} is defined as \citep{peters63,oleary09,kocsis12,gondan18}
\begin{align}
g(e,n) =& \frac{n^4}{32}  [(J_{n-2}-2eJ_{n-1}+\frac{2}{n}J_n+2eJ_{n+1}-J_{n+2})^2  + \\ \nonumber 
        &  + (1-e^2)(J_{n-2}-2J_n+J_{n+2})^2 + \frac{4}{3n^2}J_n^2 ],
\end{align}
where $J_i \equiv J_i(x)$ is the i-th Bessel function evaluated at $x=ne$ \citep{peters63}. The dominant frequency harmonic correspond to $f_p = n_p (1-e)^{3/2}\nu$, with \citep{oleary09,kocsis12}
\begin{equation}
n_p(e) = {\rm ceil}\left[1.15\frac{(1+e)^{1/2}}{(1-e)^{3/2}}\right],
\end{equation}
$ {\rm ceil}[x] $ is the nearest integer largest than $x$, and the 90\% of the GW power is emitted at frequency between $0.2f_p$ and $3f_p$ \citep{oleary09}, while it rapidly decreases outward these limiting values.

For a non-inspiraling eccentric source, the characteristic strain in each frequency bin $\Delta f$ is given by \citep{kocsis12}
\begin{equation}\label{eq:hc}
h_c^2(a,e;f) = \sum_{n=1}^{\infty} h_0^2 fT \frac{4}{n^2}g(n,e)F(f-f_n),
\end{equation}
with 
$T$ the observation time and 
\begin{equation}
F(f-f_n) = \begin{cases}
1 & {\rm if} |f-f_n|<\Delta f/2 {\rm ~and~} fT>1,\\ \nonumber
0 & {\rm otherwise}. \nonumber
\end{cases}
\end{equation}

\begin{figure}
\centering
\includegraphics[width=\columnwidth]{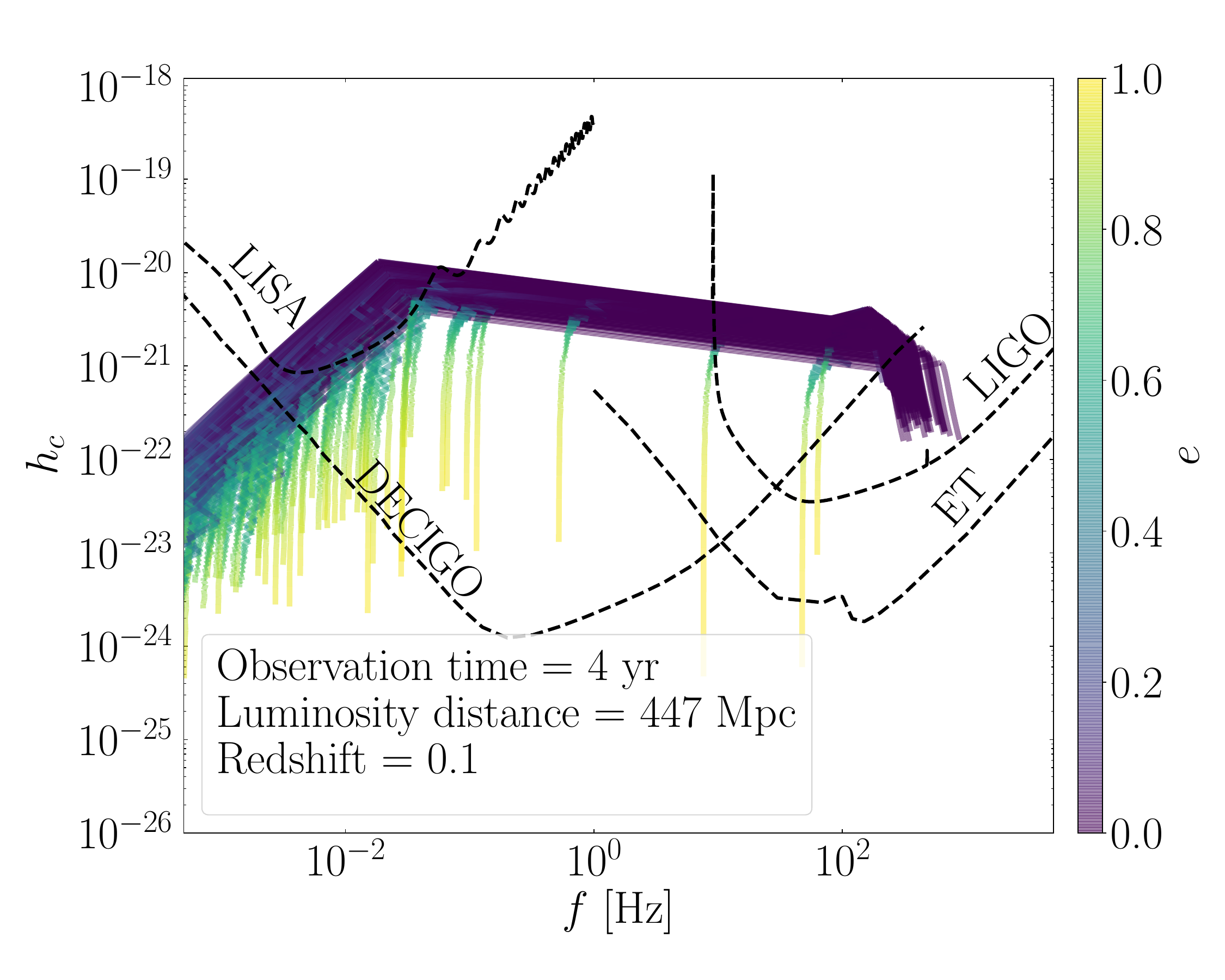}\\
\caption{Characteristic strain amplitude evolution for the dominant frequency of 150 merging BHBs in SET10, compared with the sensitivity curves for LISA, KAGRA, LIGO, DECIGO and the Einstein Telescope GW observatories. The coloured map identifies BHBs eccentricity. }
\label{sensi}
\end{figure}

Figure \ref{sensi} shows how the fundamental frequency, the corresponding characteristic strain and the eccentricity vary for 150 mergers in SET10. For clarity, we do not show other harmonics in the same plot, despite the fact that the signal is initially broadband. Nevertheless for each binary we do calculate the contribution of all harmonics which contribute $90\%$ of the total GW emitting power \citep{oleary09,kocsis12}. 
We assumed a $T=4\,$yr observation time and used equation~(\ref{eq:hc}) for a non-inspiraling source. However, for binaries with inspiral timescale in the range $f_{\rm p}/\dot{f}_{\rm p} < T$, we scaled the characteristic strain of each binary by a factor $\sqrt{(f_{\rm p}/\dot{f}_{\rm p})/T}$, in order to compensate for the fact that the binary has a limited lifetime at the characteristic frequency and the corresponding characteristic strain is suppressed. This is a conservative estimate for eccentric orbits, since in reality the binary circularizes and emits GWs not primarily only during pericentral passage, but throughout the orbit. 

The sources are assumed all at a luminosity distance $D_L = 477$ Mpc and redshift $z=0.1$. These tracks show approximately how the characteristic strain evolves as the binaries shrink and merge. The turning point evident in all the curves marks a time-to-merger of 4 yr. Note that LISA\citep{robson19} \citep{larson00}, LIGO\footnote{https://dcc.ligo.org/LIGO-T0900288/public}, KAGRA\footnote{http://gwcenter.icrr.u-tokyo.ac.jp/en/researcher/parameter}, DECIGO \citep[adapted from][]{yagi11} and the Einstein Telescope  \citep{ET08b,punturo10} sensitivity curves are also shown in Figure \ref{sensi}. 
Many sources fall in the observational LISA band at formation with high eccentricity. All stellar mass sources which merge within a Hubble time become potentially observable with LIGO, VIRGO, and KAGRA. These sources are more circular, in this case the characteristic strain in Figure \ref{sensi} follows $f^{-1/6}$. The final phase is described with a simple PhenomA prescription as described in \cite{robson19}. 

\begin{figure}
\centering
\includegraphics[width=\columnwidth]{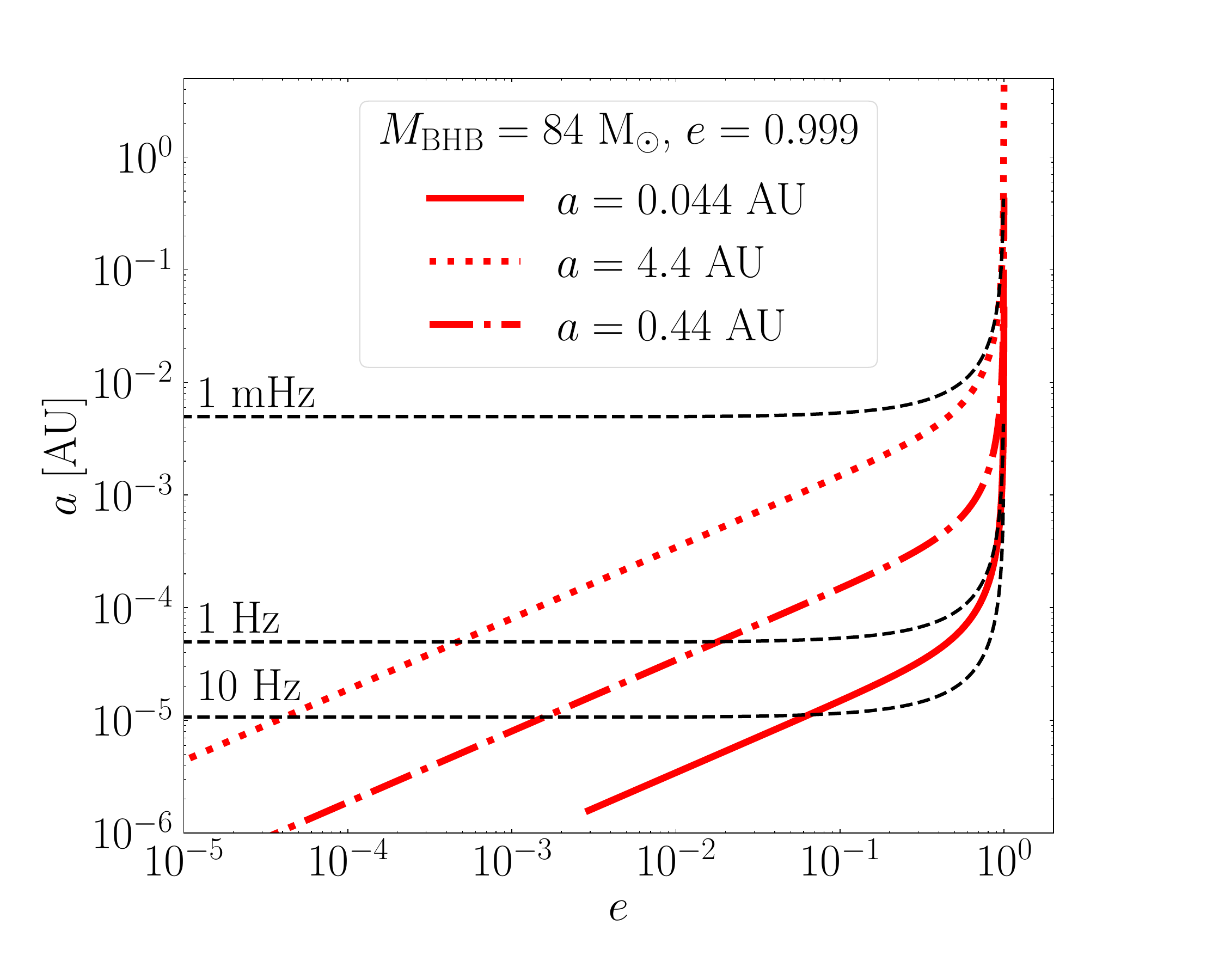}\\
\includegraphics[width=\columnwidth]{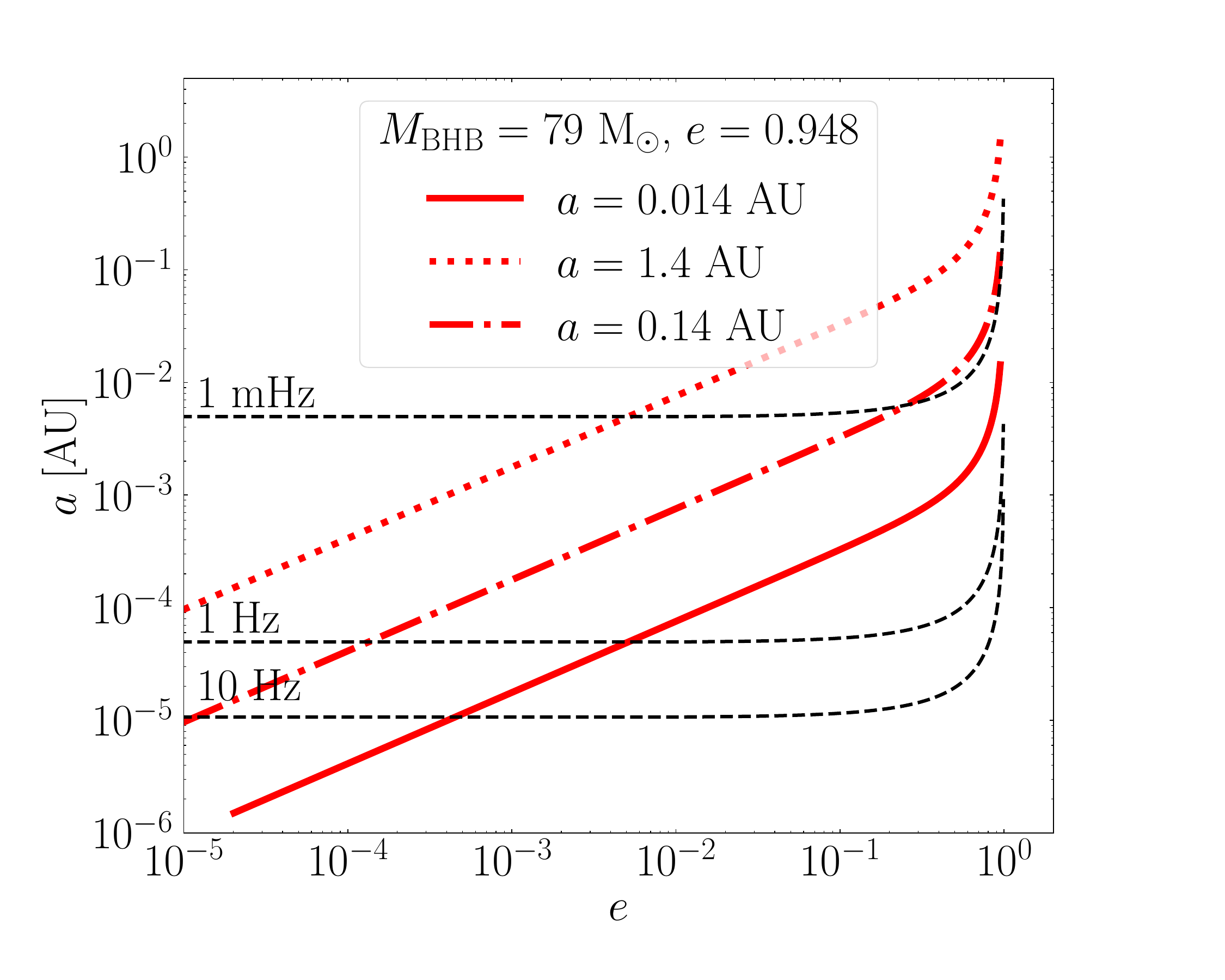}\\
\caption{Top panel: semimajor axis - eccentricity evolution for two inspiral BHBs with mass $M_\bhb = 80\Ms$, eccentricity $e = 0.999$, and corresponding semimajor axis $a = 0.044$ AU (dotted red line), $0.44$ AU (dotted-dashed red line), or $4.4$ AU (straight red line). The dashed black line marks the locus of ($a,e$) for which the peak frequency is $10^{-3}-1-10$ Hz. Bottom panel: same as above, but here the binary has a mass $M_\bhb = 79\Ms$, eccentricity $e = 0.947$, and semimajor axis $a = 0.014-0.14-1.4$ AU.}
\label{sensi2}
\end{figure}

Such a finding could help in disentangling the origin of observed GW sources. From Equation \ref{pp} we can calculate the semimajor axis of a BHB with a given eccentricity and emitting at a given frequency:
\begin{eqnarray}
a_{\gw} =& \displaystyle\frac{(GM_\bhb(1+e_\gw))^{1/3}}{(2\pi f_{\rm p})^{2/3} (1-e_\gw)} \simeq 0.003 {\rm AU} \times \\
         & \displaystyle\times \left(\frac{M_\bhb}{20\Ms}\right)^{1/3} \left(\frac{f_{\rm p}}{1{\rm mHz}}\right)^{-2/3} \frac{(1+e_\gw)^{1/3}}{(1-e_\gw)}, \nonumber
\end{eqnarray}
which implies that circular BHBs visible in the LISA frequency band has an $a_\gw$ value two/three orders of magnitude larger than those visible in the LIGO band. For reference, note that a circular BHB with mass $M_\bhb = 20\Ms$ emitting at a peak frequency of $1$ mHz has a semimajor axis $a_\gw = 0.003$ AU, while a BHB with an eccentricity $e_\gw > 0.999$ emitting at the same peak frequency has $a_\gw < 3.9$ AU. Figure \ref{sensi2} shows the evolution of semimajor axis and eccentricity during the BHB inspiral assuming two different setup: a) BHB mass  $M_\bhb = 79.4\Ms$, eccentricity $e = 0.947$ and semimajor axis $a=0.014-0.14-1.4$ AU; and b) BHB mass  $M_\bhb = 80\Ms$, eccentricity $e = 0.999$ and semimajor axis $a=0.044-0.44-4.4$\footnote{The values for $(M_\bhb,e,a)$ are taken from two mergers in SET10, we rescaled the semimajor axis assuming $l=10$ and $100$, respectively.}. The two panels highlight the importance of measuring with high precision the BHB eccentricity to get insights on the BHB properties at the time in which it "decoupled" from the dynamics of the nursing environment. For instance, measuring a BHB mass $\sim 80\Ms$ and an eccentricity $e_{\rm Hz} \sim 0.1$ at 1 Hz would imply that the BHB was extremely eccentric and tight at the decoupling ($a < 0.1$ AU and $e \sim 0.999$) thus suggesting an extremely dense birth-site. Such kind of tight and eccentric binaries would be undetected in LISA but might be clearly visible by DECIGO and DOs, as shown in Figure \ref{sensi}. Nonetheless, LISA might help in getting insights on the origin of BHB mergers that appear completely circular in LIGO and DECIGO. Observing a BHB with mass $\sim 75\Ms$ and $e \gtrsim 0.1$ at 1 mHz, for instance, would lead to decoupling parameters of $a \sim 0.1 AU$ and $e \sim 0.95$, while if the source appear circular in LISA its decoupling semimajor axis would increase by up to an order of magnitude. From Equations \ref{eq:rod}-\ref{eq:arc}, the typical semimajor axis of ejected BHB mergers in dense clusters can be written as:
\begin{equation}
a_e = 1.8\au \left(\frac{M_2}{80\Ms}\right)\left(\frac{1+q_\bhb}{2}\right)^{-1}\left(\frac{\sigma}{5{\rm km~s}^{-1}}\right)^{-2}.
\end{equation}
Comparing with Figure \ref{sensi2}, for a normal globular cluster ($\sigma \sim 5{\rm ~km~s}^{-1}$) a typical ejected merger with a mass $\sim 80\Ms$ and high eccentricity ($e\simeq 0.95$) have $a_e\sim 1.8$ AU and will appear already circular in LISA, whereas in the case of a nuclear cluster ($\sigma \sim 20{\rm ~km~s}^{-1}$) the same binary would have $a_e \sim 0.11$ AU, thus it might preserve a sizeable eccentricity while transiting in the LISA band. Although rough, this line of thinking provides a link between the observed GW signal, in particular the eccentricity, and the properties of BHB mergers forming dynamically in star clusters. LISA will provide us a view on eccentric mergers forming in low-velocity dispersion, i.e. low density, clusters, whereas LIGO, Virgo, KAGRA, and, in the future, ET, can give us insights on eccentric mergers developed in massive and dense globular clusters and galactic nuclei.

\section{Conclusion}
\label{sec:end}

In this paper we investigate the formation, evolution, and merger of BHBs formed in triples produced via BHB-BHB scattering in star clusters. Using 32,000 three-body simulations, we explored the parameter space by varying orbital parameters and initial configuration focusing the attention on the role played by the inclination of the outer perturber on the BHB evolution, in particular on the differences between prograde and retrograde configurations. Furthermore, we discuss the implications for BHB coalescence in the context of GW emission and detection. Our main results can be summarized as follows.

\begin{itemize}
\item we exploit the MOCCA Survey Database I -- a suite of Monte Carlo simulations of GCs -- to quantify the occurrence rate of BHB-BHB scattering in GCs. We find that, on average, GCs can harbour $10^3$ binary-binary scatterings over 12 Gyr lifetime, with denser clusters hosting more scattering events [Figure 1];
\item  we use 2,000 $N$-body simulations to model strong BHB-BHB scatterings at varying the host cluster velocity dispersion. Our simulations suggest that this class of interactions trigger the formation of either stable or temporary bound triples whose evolution can, in some cases, ultimately lead to the formation of a merging BHB. Around $(30-40)\%$ of triples formed this are unstable, while in several models the newly formed triple arranges in a hierarchical configuration that undergoes KL oscillations. Depending on the cluster velocity dispersion, in $0.6-10\%$ of the cases the BHB-BHB scattering and subsequent formation of short- or long-lived triples drives the formation of a BHB merger.  [Figure 2-6, Table 1];
\item we perform 1,000 simulations (SET0) of triples with initial conditions drawn accordingly to the BHB-BHB scattering experiments assuming a cluster velocity dispersion $15$ km s$^{-1}$. We find a BHB merger probability of $22.7\%$ either developing in hierarchical and mildy hierarchical ($20.8\%$) or unstable ($1.9\%$) triples. Under astrophysically motivated assumptions, we find that mergers developing this way have masses in the range $M_\bhb = 30-40\Ms$ and mass ratios peaking at $q_\bhb = 0.6$. Interestingly, we find that triples tend to evolve toward prograde configuration, being the retrograde ones prone to flip. Most of the mergers develop in triples characterised by an inclination close to $90^\circ$ [Figure 7-10];
\item for all mergers in SET0, we reconstruct the evolution of the orbital properties during the inspiral phase. We find that up to $20\%$ of the mergers retain an eccentricity above 0.1 when emitting GW at 1 mHz (the LISA observational domain), while the fraction of eccentric sources drops to $2\%$ if we consider sources emitting at 1 Hz. Around $75\%$ mergers developing from non-hierarchical triples are eccentric in the LISA band [Figure 11];
\item to investigate the role of the triple configuration in the development of mergers, we carry out a suite of 14,000 three-body simulations tailored to explore different properties of BH triples initially in non-hierarchical regime. Our results suggest that non-hierarchical triples share three evolutionary  
features: a) triple systems tend to evolve toward co-rotation, b) the orbital flip causes the formation of a tighter inner BHB, c) coplanar retrograde systems have 2.5 times more chance to decrease the inner BHB merger time by a factor 100 than coplanar prograde systems [Figure 17-18, Table 4-5];
\item we perform 16,000 three-body models to investigate the role of BH masses in the case of an inner BHB with semimajor axis either $a=1-20$ AU (8,000 simulations each). In the case of a loose BHB, we find only 15 mergers out of 8,000 simulations (merger probability $0.25\%$, while the number increased to 677 ($8.5\%$) in the case of an initially tighter binary. The majority of mergers (around 2/3) occurs in triples with an initially retrograde configuration. The merger time distribution follow a clear power-law $\mathrm{d} N / \mathrm{d} \ln t_{\gw,f} \propto t_{\gw,f}^{0.30\pm0.02}$ [Figure 14];
\item taking advantage of a convenient scaling of the simulations result, we explore the differences in the GW signal produced by merging binaries produced in triples with an inner BHB semimajor axis $a = 0.01-1$ AU. We find a non-negligible number of mergers that retain a significant eccentricity during the different phases of the inspiral. In the case of triples with an initially loose BHB, $\sim 17\%$ of mergers have eccentricity $> 0.1$ when the emitted GW signal has a peak frequency of 1 mHz, but only $0.2\%$ are eccentric in the Hz, and none of them in the 10 Hz. Initially tight BHBs, instead, appears eccentric in the mHz band in the $92\%$ of the cases, and in the 10 Hz band in the $0.5\%$ of the cases [Figure 22, 25, Table 6];  
\item we calculate the GW characteristic strain for merging BHBs and make a comparison with the sensitivity curve of several GW detectors, showing that the vast majority of mergers can be multiband sources, i.e. sources detectable with more than one detector [Figure \ref{sensi}]; 
\item we show that precise measurements of mergers eccentricity can enable us to dissect the merger evolutionary history. Depending on the BHB mass and the eccentricity measured in a given frequency band, we suggest that it is possible to argue whether the merger formed in a low- or high-velocity dispersion environment. For instance, a BHB mergers with mass $\sim 80\Ms$ observed in LISA with eccentricity $>0.1$ would suggest an orbital semimajor axis at "formation" of $a \sim 0.1$ AU, a value typical of mergers ejected in clusters with velocity dispersion $\sim 20$ km s$^{-1}$  [Figure \ref{sensi2}].

\end{itemize}

\section{Acknowledgements}

MAS acknowledges the Sonderforschungsbereich SFB 881 "The Milky Way System" (subproject Z2) of the German Research Foundation (DFG) for the financial support provided. MAS also acknowledges financial support from the Alexander von Humboldt Foundation and the Federal Ministry for Education and Research in the framework of the research project "The evolution of black holes from stellar to galactic scales". 
The authors warmly thank Abbas Askar and Mirek Giersz for useful discussions and for their help with MOCCA data handling and analysis. The authors thank Hagai Perets and Yann Gao for useful discussions.
This project benefited from support by the International Space Science Institute, Bern, Switzerland, through its International Team programme ref. no. 393 "The Evolution of Rich Stellar Populations and BH Binaries" (2017-18).
This project has received funding from the European Research Council (ERC) under the European Union's Horizon 2020 research and innovation programme ERC-2014-STG under grant agreement No 638435 (GalNUC) (to B.K.).

%\clearpage
\footnotesize{
\bibliography{paper}
}
\appendix

\section{Stability, hierarchy, and Post-Newtonian effects}

\begin{figure}
\centering
\includegraphics[width=\columnwidth]{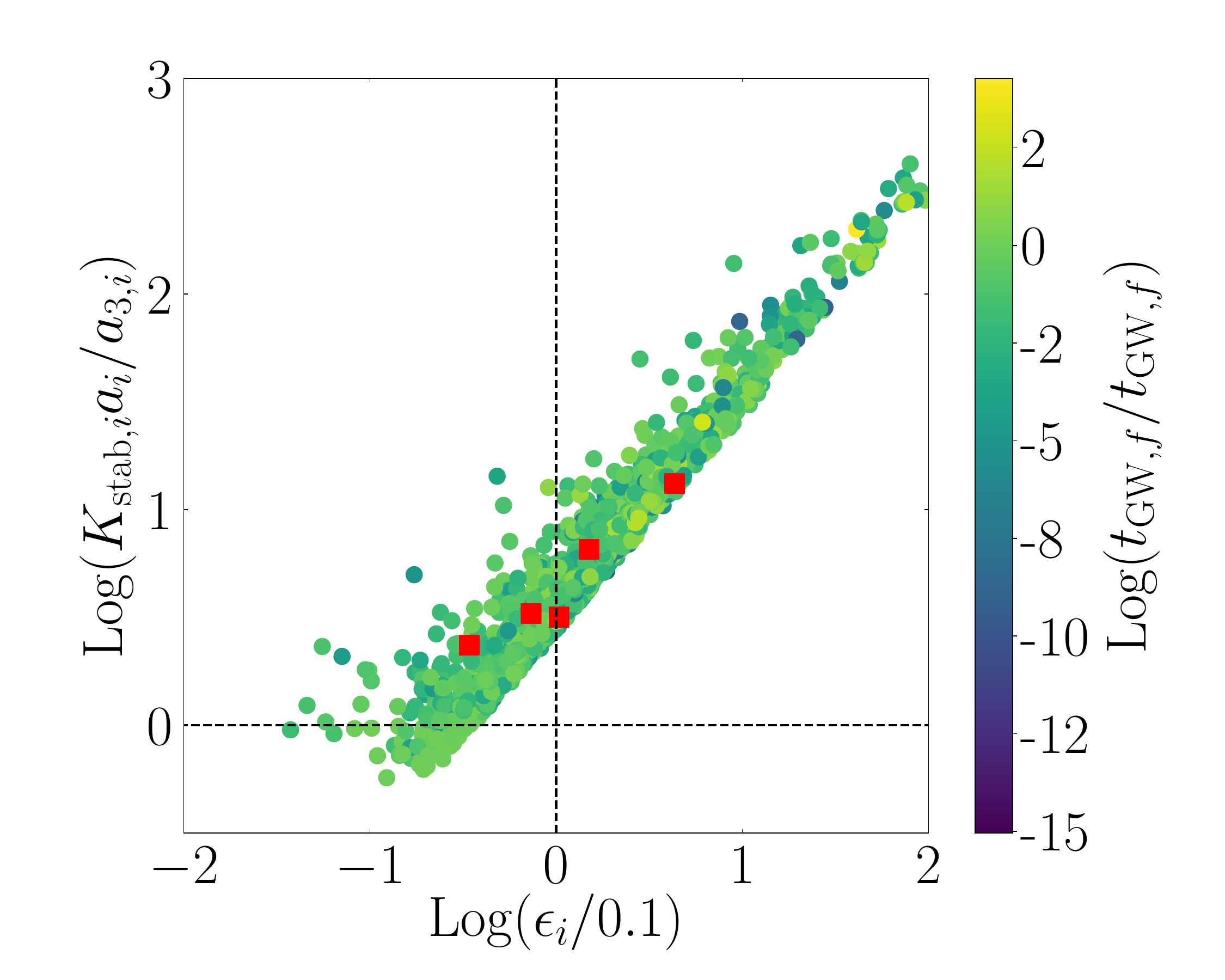}\\
\caption{
Top panel: initial value of the $K_{\rm stab}$ parameter normalized to the $a_3/a$ ratio (y-axis), and of the $\epsilon_{\rm KL}$ parameter normalized to 0.1 (x-axis) for models in SET4. The semiplane defined by $K_{\rm stab} < a_3/a$ - $\epsilon_{\rm KL} < 0.1$ contains triples in an initially stable configuration possibly prone to eccentric KL effects. Red squares mark the triples in which the inner BHB merges within 14 Gyr. The color coding marks the ratio between the initial and final value of the merger time, $t_{\gw,f}/t_{\gw,i}$.}
\label{octustab}
\end{figure}

\begin{figure*}
\centering
\includegraphics[width=8cm]{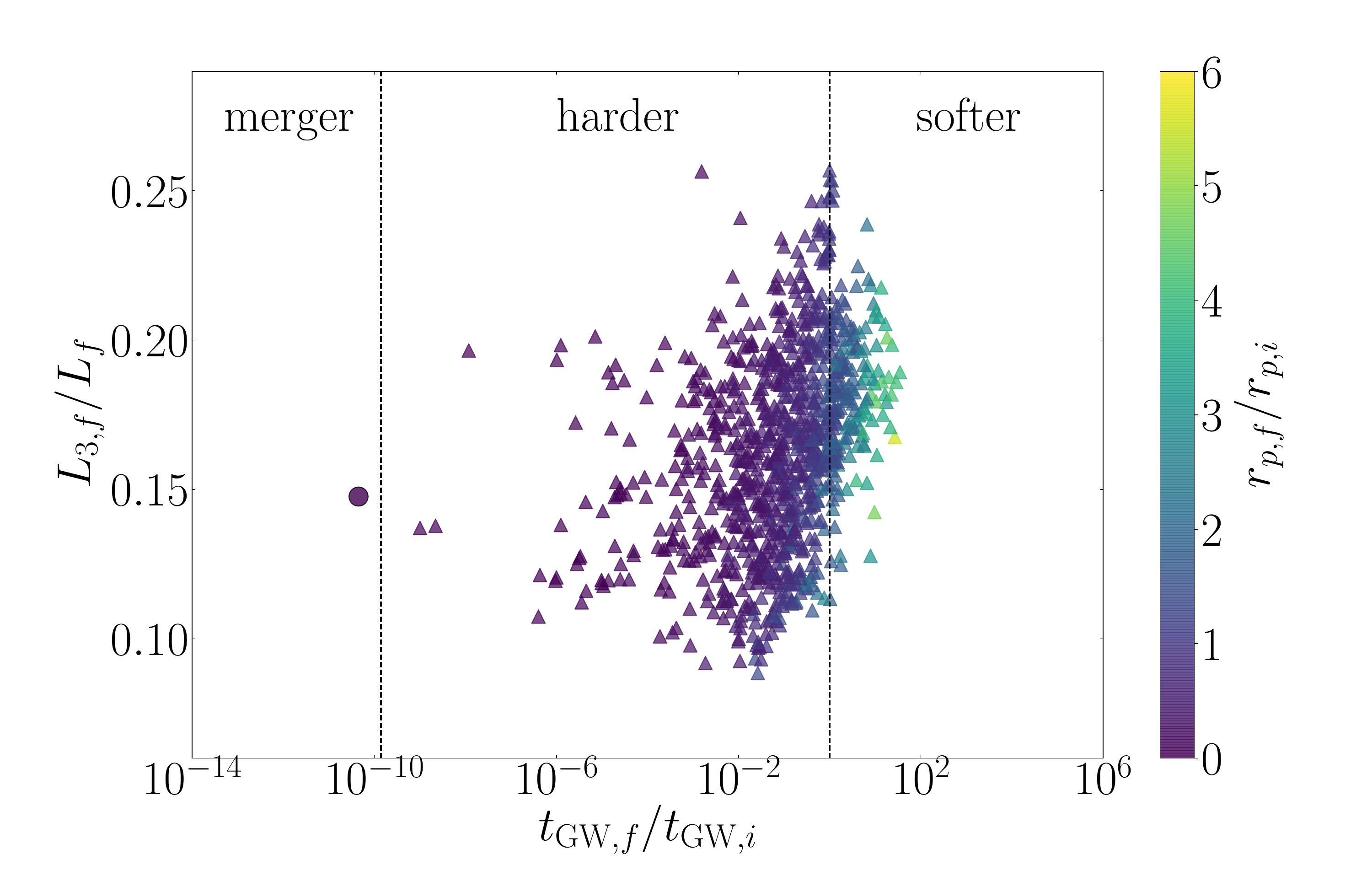}
\includegraphics[width=8cm]{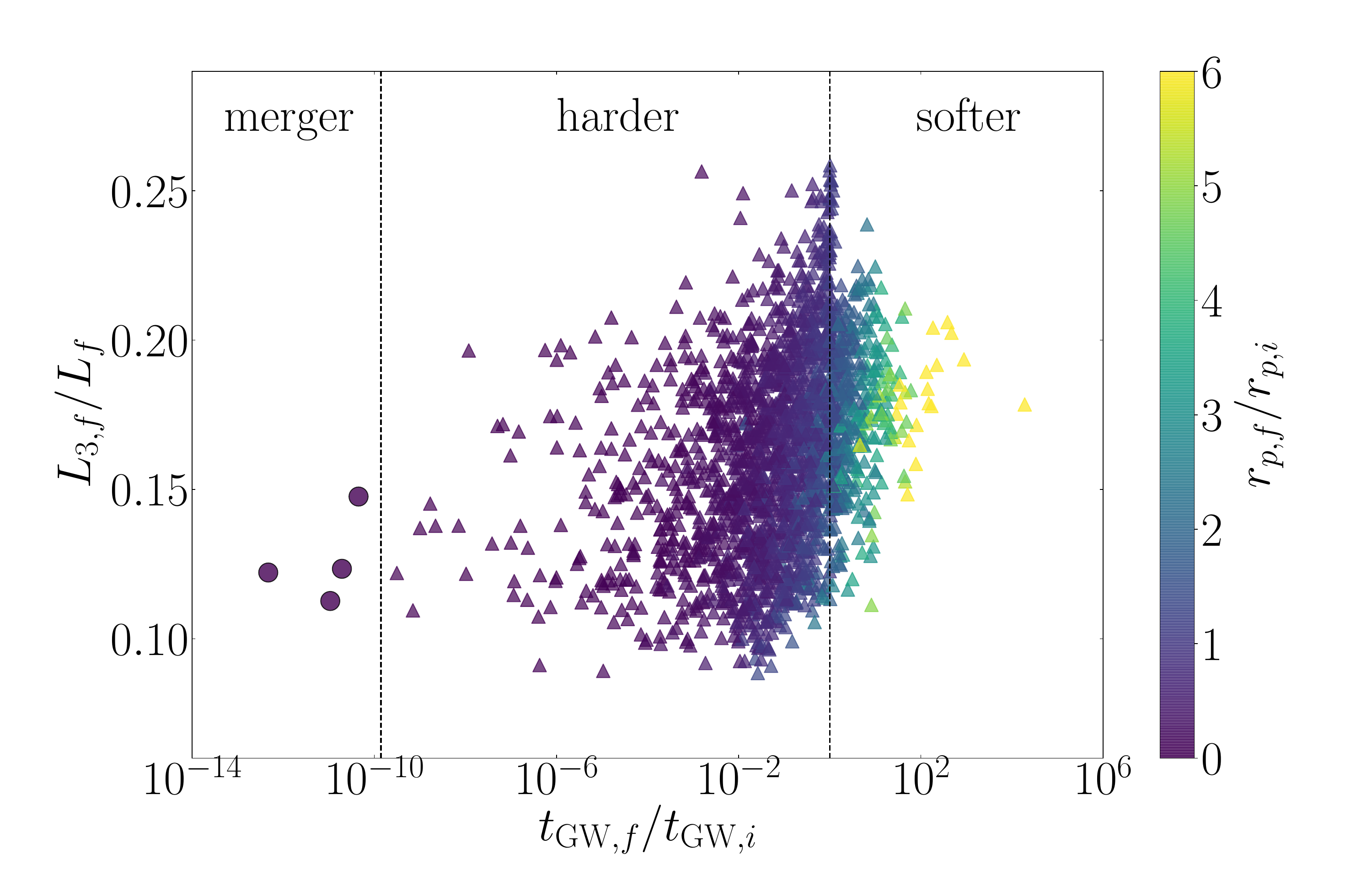}
\caption{The ratio between the angular momenta of the outer binary over that of the inner binary is compared to the ratio between the initial and final GW timescale of the inner binary for simulation SET4. The colour coding marks the ratio between the final and the initial pericentre for the inner binary. Left(right) panel refers to prograde(retrograde) models. The GW timescale may change significantly for a broad range of angular momentum ratios, but the distribution is more skewed to lower GW timescales for retrograde orbits.
}
\label{stab}
\end{figure*}

As we discussed in Section \ref{sec:GC}, a triple is, in principle, stable and prone to KL oscillations if the criteria given in Eq. \ref{eq1}-\ref{eq:epsilonKL} are fulfilled. However, while it is true that a stable system is generally hierarchical, the converse need not hold. Figure \ref{octustab} shows the quantities $K_{\rm stab}a/a_3$ and $\epsilon_{\rm oct}/0.1$ values for all the triples in SET4. Only $3.4\%$ of all models initially fulfill the two criteria, thus suggesting that the evolution of the majority of these triples is unstable and determined by chaos. Figure \ref{stab} shows the ratio between the initial value of the outer and inner binary angular momentum $L_{3,i}/L_i$ for all the initially retrograde (left panel) and prograde (right panel) models in SET4, as a function of the ratio between the final and initial value of the GW merger timescale. For each model, we also mark the ratio between the initial and final values of the inner binary pericentre. We find that the merger time decreases more, on average, for smaller values of the outer angular momentum compared to the inner binary and in particular retrograde models merge more easily than prograde ones.

\begin{figure}
\centering
\includegraphics[width=8cm]{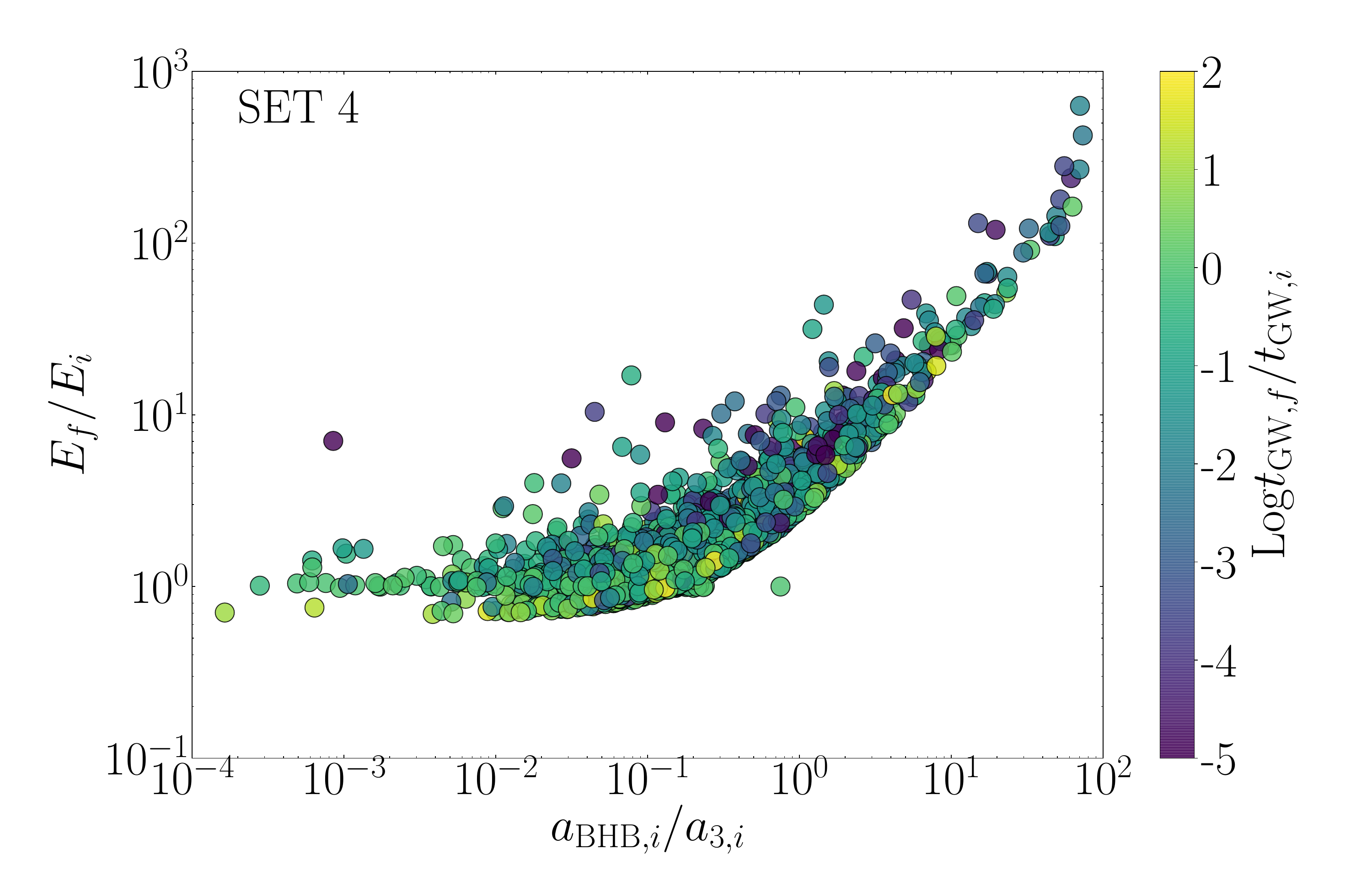}
\caption{Ratio between the final and initial value of the inner binary binding energy as a function of the ratio between the initial inner and outer semi-major axis. The colour-coding marks the ratio between the final and initial values of the coalescence timescale.}
\label{ebinds}
\end{figure}

This is in agreement with the qualitative arguments discussed in the Section~\ref{sec:num}. Figure \ref{ebinds}, which shows the final binding energy of the inner binary, normalized to its initial value, as a function of the ratio between the inner and outer semi-major axis at time zero. Figure \ref{ebinds} shows that the binding energy of the inner binary typically increases during the evolution by up to a factor 1000.

\begin{figure}
\includegraphics[width=0.8\columnwidth]{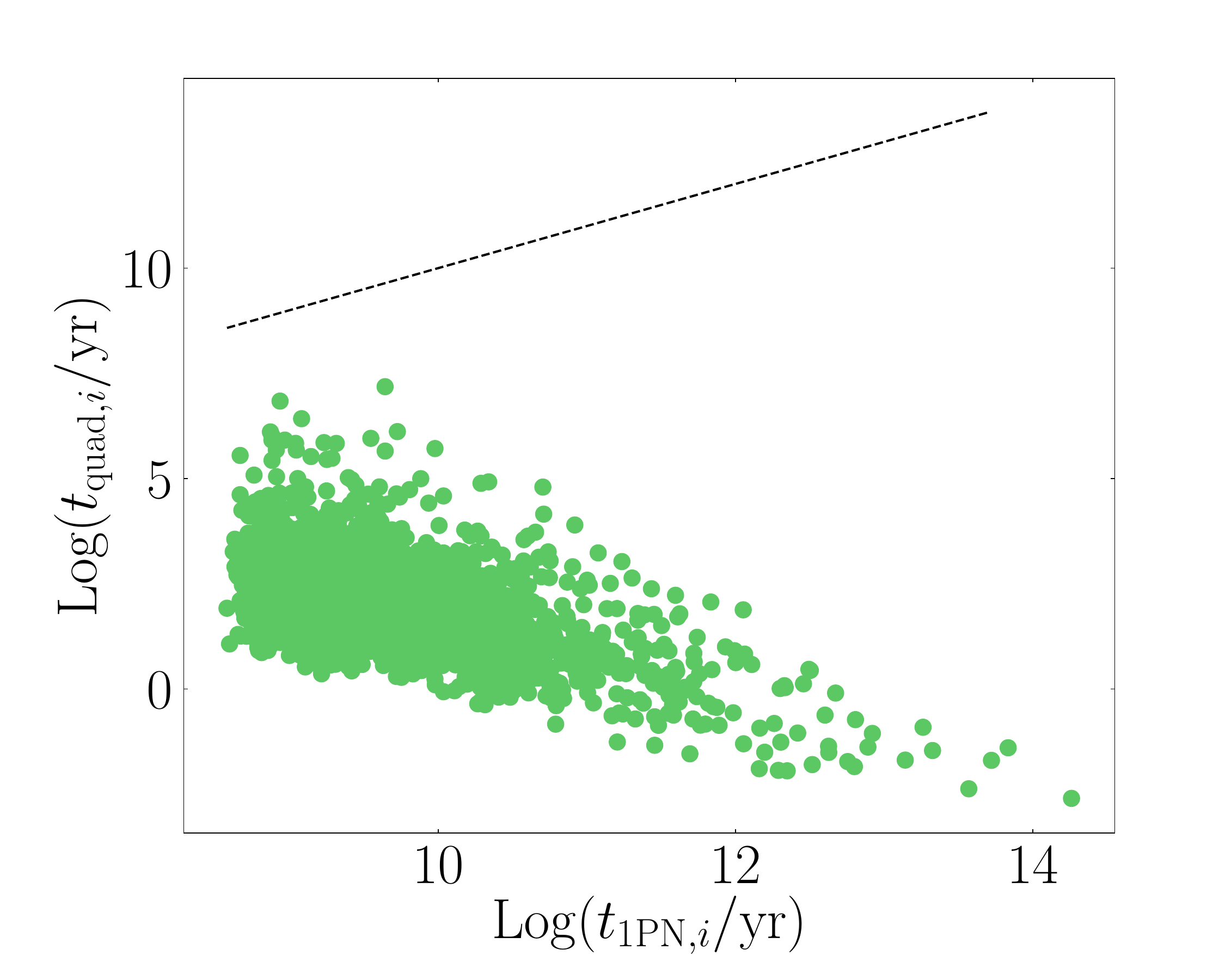}
\caption{Initial value of the precession time due to GR according to 1PN terms (y-axis) compared with the quadrupole timescale (x-axis) for all models in SET4. The dotted line represents the locus of points having $x=y$.}
\label{prec}
\end{figure}

If the inner BHB is sufficiently tight, general relativity (GR) effects have an important effect. However, we find that none of the triples in SET4 satisfy the condition for GR effects to be important. For instance GR precession can suppress KL oscillation if its typical timescale is shorter than the KL timescale \citep{hollywood1997,blaes02,antonini12,naoz13b,naoz16}, i.e.:
\begin{align}
\displaystyle
t_{\rm 1PN}  =& \frac{2\pi a_i^{5/2} c^2 (1-e_i^2) }{3 [G(M_{1,i}+M_{2,i})]^{3/2}} \\
t_{\rm quad} =& \frac{16}{30\pi} \frac{M_{1,i}+M_{2,i}+M_{3,i}}{M_{3,i}} \frac{P_{3,i}^2}{P_i} (1-e_{3,i})^{3/2}
\end{align}
In model SET4, none of the triples satisfies the limit $t_{\rm 1PN} < t_{\rm quad}$, as shown in Figure \ref{prec}, thus we expect that GR effects are negligible in affecting the BHB evolution in this type of BH triples. 

\section{Eccentricity evolution}

\begin{figure}
\centering
\includegraphics[width=8cm]{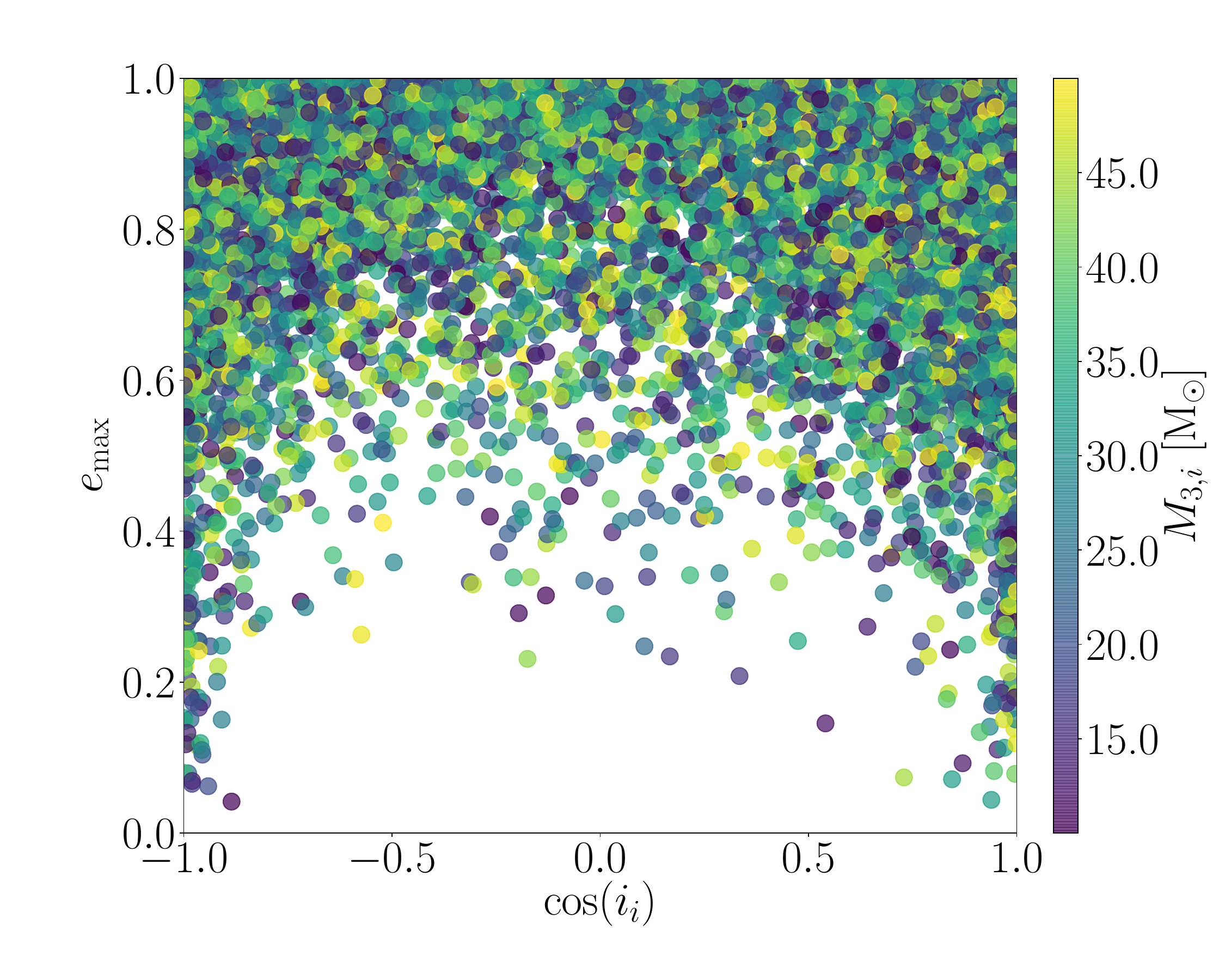}\\
\includegraphics[width=8cm]{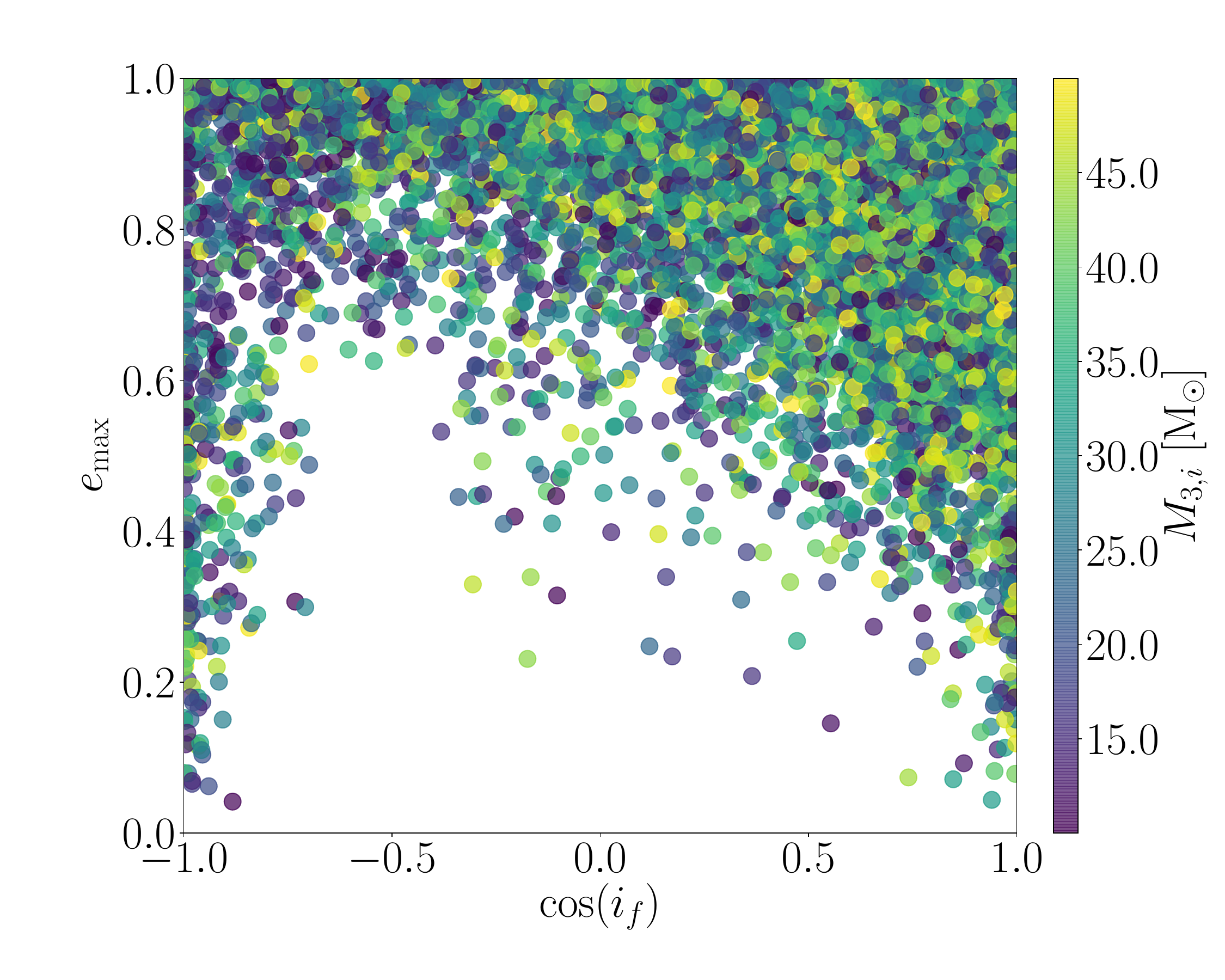}\\
\caption{Maximum eccentricity as a function of the initial (top panel) and final (bottom panel) inclination. The colour-coded map represents the mass of the outer BH.}
\label{emax1}
\end{figure}

\begin{figure}
\centering
\includegraphics[width=8cm]{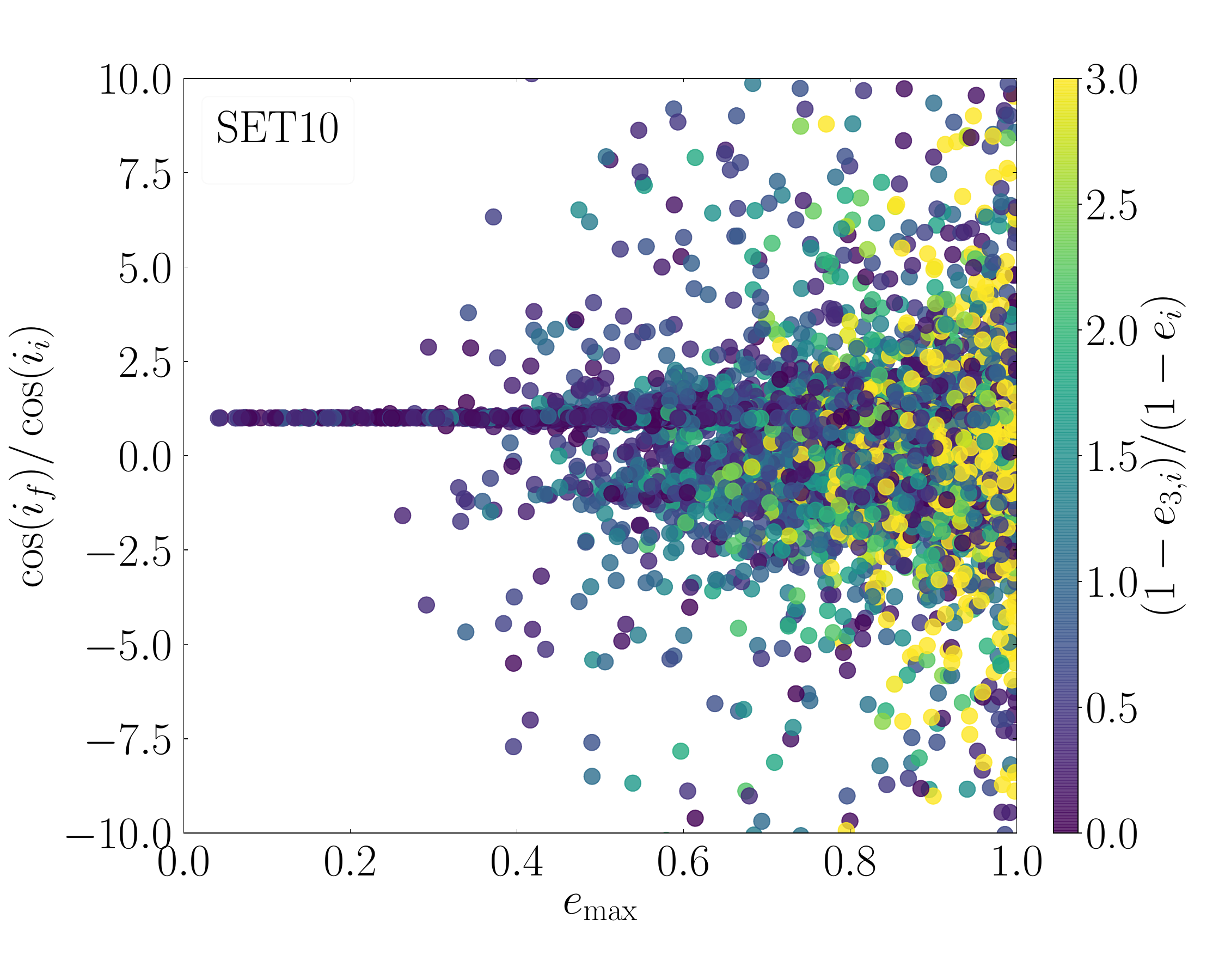}
\caption{Ratio between the initial and final value of the cosine of the inclination as a function of the maximum inner eccentricity. The colour coding identifies the ratio between the BHB final and initial values of $1-e$.}
\label{emax2}
\end{figure}

Another parameter that can be checked to obtain a clearer overview on the triple evolution is the evolution of the maximum value of the eccentricity $e_{\rm max}$. Figure \ref{emax1} shows how the maximum eccentricity varies with the final and initial inclination. The evolution of the triple leads to a decrease of models characterized by high inclinations and low eccentricities, leading to a more efficient population of high eccentricity states. 
Looking at Figure \ref{emax2}, which shows how the ratio between initial and final values of the inclination varies at increasing $e_{\rm max}$. Note that $e_{\rm max}\simeq 0.2$ marks a threshold above which the triple configuration changes significantly over time and undergoes, in some cases, a flip. Note that $e_{\rm max}$ is linked to the ratio $(1-e_3)/(1-e)$. 

\begin{figure}
\centering
\includegraphics[width=8cm]{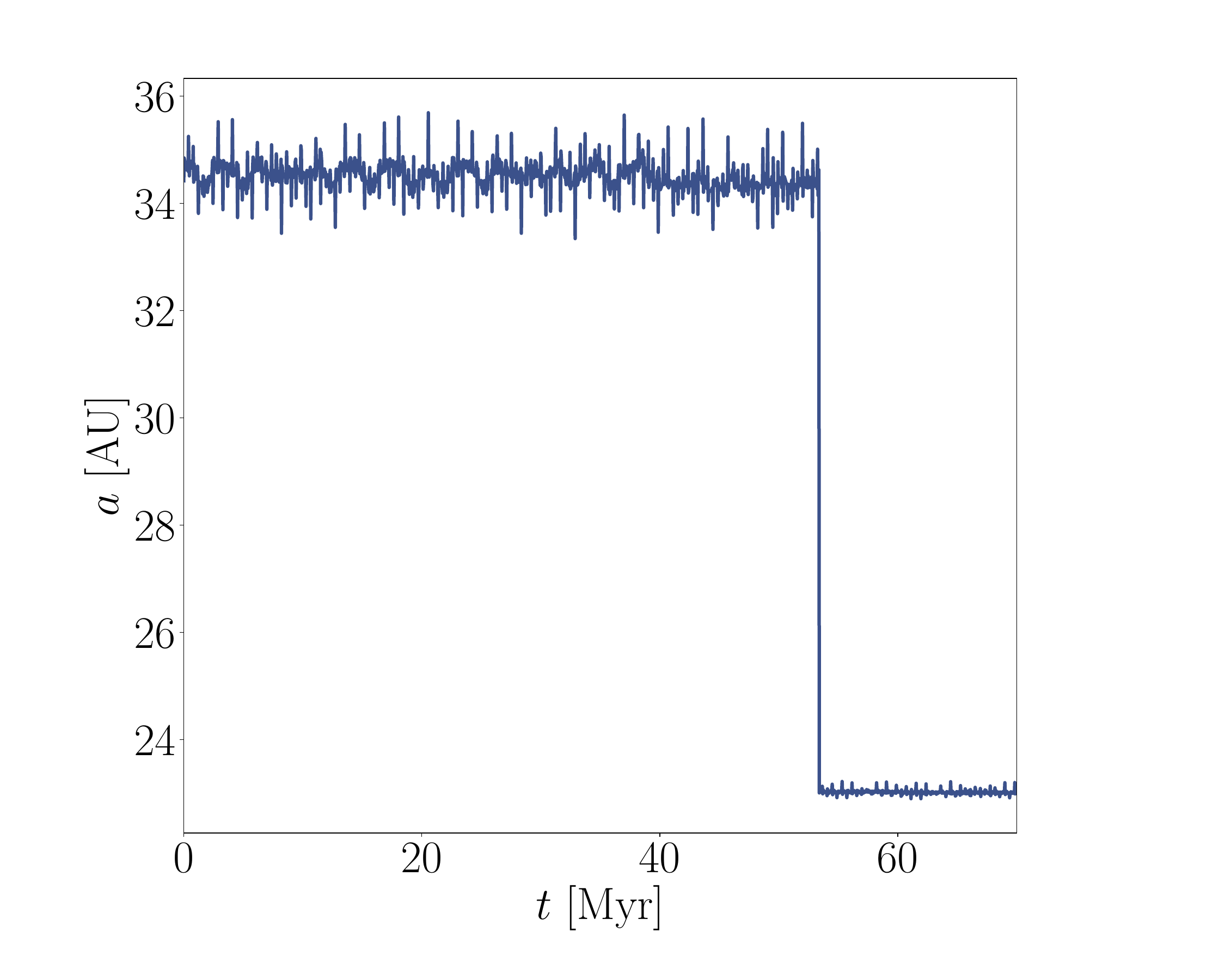}\\
\includegraphics[width=8cm]{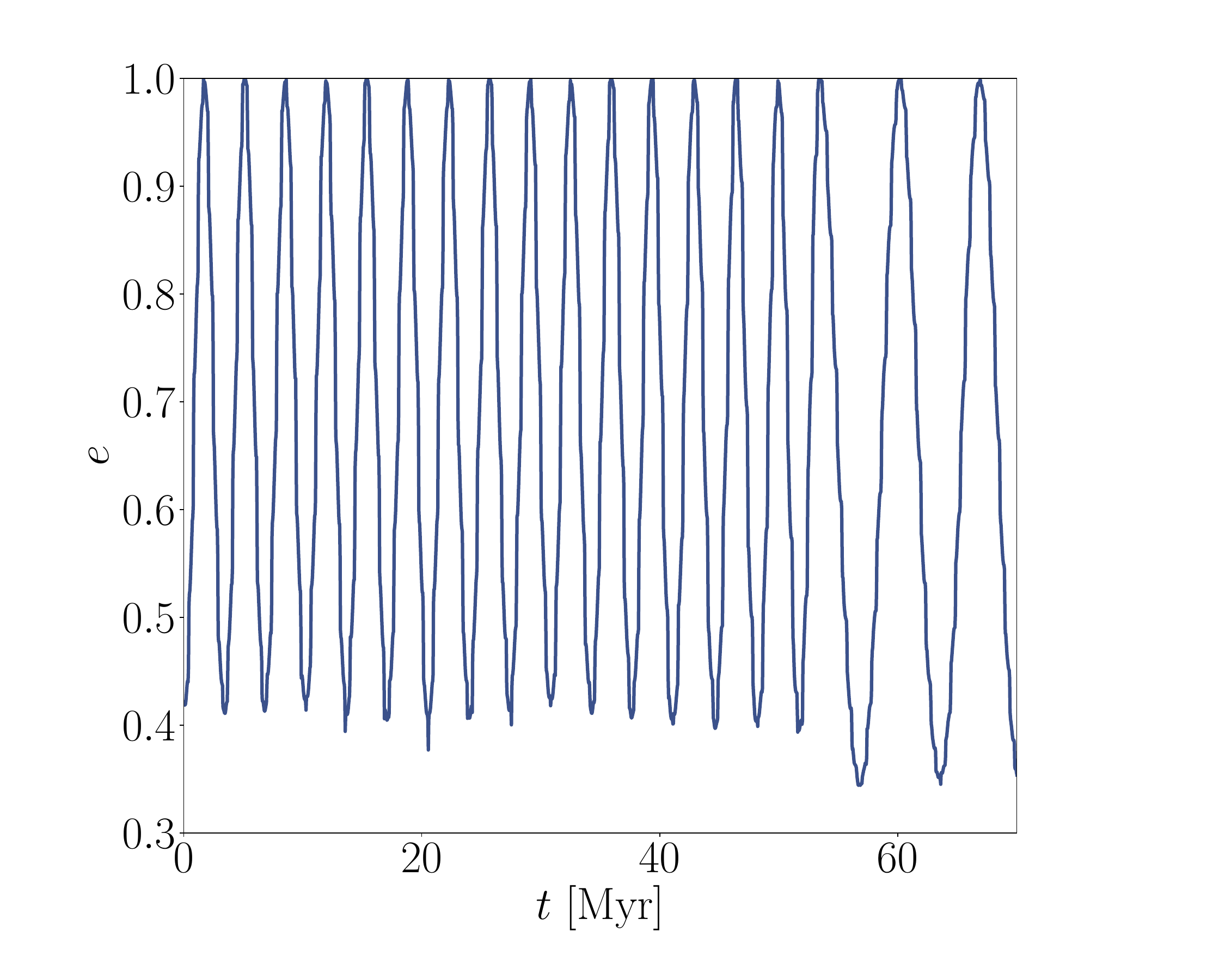}\\
\caption{Top panel: time evolution of $a$ in simulation No. 593 in SET4. Central panel: same as in top panel, but here is represented the eccentricity.Note the moment in which the shrinking efficiency peaks at $\sim 0.05$ Myr, after which the semi-major axis drops, the minimum eccentricity decreases and the eccentricity variation period increases until the merger occurs at 0.07 Myr.} 
\label{593}
\end{figure}

\begin{figure}
\centering
\includegraphics[width=8cm]{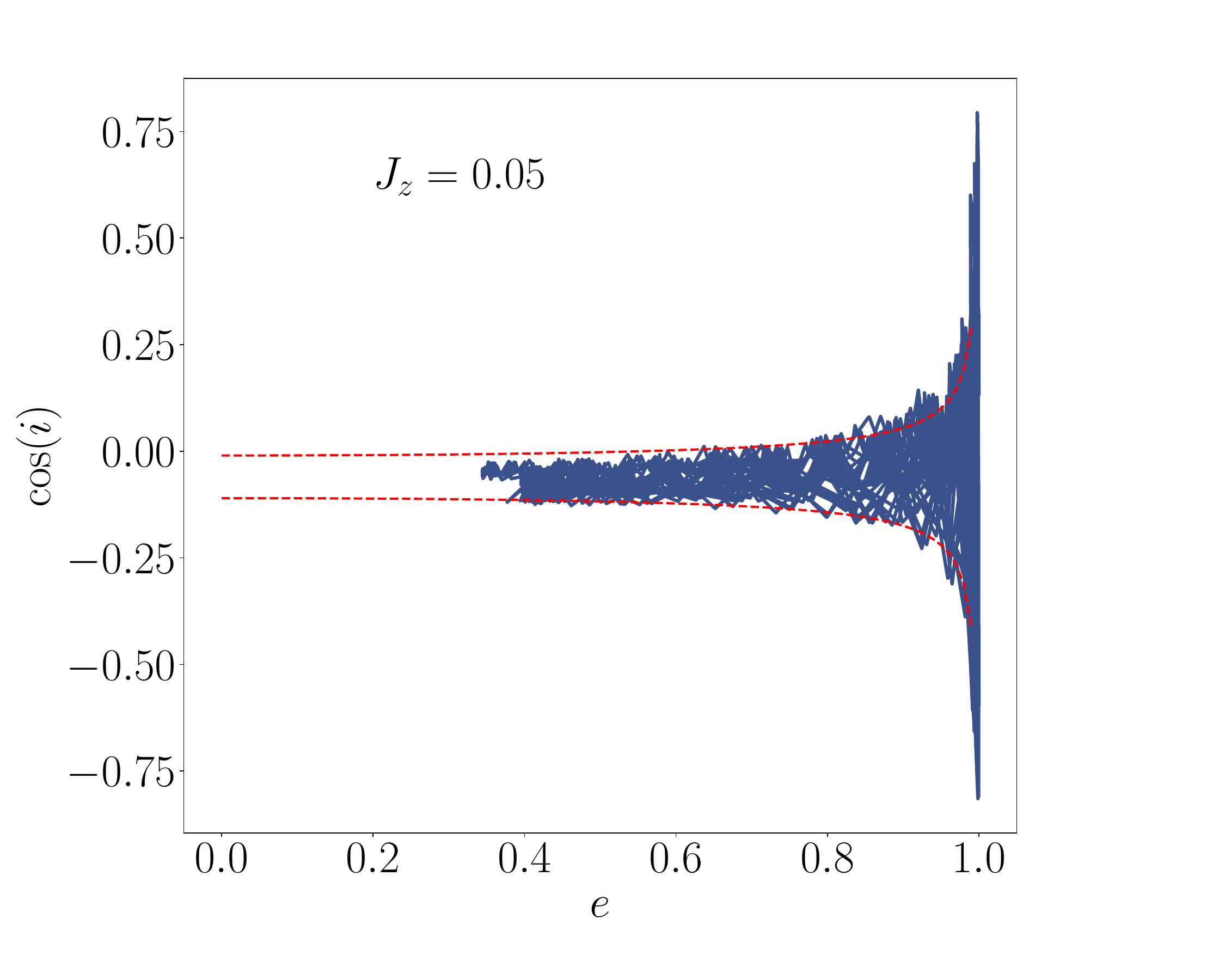}
\caption{Inclination as a function of the eccentricity in model No. 593 in SET4.}
\label{incli593}
\end{figure}

Figure \ref{593} shows the time evolution of one of the simulation in SET4. Here the outer binary orbit has initial inclination $i=90^{\circ}$ and eccentricity $e_3\sim 0.35$, while the inner binary has eccentricity $e\simeq 0.4$. The ratio between the inner and outer semi-major axis is, in this case, $a/a_3 = 0.17$. 
The perturbation induced by the outer objects impinges an oscillation in $e$, whose periodicity roughly doubles in the last stage preceding the merger event. The merger is extremely fast, taking place within $\sim 5.6\times 10^4$ yr. Also the inclination varies significantly until the merger.
The inner binary evolution draws a clear pattern in the $e$-$i$ plane, as shown in Figure \ref{incli593}, with the inclination that continuously flips from a co- to a counter-rotating configuration as the inner eccentricity approaches the unity.
The boundaries of Figure \ref{incli593}, showing how the inclination and eccentricities of the inner binary relate each other, are obtained assuming a constant value for the z component of the angular momentum $J_z = J_0\pm\sqrt(1-e^2)\cos(i)$, as expected for standard Kozai-Lidov oscillations. Note that our model start initially on a nearly perpendicular configuration ($\cos(i)\sim 0$).

\end{document}